\documentclass[12pt]{article}

\usepackage{graphicx}
 
\usepackage{amssymb}

\newcommand{\hodge}{\mbox{\rotatebox[origin=cc]{0}{$\bigstar$}}}

\begin{document}

\title{{\bf Symplectic Dirac-K\"ahler Fields}
  \raisebox{3cm}[0mm][0mm]{ 
\begin{minipage}[b]{0cm}
  \normalsize \noindent \mbox{\hspace{-0.5cm}
        MZ-TH/99-43
  }
  \\ \noindent
  \mbox{\hspace{-0.5cm}
       hep-th/9910085
   }
\end{minipage}}}

\author{\\ \\ \\ \sc M. Reuter}

\date{}

\maketitle

\vspace{-1cm}
\begin{center}
  {\it Institut f\"ur Physik, Universit\"at Mainz\\ Staudingerweg 7,
    D-55099 Mainz, Germany\\E-mail: reuter@thep.physik.uni-mainz.de}
\end{center}

\vspace{1cm}
\thispagestyle{empty}

\begin{abstract}
  For the description of space-time fermions, Dirac-K\"ahler fields
  (inhomogeneous differential forms) provide an interesting alternative
  to the Dirac spinor fields. In this paper we develop a similar concept
  within the symplectic geometry of phase-spaces. Rather than on
  space-time, symplectic Dirac-K\"ahler fields can be defined on the
  classical phase-space of any Hamiltonian system. They are equivalent
  to an infinite family of metaplectic spinor fields, i.e. spinors of
  ${\rm Sp}(2N)$, in the same way an ordinary Dirac-K\"ahler field is
  equivalent to a (finite) mulitplet of Dirac spinors. The results are
  interpreted in the framework of the gauge theory formulation of
  quantum mechanics which was proposed recently. An intriguing analogy
  is found between the lattice fermion problem (species doubling) and
  the problem of quantization in general.
\end{abstract}

\newpage

\renewcommand{\theequation}{1.\arabic{equation}}
\setcounter{equation}{0}

\section{Introduction}
In a classic paper \cite{ka} E. K\"ahler proposed a description of
fermions in terms of inhomogeneous differential forms. Rather than by
spinor fields, the fermions are represented by a set of antisymmetric
tensors in this approach. The role of the Dirac equation is taken over
by the so-called Dirac-K\"ahler equation which involves only tensor
manipulations. It imitates the $\gamma$ - matrix algebra with the help
of the Clifford product for forms.

At first sight it seems puzzling how a family of tensor fields
carrying integer spin can describe a particle of half-integer spin.
This paradox is resolved if one notes that (in 4 space-time
dimensions) a single Dirac-K\"ahler field actually corresponds to a
multiplet of 4 ordinary Dirac spinors which mix under Lorentz
transformations in a nontrivial way (``flavor mixing'').

The Dirac-K\"ahler fermions have attracted a lot of attention both from
the physics \cite{banks} - \cite{dil} and the mathematics \cite{at,graf}
point of view. In particular they made their appearance in lattice field
theory \cite{mm}. It is a well known problem that a straightforward
lattice discretization of the ordinary Dirac action does not describe
one but rather 16 fermions in the continuum limit. The reason for this
replication of fermionic states (usually refered to as the species
``doubling'' problem) is that the lattice propagator in momentum space
has poles at all 16 corners of the Brillouin zone. The Kogut-Susskind
\cite{kosu} or staggered lattice fermions were proposed as an attempt to
solve this problem. They are based on a more sophisticated lattice
action which reduces the number of fermion species from 16 to 4. Later
on it turned out \cite{banks,bj} that the Kogut-Susskind fermions are
nothing but Dirac-K\"ahler fields discretized on a hypercubic lattice.
As it deals with differential forms only, Dirac-K\"ahler theory on the
lattice can take advantage of all the mathematical tools provided by the
algebraic topology of cell complexes. In particular, by a standard
procedure, the differential forms of the continuum formulation can be
replaced by appropriate cochains on the lattice. These cochains are
functions defined on the lattice points, links, plaquettes, cubes and
hypercubes of the underlying lattice. In this manner it becomes obvious
that the extra fermion species implied by the Kogut-Susskind lattice
action and the fact that a Dirac-K\"ahler field contains 4 ordinary
Dirac fermions have a common origin.

We only mention that the species doublers on the lattice can be avoided
completely by using Wilson fermions or the nonlocal ''SLAC derivative''
\cite{mm}, for instance. Alternatively one can regard the 4 Dirac
fermions contained in one Kogut-Susskind field as 4 different physical
``flavors''. As we are interested in Dirac-K\"ahler fermions here we
shall adopt this latter point of view in the following.

Dirac-K\"ahler (DK) fields can be defined on any Riemannian\footnote{The
  pseudo-Riemannian case (Lorentzian space-times) can be dealt with in a
  completely analogous fashion.} manifold ($ {\cal M}_n,g $), i.e. on
any smooth $n$-dimensional manifold equipped with a metric $g$. From the
physics point of view this manifold represents {\it space-time}.

The main purpose of the present paper is to propose an analog of the
DK-fields which ``live'' on symplectic rather than Riemannian
manifolds. This means that we are going to study DK-fields not over
space-time but rather over a {\it phase-space}. 

A symplectic manifold ($ {\cal M}_{2 N}, \omega $) is a smooth
$2N$-dimensional manifold which is endowed with a closed, nondegenerate
2-form $\omega =\frac{1}{2} \omega_{a\, b} d \phi^{a}\wedge d\phi^{b} $.
(The $\phi^{a}, a=1,...,2N\, ,$ are local coordinates on ${\cal
  M}_{2\,N}$.) This manifold should be thought of as the phase-space of
a Hamiltonian system with $N$ degrees of freedom. The corresponding
Poisson bracket is given by $\{ \phi^{a}, \phi^{b} \} = \omega^{a\,b}$
where the matrix $(\omega^{a\,b})$ is the inverse of $( \omega_{a\,b}
)$.  Using local Darboux coordinates $\phi^{a} \equiv (p^{i},q^{i})
,i=1,...,N $, this matrix is independent of $\phi^{a},
\omega^{qp}=-\omega^{pq}=I$, and the only nonvanishing brackets are $\{
q^{i},p^{j}\}=\delta^{i\,j}$. If $\phi^{a}$ and $\widetilde \phi^{a}$
are local coordinates belonging to two overlapping charts of an atlas
covering ${\cal M}_{2 N}$ then, by the very definition of a symplectic
manifold, the coordinate transformation $\phi \rightarrow \widetilde
\phi$ is symplectic, i.e. the Jacobian matrix $( \partial
\widetilde\phi^{a} / \partial \phi^{b} ) $ is an element of ${\rm
  Sp}(2N)$ at every point of the overlap region. ${\rm Sp}(2N)$, the
group of linear canonical transformations, plays the same role for
phase-space which the Lorentz group plays for space-time. In particular,
it is the structure group of the frame bundle over ${\cal M}_{2\,N}$.
    
As for introducing DK-fields on symplectic manifolds the first question
which we must answer is what kind of spinor field should be used in
place of the ordinary Dirac spinors of relativistic field theory.  The
only natural choice here is to employ the so-called metaplectic spinors
\cite{kost}, i.e. the spinors of the metaplectic group ${\rm Mp}(2N)$.
Basically ${\rm Mp}(2N)$ is related to ${\rm Sp}(2N)$ in the same way
${\rm Spin}(n)$ is related to ${\rm SO}(n)$.  In particular, there
exists a two-to-one homomorphism between the two groups, i.e. ${\rm
  Mp}(2N)$ covers ${\rm Sp}(2N)$ twice. The construction of metaplectic
spin bundles and spinor fields over a symplectic manifold proceeds
almost literally along the same lines as in the case of space-time
spinors, the main difference being that it is ${\rm Mp}(2N)$ now which
serves as the structure group. For a detailed exposition we must refer
to the literature \cite{kost,wood}.
    
Metaplectic spinors have been used in many different contexts including
geometric quantization \cite{wood}, semi-classical approximations
\cite{lj}, Parisi-Sourlas super-symmetry \cite{pct}, string theory
\cite{green,compean}, and anyon super-conductivity \cite{castano}. Most
recently they played an important role in an approach to quantization
\cite{metaqm} which is based upon a Yang-Mills theory on phase-space
with metaplectic ``matter'' fields. This new formulation of quantum
mechanics is one of the main motivations for the present work. We shall
come back to it later on.

Let us briefly describe how one can construct representations of ${\rm
  Mp}(2N)$ \cite{meta1}. One has to associate an operator $M(S)$ to
every matrix $S \equiv ( S^{a} _{\,\,b}) \in {\rm Sp}(2N) $ in such a
way that $M(S_{1})M(S_{2})=\pm M(S_{1}S_{2})$. These operators can be
built up from a kind of ``$\gamma$-matrices'' which constitute a
symplectic Clifford algebra:
\begin{eqnarray}
\gamma^{a}\gamma^{b}-\gamma^{b}\gamma^{a}=2\,i\,\omega^{a\,b}
\label{V.1} 
\end{eqnarray}
We require $M(S)$ to satisfy the usual compatibility condition between
the vector and the spinor representation:
\begin{eqnarray}
M(S)^{-1}\gamma^{a} M(S)=S^{a}_{\,\,b}\gamma^{b} \label{V.1a}
\end{eqnarray}
Every infinitesimal ${\rm Sp}(2N)$-transformation is of the form
$S^{a}_{\,\,b}=\delta^{a}_{\,\,b}+\omega^{ac}\kappa_{cb}$ with symmetric
coefficients $\kappa_{ab}$. Inserting this together with the ansatz
$M(S)=1-\frac{i}{2} \kappa_{ab} \Sigma_{\rm meta}^{ab} $ into the
compatibility condition it is easy to show that the latter is solved by
\begin{eqnarray}
\Sigma_{\rm meta}^{ab}=\frac{1}{4}
(\gamma^{a}\gamma^{b}+\gamma^{b}\gamma^{a}) \label{V.2}
\end{eqnarray}
and that these generators satisfy the ${\rm Sp}(2N)$ commutator relations
\cite{meta1}. Thus every representation of the symplectic Clifford
algebra gives rise to a representation of ${\rm Mp}(2N)$. 

The most obvious difference between the metaplectic and the space-time
spinors is that the symplectic Clifford algebra involves a commutator
rather than an anticommutator. As an immediate consequence, this algebra
has no finite dimensional matrix representations, and metaplectic
spinors are necessarily infinite component objects. What is meant by a
``metaplectic representation'' is a representation in which $\gamma^{a}$
is a hermitian operator on an infinite dimensional Hilbert space ${\cal
  V}$. Hence the operators $M$ obtained by exponentiating the generators
(\ref{V.2}) give rise to a unitary representation. (See
\cite{meta1,meta2}, for further details.)

The symplectic Clifford algebra (\ref{V.1}) admits a rather intriguing
reinterpretation which is also at the heart of the new approach to
quantization \cite{metaqm} mentioned above. Assume we are given a
quantum mechanical system with a Hilbert space ${\cal V}$ along with $N$
position and momentum operators $\widehat x^{i}$ and $\widehat \pi^{i} $
acting on it. They satisfy the canonical commutator relations $[
\widehat x^{i},\widehat\pi^{j} ]=i\,\hbar\,\delta^{ij}$. By virtue of
the identification $\gamma^{i}=\kappa \widehat\pi^{i},\quad
\gamma^{N+i}=\kappa\widehat x^{i} $ for $i=1,...,N$ and with the
constant $\kappa\equiv \sqrt{2 / \hbar} $ it is obvious that the
``symplectic Clifford algebra'' (\ref{V.1}) is actually nothing but the
canonical commutation relations for the $\widehat x$-$\widehat
\pi$-auxiliary quantum system. We call it an ``auxiliary'' system
because it should not be confused with the actual physical system under
consideration, the one whose (curved) phase-space is ${\cal M}_{2N}$.
(The classical phase-space pertaining to the auxiliary system is simply
$\mathbf{R}^{2N}$ equipped with the standard symplectic structure.)

The metaplectic spin bundles are bundles over ${\cal M}_{2N}$ with the
typical fiber ${\cal V}$ and the structure group ${\rm Mp}(2N)$
\cite{kost}. At each point $\phi$ of ${\cal M}_{2N}$ a local copy of
${\cal V}$, denoted ${\cal V}_{\phi}$, is attached. Metaplectic spinor
fields are sections through these bundles. Locally they are simply
functions which assume values in ${\cal V}$:
\begin{eqnarray}
\psi :\quad {\cal M}_{2N} \rightarrow {\cal V}, \qquad
 \phi \mapsto | \psi \rangle_{\phi} \in {\cal V}_{\phi} \label{V.3}
\end{eqnarray}
The notation $| \psi \rangle_{\phi}$ means that the spinor $| \psi
\rangle \in {\cal V}$, ``lives'' in the local Hilbert space at $\phi$.
Upon introducing a basis $\{ |\alpha \rangle \}$ in ${\cal V}$ we write
$\psi^{\alpha}(\phi)\equiv \langle \alpha |\psi \rangle_{\phi} $ for its
components. Here $\alpha$ is an infinite dimensional generalization of a
spinor index. If we take $\{ | \alpha \rangle \} $ to be the $\widehat
x$-eigenbasis, for instance, then $\alpha \equiv
(\alpha^{1},...,\alpha^{N} )\in \mathbf{R}^{N}$.  (See
\cite{meta1,meta2} for details.)

In the present paper we shall focus on the local aspects of the bundles
involved. We only mention that on certain manifolds there are
topological obstructions which prevent them from carrying globally well
defined metaplectic spinor fields \cite{kost}. In ref. \cite{meta2} we
characterized these obstructions using methods from quantum field
theory.

\vspace{3mm}
Let us come back to the main question which we are trying to answer
in this paper: {\it Do there exist ``symplectic Dirac-K\"ahler
  fields'' which are related to the metaplectic spinors in the same way
  the ordinary Dirac-K\"ahler fields are related to Dirac spinors?}

Apart from being interesting in its own right, this question is of
obvious physical relevance. The fascinating property of metaplectic
spinor fields is that, {\it on a purely group theoretical basis,} they
introduce aspects of quantum mechanics into the geometry of classical
phase-spaces. By pure representation theory one is led to the
auxiliary quantum system in the local Hilbert spaces
${\cal V}_{\phi}$. In ref.  \cite{metaqm} we explained in detail how
these auxiliary systems relate to the actual physical quantum system
with the classical phase-space ${\cal M}_{2N}$. Using this as our
starting point, we showed that it is
possible to replace conventional canonical quantization by two new
rules with a more transparent physical and geometrical meaning.

Classical mechanics and classical statistical mechanics are geometric
theories which are conveniently described in the language of
symplectic geometry. Only tensor fields are needed to formulate them.
Quantum mechanics, on the other hand, has a natural interpretation in
terms of spinor fields on phase-space. Thus, in a sense, the very
process of quantization is tantamount to a transition from tensors to
spinors. But this is precisely what Dirac-K\"ahler theory is about:
its basic fields are tensors which, however, are equivalent to a
multiplet of spinors.

Before embarking on the detailed constructions let us briefly outline
the strategy for finding the ``symplectic DK-fields'' which we shall
follow in this paper.

Our main tools are two types of auxiliary quantum systems with Hilbert
spaces ${\cal V}$ and ${\cal V}^{\rm F}$, respectively. We mentioned
already the (bosonic) $\widehat x$-$\widehat \pi$-system on ${\cal V}$
whose canonical operators realize the metaplectic $\gamma$-matrices
$\gamma^{a}$. We also need a similar fermionic system with a
(finite-dimensional) Hilbert space ${\cal V}^{\rm F}$ and a set of
operators $\widehat \chi^{\mu}, \mu=1,...,n$, satisfying the canonical
anticommutator relations $\widehat \chi^{\mu} \widehat
\chi^{\nu}+\widehat \chi^{\nu} \widehat \chi^{\mu} = \hbar \delta^{\mu
  \nu}$. The $SO(n)$-Dirac matrices $\gamma^{\mu}$ are treated as a
special realization of this algebra.

An important technical ingredient is the Weyl symbol calculus
\cite{flato}-\cite{winf}. Let ${\cal L} ({\cal V})$ and ${\cal L} ({\cal
  V}^{\rm F})$ be the spaces of linear operators on ${\cal V}$ and
${\cal V}^{\rm F}$, respectively. It is possible to uniquely
characterize every operator $\widehat b \in {\cal L} ({\cal V})$ and
$\widehat f \in {\cal L} ({\cal V}^{F})$ in terms of classical
phase-functions (symbols) $b(y)$ and $f(\theta)$. Here $y$ and $\theta$
are coordinates on the (flat) classical phase-spaces which belong to the
auxiliary systems. In the bosonic case, $y\equiv (y^{a})\in
\mathbf{R}^{2N}$ is a vector with commuting entries, while $\theta
\equiv (\theta^{\mu} )$ is a set of $n$ anti-commuting Grassmann
numbers. The space of all bosonic (fermionic) symbol functions, equipped
with certain algebraic structures, is referred to as the bosonic
(fermionic) Weyl algebra ${\cal W} ({\cal W}^{\rm F} )$.

Given a space-time manifold ${\cal M}_{n}$, we consider fields on this
manifold which assume values in ${\cal V}^{\rm F}, {\cal L} (
{\cal V}^{\rm F})$ and $ {\cal W}^{\rm F}$, respectively. In an obvious
notation, we denote them $\psi^{\alpha}(x), \widehat F (x), $ and $F(x,
\theta)$.

Similarly, given a phase-space manifold ${\cal M}_{2N}$, we define
fields $\psi^{\alpha}(\phi), \widehat B (\phi)$, and $B(\phi,y)$ which
assume values in ${\cal V}, {\cal L} ( {\cal V} )$ and $
{\cal W}$, respectively.

In the first part of this paper we shall reformulate standard
Dirac-K\"ahler theory in terms of the fermionic Weyl symbol calculus. We
shall see that $\psi^{\alpha}(x)$ is an ordinary Dirac spinor and that
$F(x,\theta)$ can be identified with a DK-field. The Grassmann
variables $\theta^{\mu}$ will play the role of the basis differentials
$d x^{\mu}$.

This first part of the investigation is quite interesting in its own
right. For instance, we shall discover that the Clifford product which
is at the heart of DK-theory is basically the same thing as the star
product of the fermionic Weyl symbol calculus. As a consequence, ${\cal
  W}^{\rm F}$ turns out to be an Atiyah-K\"ahler algebra \cite{at,graf}.

In the second part of this paper we investigate in detail what happens
to the standard DK-theory, reformulated in terms of fermionic Weyl
symbols, when we replace fermionic symbols by bosonic ones everywhere.
This means that we switch from the $\widehat \chi^{\mu}$- to the
$\widehat x$-$\widehat \pi $-system.  Then $\psi^{\alpha}(\phi)$ is a
metaplectic spinor field, and by analogy with the fermionic setting we
shall argue that $B(\phi, y) $ is the ``symplectic DK field'' which we
are looking for. Schematically our approach can be summarized as
follows:
\begin{eqnarray}
{\rm DK-fields} &\,&\Longleftrightarrow \,\,\, {\rm fermionic\,\, symbols}
\nonumber \\
&\,& \hspace{2cm} \mbox{$\downarrow$} \nonumber \\
{\rm symplectic\,\, DK-fields} &\,&\Longleftrightarrow \,\,\, {\rm
  bosonic\,\, symbols} 
\end{eqnarray}

The rest of this paper is organized as follows: In the second half of
this introduction we discuss some aspects of standard DK-theory which
will be important later on. Then, in Section 2, we reformulate this
theory in terms of fermionic Weyl symbols. Particular attention is paid
to the decomposition of DK-fields as a set of Dirac spinors. The
construction of the symplectic DK-fields is performed in Section 3. We
investigate in detail which properties of $SO(n)$ DK-fields can be
translated to the ${\rm Sp}(2N)$-case and which cannot. Section 4
contains a summary and various remarks on the quantization problem in
the light of the present work. Some material needed as a background for
Section 2 is relegated to the appendix.

As for its mathematical rigor, the style of this paper is informal.
Occasionally the language of fiber bundles is used as a convenient tool
but we are mostly interested in the local properties of the bundles
involved and no pretense is made as for a rigoros and complete
discussion of the global aspects.

\vspace{1cm}
\begin{center}
{\large  ------\qquad DK fields on space-time\qquad ------}
\end{center}

Let us start with an arbitrary (curved) $n$-dimensional Riemannian
manifold $({\cal M}_{n},g)$. Upon introducing local coordinates
$x^{\mu}$, the tangent space $T_{x}{\cal M}_{n}$ and the cotangent space
$T_{x}^{\ast}{\cal M}_{n}$ at the point $x$ of ${\cal M}_{n}$ are
spanned by the basis vectors $\partial_{\mu}\equiv \partial / \partial
x^{\mu}$ and $dx^{\mu}, \mu=1,\cdots,n$, respectively. These spaces
constitute the fibers of the (co-)tangent bundle over ${\cal M}_{n}$.
Replacing $T^{\ast}{\cal M}_{n}$ by its $p$-fold tensor power we obtain
the bundle of (covariant) tensors of rank $p$. Restricting ourselves to
completely antisymmetric tensors we are led to the exterior algebra
$\bigwedge (T^{\ast}_{x}{\cal M}_{n})= \bigoplus_{p=0}^{n} \bigwedge^{p}
(T_{x}^{\ast}{\cal M}_{n}).$ Its elements are the inhomogeneous
differential forms
\begin{eqnarray}
\Phi(x)&=&\sum_{p=0}^{n} \Phi^{(p)}(x),\qquad \Phi^{(p)}(x)\in 
\bigwedge \nolimits ^{p}(T_{x}^{\ast} {\cal M}_{n})\qquad , \nonumber 
\\
\Phi^{(p)}(x)&=&\frac{1}{p!} F^{(p)}_{\mu_{1}\cdots\mu_{p}}(x) 
dx^{\mu_{1}}\wedge \cdots\wedge dx^{\mu_{p}} \label{1.1}
\end{eqnarray}
where $ F^{(p)}_{\mu_{1}\cdots\mu_{p}}$ are completely antisymmetric
coefficients. The corresponding algebra multiplication is the wedge
product ``$\wedge$''.

Since we have a metric $g=g_{\mu\nu}(x)dx^{\mu}\otimes dx^{\nu}$ at
our disposal which gives rise to an analogous bilinear form
$g'=g^{\mu\nu}(x)\partial_{\mu}\otimes\partial_{\nu}$ for the
cotangent bundle we can promote the fibers $\bigwedge
(T^{\ast}_{x}{\cal M}_{n})$ of the exterior algebra bundle to an
Atiyah-K\"ahler algebra ${\bf AK}(T^{\ast}_{x}{\cal M}_{n}, g')$
\cite{ka,at,graf}.

Quite generally, the Atiyah-K\"ahler algebra ${\bf AK}(V,Q)$
corresponding to an arbitrary vector space $V$ equipped with a quadratic
form Q consists of the elements of the exterior algebra over $V$,
$\bigwedge (V)=\bigoplus_{p} \bigwedge^{p}(V)$, for which the following
three products are defined:
\begin{itemize}
\item the exterior product ``$\wedge$''
\item the inner product $(\cdot, \cdot)$ induced by Q
\item the Clifford product ``$\vee$''
\end{itemize}
The three products are required to be distributive with respect to the
addition and to satisfy the relation
\begin{eqnarray}
a \vee b=a\wedge b + (a,b) \label{1.2}
\end{eqnarray}
for all $a,b\in \bigwedge^{1}(V)$. The Clifford product is associative
by definition. Hence the basic rule (\ref{1.2}) is sufficient in order
to work out the $\vee$-product of two arbitrary elements in
$\bigwedge(V)$. Below we shall give a closed formula for this product.

The Atiyah-K\"ahler algebra combines the notions of an exterior
algebra, a Grassmann algebra and a Clifford algebra in an consistent
manner. If we omit the Clifford product it reduces to the Grassmann
algebra $\bigwedge (V,Q)$, while omitting both $\vee$ and $(\cdot,
\cdot)$ yields the exterior algebra $\bigwedge (V)$. Without the
structure of the $\wedge$-product it becomes a Clifford algebra
because (\ref{1.2}) entails $a\vee b+b\vee a =2(a,b)$ \cite{graf}.

In the case at hand, $V=\bigwedge(T^{\ast}_{x}{\cal M}_{n})$ and $Q=g'$.
This means that for two basis 1-forms the inner product is given by
$(dx^{\mu},dx^{\nu})=g'(dx^{\mu},dx^{\nu})=g^{\mu\nu}$ and similarly for
higher forms; for instance, $(dx^{\mu}\wedge dx^{\nu},dx^{\rho}\wedge
dx^{\sigma})=g^{\mu\rho}g^{\nu \sigma}- g^{\mu \sigma}g^{\nu\rho}$.

A bundle over ${\cal M}_{n}$ with typical fiber ${\bf AK}
(T_{x}^{\ast}{\cal M}_{n}, g')$ is called an Atiyah-K\"ahler bundle and
sections through such bundles are referred to as Dirac-K\"ahler fields.
Locally they are described by a collection of antisymmetric tensor
fields $\lbrace F^{(p)}_{\mu_{1}\cdots\mu_{p}}, p=0,\cdots n \rbrace$.
The three products defined in the fiber give rise to analogous products
on the space of sections, for instance $(\Phi_{1} \vee
\Phi_{2})(x)\equiv \Phi_{1}(x) \vee \Phi_{2}(x)$. Of course, also all
the other operations of the conventional exterior calculus can be
applied to Dirac-K\"ahler fields: the exterior derivative $d$, the
coderivative $d^{\dagger}$, or the contraction with a vector field $v,
{\bf i}(v)$, to mention just a few.
  
In our case the relations defining the Clifford product assume the
following form when expressed in terms of the generating elements:
\begin{eqnarray}
1\vee 1=1,\qquad &\,& 1\vee dx^{\mu}=dx^{\mu}\vee 1=dx^{\mu} \nonumber \\
dx^{\mu}\vee dx^{\nu} &=&dx^{\mu} \wedge dx^{\nu} + g^{\mu\nu} \label{1.3}
\end{eqnarray}
By virtue of the postulated associativity of the $\vee$-product, these
relations are sufficient in order to determine the Clifford product of
two arbitrary differential forms. One finds \cite{ka,bj}
\begin{eqnarray}
\Phi_{1} \vee \Phi_{2}=\sum_{p=0}^{n} \frac{(-1)^{p(p-1) / 2}}{p!} 
({\cal A}^{p}\, e_{\mu_{1}}\neg 
\cdots e_{\mu_{p}}\neg \Phi_{1}) \wedge (e^{\mu_{1}}\neg \cdots e^{\mu_{p}}
\neg \Phi_{2}) \label{1.4}
\end{eqnarray}
with $e_{\mu}\neg \equiv {\bf i} (\partial_{\mu}), \quad e^{\mu}\neg
\equiv g^{\mu\nu} {\bf i} (\partial_{\nu}) \label{1.5}$ where ${\bf i}
(\partial_{\mu})$ denotes the contraction with the basis vector
$\partial_{\mu}$. It is an anti-derivation with the properties
\begin{eqnarray}
{\bf i}(\partial_{\mu})1& = &0 \nonumber \label{1.6},\qquad
{\bf i}(\partial_{\mu})dx^{\nu} =  \delta^{\nu}_{\mu} \\
{\bf i}(\partial_{\mu})(\Phi_{1} \wedge \Phi_{2})& = &({\bf i} 
(\partial_{\mu})\Phi_{1})\wedge \Phi_{2} + ({\cal A} 
\Phi_{1})\wedge {\bf i} (\partial_{\mu})\Phi_{2} \nonumber
\end{eqnarray}
In writing down eqs. (\ref{1.4}) and (\ref{1.6}) we used the ``main
automorphism'' ${\cal A}$, a linear map whose action on the DK-field
(\ref{1.1}) is defined as
\begin{eqnarray}
{\cal A}\Phi=\sum^{n}_{p=0} (-1)^{p}\, \Phi^{(p)} \label{1.7}
\end{eqnarray}
Later on we shall also need the ``main antiautomorphism'' $\cal B$ which
acts according to
\begin{eqnarray}
{\cal B}\Phi=\sum^{n}_{p=0}(-1)^{p(p-1)/2} \,\Phi^{(p)} \label{.1.8}
\end{eqnarray}
Obviously, ${\cal A}^{2}={\cal B}^{2}=1,\, {\cal AB=BA}$, and also
\begin{eqnarray}
{\cal A}(\Phi_{1} \wedge \Phi_{2}) =({\cal A} \Phi_{1} )\wedge 
({\cal A} \Phi_{2}) \nonumber \\
{\cal B}(\Phi_{1} \wedge \Phi_{2}) =({\cal B}\Phi_{2} )\wedge 
({\cal B} \Phi_{1}) \label{1.9}
\end{eqnarray}
 for any pair of DK-fields.

As an important special case of (\ref{1.4}) we note for later use that 
\begin{eqnarray}
dx^{\mu} \vee \Phi=dx^{\mu}\wedge \Phi + e^{\mu}\neg \Phi \label{1.9a}
\end{eqnarray}

Let us look at the physical interpretation of the DK-fields now. From
now on we shall specialize the discussion to a flat space-time ${\cal
  M}_{n}={\mathbf R}^{n}$ with the metric
$g_{\mu\nu}=\delta_{\mu\nu}$. The generalization of a curved manifold
and/or a manifold with Lorentzian signature would be straightforward,
but we shall avoid these technical complications here since they are
not important for the point we would like to make.

The interpretation of a DK-field as a multiplet of Dirac spinors is
based upon the following two logically independent observations.
\begin{enumerate}
\item[{\bf (i)}]\noindent From (\ref{1.3}) we obtain for the
  antisymmetrized Clifford product of two basis differentials
\begin{eqnarray}
dx^{\mu}\vee dx^{\nu} + dx^{\nu}\vee dx^{\mu}=2 \delta^{\mu\nu}
\label{1.10} 
\end{eqnarray}
This relation should be compared to the one satisfied by the euclidean
Dirac matrices $\gamma^{\mu}$:
\begin{eqnarray}
\gamma^{\mu}\gamma^{\nu}+\gamma^{\nu}\gamma^{\mu}=2 \delta^{\mu\nu}
\label{1.11} 
\end{eqnarray}
We conclude that the Clifford left multiplication with $dx^{\mu}$
defines a representation of the algebra of $\gamma$-matrices in the
space of (complex) inhomogeneous differential forms: $\gamma^{\mu}\,
\widehat = \,dx^{\mu}\vee $. This representation is reducible though.
Assuming $n$ even from now on, a Dirac spinor has $2^{n/2}$ complex
components, and an irreducible representation of the algebra
(\ref{1.11}) is in terms of $2^{n/2}\times 2^{n/2}$ matrices. On the
other hand, the dimension of the exterior algebra is $2^{n}$, i.e. a
DK-field $\Phi$ has $2^{n}$ independent complex component fields. We
shall see in a moment that the space $\cal K$ of all DK-fields $\Phi$
can be decomposed into $2^{n/2}$ subspaces ${\cal K}^{(\alpha)}$ which
are invariant under Clifford left multiplication, ${\cal
  K}=\oplus_{\alpha=1}^{k} {\cal K}^{(\alpha)}, k\equiv 2^{n/2}$. On
${\cal K}^{(\alpha)}, dx^{\mu} \vee$ gives rise to an irreducible
representation of the algebra (\ref{1.11}).

\item[{\bf (ii)}] \noindent From the exterior derivative $d$ and its
  adjoint, the coderivative $d^{\dagger}$, we can form the socalled
  Dirac-K\"ahler operator $d-d^{\dagger}$ which has the property that it
  squares to the Laplacian:
\begin{eqnarray}
(d-d^{\dagger})^{2}=-(dd^{\dagger}+d^{\dagger}d)=
\partial_{\mu}\partial^{\mu}  
\label{1.12}
\end{eqnarray}
It shares this property with the Dirac operator
$\gamma^{\mu}\partial_{\mu}$ and hence some relationship among the two
might be expected. In fact, it turns out that the Dirac-K\"ahler
operator can be expressed in terms of a Clifford multiplication from the
left:
\begin{eqnarray}
(d-d^{\dagger})\Phi(x)=dx^{\mu}\vee\partial_{\mu}\Phi(x) \label{1.13}
\end{eqnarray}
Since we know already that $dx^{\mu}\vee$ corresponds to a
$\gamma$-matrix and leaves the spaces ${\cal K}^{(\alpha)}$ invariant, 
we see that the Dirac-K\"ahler equation
\begin{eqnarray}
(d-d^{\dagger}+m)\Phi=0 \label{1.14}
\end{eqnarray}
decomposes to a set of equations
$(d-d^{\dagger}+m)\Phi^{(\alpha)}=0,\quad \Phi^{(\alpha)}\in {\cal
  K}^{(\alpha)}$, each of which is equivalent to an ordinary Dirac
equation $(\gamma^{\mu}\partial_{\mu}+m)\psi=0$.
\end{enumerate}
\vspace{3mm}

Following Becher and Joos \cite{bj} we can construct the invariant
subspaces ${\cal K}^{(\alpha)}$ as follows.  We introduce a new basis
$\{ Z_{\alpha \beta}\}$ in $\cal K$ whose elements are labeled by a pair
of indices $\alpha, \beta=1,\cdots,2^{n/2}$ and which are required to
satisfy\footnote{We use the notation $\mu,\nu,\cdots=1,\cdots,n$ for
  Lorentz indices and $\alpha,\beta,\gamma,\cdots=1,\cdots,2^{n/2}$ for
  spinor indices.}
\begin{eqnarray}
dx^{\mu}\vee Z_{\alpha\beta}=\sum_{\gamma=1}^{2^{n/2}} (\gamma^{\mu
  T})_{\alpha \gamma}\, Z_{\gamma\beta} \label{1.15}
\end{eqnarray}
where the euclidean Dirac matrices $\gamma^{\mu}$ are in the irreducible
$2^{n/2}$-dimensional representation.  They satisfy (\ref{1.11}) and are
assumed to be hermitian, $\gamma_{\mu}=\gamma_{\mu}^{\dagger}$.
Frequently we shall regard $Z\equiv(Z_{\alpha\beta})$ as a matrix or,
more precisely, as an inhomogeneous differential form which assumes
values in the space of spinor matrices. Then (\ref{1.15}) reads
\begin{eqnarray}
dx^{\mu}\vee Z=\gamma^{\mu T} Z \label{1.15a}
\end{eqnarray}
This equation is satisfied by
\begin{eqnarray}
Z=\sum_{p=0}^{n} \frac{1}{p!}
\,\gamma_{\mu_{1}}^{T}\cdots\gamma_{\mu_{p}}^{T} \,\, dx^{\mu_{1}}\wedge
\cdots\wedge dx^{\mu_{p}} \label{1.16}
\end{eqnarray}
Every DK-field $\Phi$ can be expanded in the basis $\lbrace
Z_{\alpha\beta}\rbrace$:
\begin{eqnarray}
\Phi(x)=\sum_{\alpha,\beta} \psi_{\alpha}^{(\beta)}(x)\,Z_{\alpha\beta}
\label{1.17} 
\end{eqnarray}
Hence it follows immediately from (\ref{1.15}) that the invariant
subspaces ${\cal K}^{(\alpha)}$ are spanned by
\begin{eqnarray}
\Phi^{(\beta)}\equiv \sum_{\alpha}
\psi^{(\beta)}_{\alpha}\,Z_{\alpha\beta}\in{\cal K}^{(\beta)},\quad \beta 
\quad{\rm fixed}. \label{1.18}
\end{eqnarray}
In fact, one has
\begin{eqnarray}
dx^{\mu}\vee \Phi^{(\beta)}=\sum_{\alpha}\left(\sum_{\delta}
\gamma^{\mu}_{\alpha\delta}\psi_{\delta}^{(\beta)}\right)\,Z_{\alpha\beta}
\label{1.19} 
\end{eqnarray}
which shows that on ${\cal K}^{(\beta)}$ Clifford left-multiplication
with $dx^{\mu}$ is equivalent to acting with the Dirac matrix
$\gamma^{\mu}$ on the spinor
$\psi^{(\beta)}\equiv\{\psi^{(\beta)}_{\alpha};\alpha=1,\cdots,2^{n/2}\}$.
For every fixed value of $\beta,\psi^{(\beta)}$ is an ordinary
$2^{n/2}$-component Dirac field. By virtue of the orthogonal
decomposition $\Phi=\sum_{\beta}\Phi^{(\beta)}$, a DK-field describes a
multiplet of $2^{n/2}$ Dirac fields.

It is convenient to combine the expansion coefficients
$\psi^{(\beta)}_{\alpha}$ into a spinor matrix $\widehat \psi$,
\begin{eqnarray}
(\widehat \psi)_{\alpha\beta}\,\equiv \,\psi_{\alpha}^{(\beta)}\quad ,
\label{1.20} 
\end{eqnarray}
so that (\ref{1.17}) reads
\begin{eqnarray}
\Phi(x)={\rm{Tr}}\left[\widehat \psi(x) Z^{T}\right] \label{1.21}
\end{eqnarray}
Writing $\widehat \psi[\Phi]$ for the matrix related to a given DK-field
$\Phi$, eq.(\ref{1.19}) amounts to
\begin{eqnarray}
\widehat\psi[dx^{\mu}\vee\Phi]=\gamma^{\mu}\, \widehat\psi[\Phi]
\label{1.22} 
\end{eqnarray}

\vspace{3mm} Occasionally one finds a slightly different approach in the
literature \cite{banks}. One assumes that the inhomogeneous form
(\ref{1.1}) is given and one uses its coefficient functions
${F^{(p)}_{\mu_{1}\cdots\mu_{p}}}$ in order to construct a spinor matrix
$\widehat F$ by simply replacing $dx^{\mu}\rightarrow\gamma^{\mu}$
everywhere:
\begin{eqnarray}
\widehat F\equiv \widehat F[\Phi]\equiv \sum_{p=0}^{n}\frac{1}{p!}\,
{F^{(p)}_{\mu_{1}\cdots\mu_{p}}\,\,
  \gamma^{\mu_{1}}\cdots\gamma^{\mu_{p}}} \label{1.23} 
\end{eqnarray}
Then one verifies that the map $\Phi\mapsto \widehat F [\Phi]$ satisfies
\begin{eqnarray}
\widehat F [dx^{\mu}\vee\Phi]=\gamma^{\mu} \widehat F [\Phi] ,
\label{1.24} 
\end{eqnarray}
a property it has in common with $\widehat \psi$. Hence we might expect
that these two matrix-valued fields are related. Indeed, it turns out
that they coincide up to a constant factor. To see this, one inserts the
expansions (\ref{1.1}) and (\ref{1.16}) into (\ref{1.21}) and obtains
the following formula for the coefficients of $\Phi,\,
{F^{(p)}_{\mu_{1}\cdots\mu_{p}}}$, as a function of $\widehat \psi$:
\begin{eqnarray}
F^{(p)}_{\mu_{1}\cdots\mu_{p}}(x)=(-1)^{p(p-1)/2}\,
{\rm{Tr}}\left[\widehat\psi(x)
  \gamma_{[\mu_{1}}\cdots\gamma_{\mu_{p}]}\right]  
\label{1.25} 
\end{eqnarray}
Because of the orthogonality and completeness relations enjoyed by the
Dirac matrices, eq.(\ref{1.25}) has a unique solution for $\widehat\psi$
as a function of the coefficients ${F^{(p)}_{\mu_{1}\cdots\mu_{p}}}$
which define $\Phi$. One finds that $\widehat \psi$ and $\widehat F $
are essentially the same thing:
\begin{eqnarray}
\widehat
\psi(x)&=&2^{-n/2}\,\sum_{p=0}^{n}\frac{1}{p!}\,
F^{(p)}_{\mu_{1}\cdots\mu_{p}}(x) 
\,\,\gamma^{\mu_{1}}\cdots  
\gamma^{\mu_{p}} \\ &=&2^{-n/2}\, \widehat F (x) \label{1.26}
\end{eqnarray}
This formula together with (\ref{1.20}) gives us a practical tool to
compute the projection of $\Phi$ on the invariant subspaces ${\cal
  K}^{(\alpha)}$.

In standard discussions of Dirac-K\"ahler theory, because of the simple
proportionality of $\widehat \psi$ and $\widehat F$, there is no need
for a conceptual distinction between the two matrices. In order to
establish their equivalence only familiar identities involving
$\gamma$-matrices such as
\begin{eqnarray}
{\rm{Tr}}\left[\gamma^{[\mu_{1}}\cdots\gamma^{\mu_{p}]}
  (\gamma_{[\nu_{1}}\cdots\gamma_{\nu_{q}]})^{\dagger}
\right]=2^{n/2}p!\,  
\delta^{pq}\,\delta^{\mu_{1}}_{[\nu_{1}}\cdots
\delta^{\mu_{p}}_{\nu_{q}]} \label{1.27}
\end{eqnarray}
are needed. In the symplectic case, the situation will be more
complicated and we have to distinguish more carefully $\widehat \psi$
which arises from the construction of left-invariant subspaces and
$\widehat F$ which obtains by replacing $dx^{\mu}\rightarrow
\gamma^{\mu}$ in $\Phi$. A priori it is not clear that the two objects
can easily be related to each other since the metaplectic
$\gamma$-``matrices'' are infinite dimensional. Hence the question
whether there are trace identities analogous to (\ref{1.27}) is a
nontrivial issue.

\renewcommand{\theequation}{2.\arabic{equation}}
\setcounter{equation}{0}

\section{Dirac-K\"ahler fields and fermionic Weyl\\ symbols}

In this section we describe the relation between the conventional
Dirac-K\"ahler fermions and the Weyl symbol calculus. In subsection 2.1
we summarize various properties of the fermionic Weyl symbol calculus
and discuss a number of special aspects and applications which will be
relevant. In section 2.2 we show that the fermionic Weyl algebra ${\cal
  W}^{\rm F}$ is an Atiyah-K\"ahler algebra, and in section 2.3 we
introduce ${\cal W}^{\rm F}$-valued fields over space-time. In 2.4 we
demonstrate that they can be identified with Dirac-K\"ahler fields. They
carry a reducible representation of the Clifford algebra. The
decomposition of ${\cal W}^{\rm F}$ into invariant subspaces which carry
an irreducible representation is performed in subsection 2.5.

\subsection{The fermionic Weyl algebra}

We consider a set of operators $\widehat\chi^{\mu}, \mu=1,\cdots,n$,
which satisfy the canonical anticommutation relations
\begin{eqnarray}
\widehat\chi^{\mu}\,\widehat\chi^{\nu}+
\widehat\chi^{\nu}\,\widehat\chi^{\mu}=\hbar\,\delta^{\mu\nu} 
\label{2.1} 
\end{eqnarray}
We could think of the $\widehat \chi$'s as world line fermions which
represent the spin of a relativistic particle, for instance
\cite{pct,bm}.  The most general operator we can construct by forming
linear combinations of products of $\widehat\chi$'s has the structure
\begin{eqnarray}
\widehat f
=\sum_{p=0}^{n}\frac{1}{p!}\,f^{(p)}_{\mu_{1}\cdots\mu_{p}}\,
\widehat\chi^{\mu_{1}}\widehat\chi^{\mu_{2}}
\cdots\widehat\chi^{\mu_{p}}  
\label{2.2}
\end{eqnarray}
with arbitrary (complex-valued) constants
$f^{(p)}_{\mu_{1}\cdots\mu_{p}}$. 

We would like to establish a linear one-to-one correspondence between
the operators (\ref{2.2}) and functions $f$ depending on Grassmann
numbers $\theta^{1},\theta^{2},\cdots,\theta^{n}$ with
$\theta^{\mu}\theta^{\nu}+\theta^{\nu}\theta^{\mu}=0$. The function
$f(\theta)$ which characterizes the operator $\widehat f$ is called the
symbol of $\widehat f$: $f={\rm symb}(\widehat f)$. There are many
``symbol maps'' which relate operators to classical functions. Here we
are interested in the Weyl symbol which is defined as follows. Given the
operator (\ref{2.2}) we define
\begin{eqnarray}
f(\theta)=[{\rm symb}(\widehat
f)](\theta)=\sum_{p=0}^{n}\frac{1}{p!}\,
f^{(p)}_{\mu_{1}\cdots\mu_{p}}\,\theta^{\mu_{1}}
\theta^{\mu_{2}}\cdots\theta^{\mu_{p}}  
\label{2.3}
\end{eqnarray}
which,
for a given ordering,
is a well defined map from operators to functions. In particular,
${\rm symb} (\widehat\chi^{\mu})$ $=\theta^{\mu}$ and ${\rm symb}(I)=1$
where $I$ is the unit operator. The inverse mapping is not well defined
yet, because in (\ref{2.3}) we can add to
$f^{(p)}_{\mu_{1}\cdots\mu_{p}}$ arbitrary tensors which are symmetric
in at least one index pair. This does not change $f(\theta)$, but it
does change $\widehat f$.  Specifying a unique operator $\widehat f$ for
a given $f(\theta)$ amounts to picking a particular operator ordering
prescription. In fact, $f(\theta)$ can be regarded as a classical phase
function of a mechanical system with Grassmann-odd phase-space
coordinates $\theta^{\mu}$, and the $\widehat\chi^{\mu}$'s are the
corresponding quantum operators. We shall employ the Weyl correspondence
rule which means that $\widehat f$ follows from $f(\theta)$ by
substituting $\theta^{\mu}\rightarrow\widehat\chi^{\mu}$ in (\ref{2.3})
and writing all operator products in Weyl ordered, i.e. completely
antisymmetrized form. For instance, the product
$\theta^{\mu}\theta^{\nu}$ yields the operator $[\widehat
\chi^{\mu}\widehat \chi^{\nu}]_{{\rm Weyl}}=\frac{1}{2}(\widehat
\chi^{\mu}\widehat \chi^{\nu}-\widehat \chi^{\nu}\widehat
\chi^{\mu})\equiv \widehat \chi^{[\mu}\widehat \chi^{\nu]}$. For an
arbitrary monomial,
\begin{eqnarray}
{\rm symb}^{-1}(\theta^{\mu_{1}}\cdots\theta^{\mu_{p}})=\widehat
\chi^{[\mu_{1}}\cdots\widehat \chi^{\mu_{p}]} \label{2.4}
\end{eqnarray}
where the square brackets indicate complete antisymmetrization. 

In the following we shall require the constants
$f^{(p)}_{\mu_{1}\cdots\mu_{p}}$ appearing in the series expansion of
the symbol $f(\theta)$ to be completely antisymmetric tensors. Then the
operator $\widehat f$ associated to the series (\ref{2.3}) is obtained
by simply replacing $\theta^{\mu}\rightarrow \widehat \chi^{\mu}$ in
this series, and this leads us back to the operator (\ref{2.2}).
  
  If $n$ is odd, the inverse symbol map is still not uniquely defined,
  because in this case the operator $\widehat \chi^{1}\widehat
  \chi^{2}\cdots\widehat \chi^{n}$ commutes with all operators and is
  proportional to the identity therefore. By multiplying any operator by
  $\widehat \chi^{1}\widehat \chi^{2}\cdots\widehat \chi^{n}$ if
  necessary one can represent all operators by even symbols. This
  prescription makes
  the correspondence between operators and symbols bijective. (See
  \cite{bm,ht} for further details.)
  
  If a string of operators is not contracted with an antisymmetric
  tensor we must reorder it before we can use
\begin{eqnarray}
{\rm symb}(\widehat
\chi^{[\mu_{1}}\cdots\widehat\chi^{\mu_{p}]})=
\theta^{\mu_{1}}\cdots\theta^{\mu_{p}}  
\label{2.5}
\end{eqnarray}
in order to read off its symbol. For instance,
\begin{eqnarray}
{\rm symb}(\widehat \chi^{\mu}\widehat\chi^{\nu})= {\rm
  symb}\left[\widehat \chi^{[\mu}\widehat \chi^{\nu]}+
  \frac{\hbar}{2}\delta^{\mu\nu}\right] 
=\theta^{\mu}\theta^{\nu}+ \frac{\hbar}{2}\delta^{\mu\nu} \label{2.6} 
\end{eqnarray}

The symbols $f(\theta)$ are functions of the same type as those
considered in Appendix A, to which the reader might turn at this point.
Among other things, various linear operations on such functions are
discussed there which are particularly useful in the context of the
symbol calculus. This includes the ``main automorphism'' $\cal A$, the
``main antiautomorphism'' $\cal B$, the Hodge operator $\ast$ and the
modified Hodge operator $\hodge$.

While we allow for complex coefficients
$f^{(p)}_{\mu_{1}\cdots\mu_{p}}$, we assume that the operators
$\widehat\chi^{\mu}$ are hermitian, $\widehat
\chi^{\mu}=(\widehat\chi^{\mu})^{\dagger}$, and that $\theta^{\mu}$ is
real, $\bar \theta^{\mu}=\theta^{\mu}$. Hence it follows that
\begin{eqnarray}
{\rm symb} ({\widehat f}^{\dagger}) =\overline{{\rm symb} (\widehat f)}
\label{2.7} 
\end{eqnarray}
where the overbar means complex conjugation.

There is a simple integral formula for the operator $\widehat f$
 associated to a given Weyl symbol $f(\theta)$:
\begin{eqnarray}
\widehat f=\int \widehat \Omega (\rho)\, \widetilde f(\rho)\, d^{n}\rho
\label{2.8} 
\end{eqnarray}
Here $\widetilde f (\rho)$ is the Fourier transform of $f(\theta)$ as
defined in eq.(\ref{A.22}) of the Appendix, and
\begin{eqnarray}
{\widehat \Omega} (\rho)\equiv {\rm exp} \left(-i\, {\widehat \chi}^{\mu}
  \rho_{\mu} \right) \label{2.9}
\end{eqnarray}
is the fermionic analogue of the Weyl operators which implement
translations on phase-space. Using the identities of appendix A one can
verify that eqs. (\ref{2.4}) and (\ref{2.8}) are indeed equivalent.

An important concept is the ``star product''\footnote{Both in the
  fermionic and the bosonic case we keep using the traditional name
  ``star product'' even though we write `$\circ$' instead of the usual
  symbol `$\ast$'.  Following refs. \cite{fed,fedbook,metaqm} this
  notation indicates that we are dealing with a {\it fiberwise} twisted
  product which should not be confused with the $\ast\equiv\ast_{\cal
    M}$-product which would refer to the {\it base} of the Weyl algebra
  bundles we are going to construct in section 2.2. It is the
  $\ast_{\cal M}$- product rather than the $\circ$-product which is
  needed for the deformation quantization \cite{flato} of physical
  systems on the phase-space $\cal M$. In the present paper, the
  $\ast_{\cal M}$-product plays no role, however. (Note also that
  `$\ast$' stands for the Hodge operator in our case.)}  or ``twisted
product'' which mimics the multiplication of operators at the level of
symbols. It satisfies
\begin{eqnarray}
{\rm symb}(\widehat f\,\widehat g)={\rm symb}(\widehat f)\circ {\rm
  symb}(\widehat g) \label{2.10}
\end{eqnarray}
for all operators $\widehat f$ and $\widehat g$. As a consequence, the
$\circ$-product is associative, distributive with respect to $+$,
but not commutative. It is a deformation of the pointwise product of
functions to which it reduces in the limit $\hbar\rightarrow 0$. From
\begin{eqnarray}
{\rm symb}(I)=1, \,\,{\rm symb}(\widehat \chi ^{\mu})=\theta^{\mu}
\label{2.11} 
\end{eqnarray}
and eq.(\ref{2.6}) it follows that 
\begin{eqnarray}
1\circ 1=1,\,\,&\,& 1\circ \theta^{\mu}=\theta^{\mu}\circ 1=\theta^{\mu} 
\nonumber \\
\theta^{\mu} \circ \theta^{\nu}&=&\theta^{\mu}\theta^{\nu}
+\frac{\hbar}{2} \delta^{\mu\nu} \label{2.12}
\end{eqnarray}
By virtue of its postulated distributivity and associativity, the
relations (\ref{2.12}) characterize the $\circ$-product uniquely. They
are sufficient to work out the product $f\circ g$ of arbitrary functions
$f$ and $g$.

Eq.(\ref{2.7}) implies that complex conjugation changes the order of
the factors in a star product:
\begin{eqnarray}
\overline{f_{1} \circ f_{2}}=\bar f_{2} \circ \bar f_{1} \label{2.13}
\end{eqnarray}

The space of functions $f(\theta)$ equipped with the $\circ$-product
will be referred to as the {\it fermionic Weyl algebra} ${\cal W}^{\rm
  F}$.

In the literature \cite{bm,ht} one finds the following integral
representation for the $\circ$-product of two arbitrary functions:
\begin{eqnarray}
&\,&(f_{1}\circ f_{2})(\theta)= \nonumber \\
&\,&\epsilon_{n}\left(\frac{\hbar}{2i}\right)^{n} {\huge \int} {\rm exp}
\left[  
-\frac{2}{\hbar}(\theta_{1}\theta +\theta \theta_{2} +\theta_{2}
\theta_{1})\right] f_{1}(\theta_{1})f_{2} 
(\theta_{2})\,d^{n}\theta_{1}d^{n}\theta_{2} \label{2.14}
\end{eqnarray}
where\footnote{Indices are raised and lowered with
  $g_{\mu\nu}=\delta_{\mu\nu}$.} $\theta_{1}\theta\equiv
\theta_{1}^{\mu} \theta_{\mu}$, etc. and with $\epsilon_{n}$ defined as
in eq.(\ref{A.23}). For our purposes, various alternative
representations of the star product are needed. They are derived in
appendix B. The first one reads
\begin{eqnarray}
(f_{1}\circ f_{2})(\theta)=\sum_{p=0}^{n}
\left(\frac{\hbar}{2}\right)^{p}  \frac{(-1)^{p(p-1)/2}}{p!}\left[
  {\cal A}^{p} 
\frac{\partial}{\partial \theta^{\mu_{1}}} \frac{\partial}{\partial
  \theta^{\mu_{2}}}\cdots\frac{\partial}{\partial  
\theta^{\mu_{p}}} f_{1}(\theta)\right] \times \nonumber \\
\hspace{5cm}\left[\frac{\partial}{\partial
    \theta_{\mu_{1}}}\frac{\partial}{\partial \theta_{\mu_{2}}} 
\cdots \frac{\partial}{\partial \theta_{\mu_{p}}} f_{2}(\theta)\right]
\label{2.15} 
\end{eqnarray}
with the automorphism ${\cal A}: {\cal W}^{F}\rightarrow {\cal W}^{F}$
defined in appendix A. An equivalent form involving both left and right
derivatives is
\begin{eqnarray}
&\,&(f_{1}\circ f_{2})(\theta)=\nonumber \\ 
&\,&\sum_{p=0}^{n}
\left(\frac{\hbar}{2}\right)^{p}
\frac{1}{p!} f_{1}(\theta) \stackrel{\longleftarrow}{\frac{\partial}
{\partial 
\theta^{\mu_{p}}}}
\stackrel{\longleftarrow}{\frac{\partial}{\partial
    \theta^{\mu_{p-1}}}}\cdots
\stackrel{\longleftarrow}{\frac{\partial}{\partial \theta^{\mu_{1}}}}
\frac{\partial}{\partial \theta_{\mu_{1}}}\frac{\partial}{\partial 
\theta_{\mu_{2}}}
\cdots \frac{\partial}{\partial \theta_{\mu_{p}}} f_{2}(\theta)
\label{2.16} 
\end{eqnarray}
The most compact representation reads
\begin{eqnarray}
(f_{1}\circ f_{2})(\theta)=f_{1}(\theta) {\rm exp} \left[
  \frac{\hbar}{2} 
\stackrel{\longleftarrow}{\frac{\partial}{\partial 
\theta^{\mu}}}
\stackrel{\longrightarrow}{\frac{\partial}{\partial \theta_{\mu}}}
\right] f_{2}(\theta) \label{2.17}
\end{eqnarray}
where $\stackrel{\longleftarrow}{\frac{\partial}{\partial
    \theta^{\mu}}}$ is a right derivative acting on $f_{1}$ and
$\stackrel{\longrightarrow}{\frac{\partial}{\partial \theta^{\mu}}}$ a
left derivative acting on $f_{2}$. The result (\ref{2.17}) looks
surprisingly simple and is completely analogous to its bosonic
counterpart. All the complicated sign factors which appeared during the
calculation, either explicitly or hidden in the $\cal A$-automorphism,
conspired to disappear from the final result.

Depending on the problem at hand one or another of the above
representations is the most convenient one. Eq.(\ref{2.15}) we shall
relate to K\"ahler's formula for the Clifford product shortly. From
eq.(\ref{2.16}) one immediately reads off the important special cases
\begin{eqnarray}
\theta^{\mu}\circ
f(\theta)=\theta^{\mu}f(\theta)+ \frac{\hbar}{2}\,
\frac{\partial}{\partial 
  \theta_{\mu}} f(\theta) \label{2.18}\\ 
f(\theta) \circ \theta^{\mu}=f(\theta) \theta^{\mu}+\frac{\hbar}{2}\,
f(\theta) 
\stackrel{\longleftarrow}{\frac{\partial}{ \partial \theta_{\mu}}}
\label{2.19} 
\end{eqnarray}
In order to calculate the product of two $\delta$-functions, which we
shall need later on, eq.(\ref{2.14}) is most suitable:
\begin{eqnarray}
(\delta\circ
\delta)(\theta)=(-1)^{n(n-1)/2}\left( \frac{\hbar}{2}\right)^{n}
\label{2.20} 
\end{eqnarray}

Up to now we regarded the $\widehat \chi^{\mu}$'s as abstract operators.
Let us look at concrete representations on some finite dimensional
vector space ${\cal V}^{\rm F}$. If ${\Gamma_{\mu} }$ is a set of
hermitian matrices which satisfy the Clifford algebra relations
\begin{eqnarray}
\Gamma^{\mu}\,\Gamma^{\nu}+\Gamma^{\nu}\,
\Gamma^{\mu}=2\,\delta^{\mu\nu} \label{2.21} 
\end{eqnarray}
then
\begin{eqnarray}
\widehat \chi ^{\mu}=\sqrt{\frac{\hbar}{2}}\, \Gamma^{\mu} \label{2.22} 
\end{eqnarray}
satisfies the canonical anticommutation relations (\ref{2.1}). Here
$\Gamma^{\mu}$ denotes the Dirac matrices in an arbitrary, possibly
reducible representation. The notation $\gamma^{\mu}$ is reserved for
the (essentially unique) irreducible representation on ${\cal V}^{\rm
  F}={\mathbf C}^{k}, k\equiv 2^{n/2}$, if $n$ is even. The operators
$\widehat f: {\cal W}^{\rm F}\rightarrow {\cal W}^{\rm F}$ of
eq.(\ref{2.2}) are $k\times k$ matrices then. The space of these
operators will be denoted by ${\cal L}({\cal W}^{\rm F})$.

The identification (\ref{2.22}) must be interpreted with some care. In
setting up the symbol calculus one adopts the rule that the operators
$\widehat \chi ^{\mu}$ anticommute with numbers of odd Grassmann parity.
On the other hand, the entries of the matrices $\Gamma^{\mu}$ are
ordinary complex numbers, so $\Gamma^{\mu}$ commutes with all elements
of the Grassmann algebra.

Next we list a few properties of the operators $\widehat \Omega$ which
we shall need shortly. These operators are reminiscent of the (bosonic)
Weyl operators. However, as they stand, they are not unitary but rather
hermitian, $\Omega( \rho)=\Omega(\rho)^{\dagger}.$ This is due to the
fact that $\widehat \chi ^{\mu}$ anticommutes with the Grassmann-odd
$\rho_{\mu}$'s .  Actually it is the operators $\widehat\Omega (i
\rho/\hbar)={\rm exp}(\widehat \chi ^{\mu}\rho_{\mu} /\hbar)$ which play
the role of the Weyl operators on a fermionic phase-space. They are
unitary, $\widehat\Omega (i \rho/\hbar)^{\dagger}=\widehat\Omega (i
\rho/\hbar)^{-1}$, and they shift $\widehat \chi ^{\mu}$ by $\rho^{\mu}$
times the unit operator:
\begin{eqnarray}
\widehat\Omega (i \rho/\hbar)^{\dagger}\, \widehat \chi
^{\mu}\,\widehat\Omega (i \rho/\hbar) =\widehat\chi ^{\mu} +\rho^{\mu}
\label{2.23} 
\end{eqnarray}
This leads to a projective representation of the translation group since
\begin{eqnarray}
\widehat\Omega (\rho_{1})\,\widehat\Omega (\rho_{2})={\rm exp}
(\frac{\hbar}{2}\rho^{\mu}_{1}\rho_{2\mu})\,\widehat\Omega
(\rho_{1}+\rho_{2}) \label{2.24}
\end{eqnarray}
The derivative of $\widehat\Omega (\rho)$ can be written in either of
the two forms
\begin{eqnarray}
\frac{\partial}{\partial \rho_{\mu}}\,\widehat\Omega (\rho)& = &
\widehat\Omega (\rho)\,\left[ i\widehat \chi ^{\mu}
  +\frac{\hbar}{2}\rho^{\mu}\right] 
\label{2.25} \\ & = & \left[ i \widehat \chi ^{\mu} - \frac{\hbar}{2}
\rho^{\mu}\right]\,\widehat\Omega (\rho) \label{2.26}
\end{eqnarray}

When we replace in $\widehat\Omega$ the operators $\widehat \chi ^\mu$
by the Dirac matrices via (\ref{2.22}) we are led to
\begin{eqnarray}
\check \Omega (\rho)={\rm exp}(-i\sqrt{\hbar
  /2}\,\Gamma^{\mu}\rho_{\mu}) \label{2.27} 
\end{eqnarray}
The properties of $\check \Omega$ are slightly different from those of
$\widehat \Omega $ because $\Gamma^{\mu}$ commutes with $\rho_{\mu}$.
The $\check \Omega $'s are unitary matrices,
\begin{eqnarray}
\check \Omega (\rho)^{\dagger}=\check \Omega (-\rho)= \check \Omega
(\rho)^{-1} \label{2.28}
\end{eqnarray}
with the composition law
\begin{eqnarray}
\check\Omega (\rho_{1})\check\Omega (\rho_{2})={\rm exp}
(-\frac{\hbar}{2}\rho^{\mu}_{1}\rho_{2\mu})\,\check\Omega
(\rho_{1}+\rho_{2}) \label{2.29}
\end{eqnarray}
The expressions for their derivative are
\begin{eqnarray}
\frac{\partial}{\partial \rho_{\mu}}\check\Omega (\rho)& = &
\check\Omega (-\rho)\left[ -i\sqrt{\frac{\hbar}{2}} \Gamma^{\mu}
-\frac{\hbar}{2}\rho^{\mu}\right] \label{2.30} \\ & = & \left[
-i\sqrt{\frac{\hbar}{2}} \Gamma^{\mu} +
\frac{\hbar}{2}\rho^{\mu}\right]\,\check\Omega (\rho) \label{2.31}
\end{eqnarray}
We shall need these relations when we decompose the reducible
Dirac-K\"ahler representation.

An interesting example where one can see the symbol calculus at work is
the generalization of the chirality matrix $\gamma_{5}$ in 4 dimensions.
We assume that $n$ is even in the remainder of this subsection and
employ the Dirac matrices $\gamma^{\mu}$ in the $2^{n/2}$-dimensional
representation. From (\ref{1.11}) and
$(\gamma^{\mu})^{\dagger}=\gamma^{\mu}$ it follows that the matrix
\begin{eqnarray}
\gamma_{n+1}\equiv-i^{n(n-1)/2}\,\gamma^{1}\gamma^{2}\cdots\gamma^{n}
\label{2.32} 
\end{eqnarray}
satisfies $\gamma^{\mu}\gamma_{n+1}=-\gamma_{n+1}\gamma^{\mu}$,
\begin{eqnarray}
\gamma^{2}_{n+1}=1 \qquad {\rm and} \qquad
\gamma^{\dagger}_{n+1}=\gamma_{n+1} \label{2.33}
\end{eqnarray}
in all even dimensions. The sign of (\ref{2.32}) is chosen such that for
$n=4$
\begin{eqnarray}
\gamma_{5}=\gamma^{1}\gamma^{2}\gamma^{3}\gamma^{4} \label{2.34}
\end{eqnarray}
We identify
\begin{eqnarray}
\gamma^{\mu}=\kappa \,\widehat\chi ^{\mu} \label{2.35}
\end{eqnarray}
where
\begin{eqnarray}
\kappa\equiv \sqrt{\frac{2}{\hbar}} \label{2.36}
\end{eqnarray}
and interpret $\gamma_{n+1}$ as the matrix representation of the
abstract operator
\begin{eqnarray}
\widehat
G_{n+1}=-i^{n(n-1)/2}\,\kappa^{n}\,
\widehat\chi^{1}\widehat\chi^{2}\cdots\widehat\chi^{n}  
\label{2.37}
\end{eqnarray}
Its symbol ${\rm symb}(\widehat G_{n+1})\equiv G_{n+1}$ follows directly
from (\ref{2.5}) if we note that $\widehat\chi^{1}\cdots
\widehat\chi^{n}=\epsilon_{\mu_{1}\cdots\mu_{n}}
\widehat\chi^{[\mu_{1}}\cdots\widehat\chi^{\mu_{n}]}/(n!)$:
\begin{eqnarray}
G_{n+1}(\theta)& = &-i^{n(n-1)/2}\,\kappa^{n}\,
\theta^{1}\theta^{2}\cdots\theta^{n} \\ 
& = &-(-i)^{n(n-1)/2}\,\kappa^{n}\,\theta^{n}\theta^{n-1}\cdots\theta^{1}
\nonumber \label{2.38}
\end{eqnarray}
Hence, up to a constant, $\gamma_{n+1}$ is represented by the
$\delta$-function:
\begin{eqnarray}
G_{n+1}(\theta)= -(-i)^{n(n-1)/2}\,\kappa^{n}\,\delta(\theta) \label{2.39}
\end{eqnarray}
As a consequence of eqs. (\ref{2.20}) and (\ref{A.21}), this function
satisfies
\begin{eqnarray}
G_{n+1}\circ G_{n+1}=1\quad {\rm and}\quad \overline{G_{n+1}}= G_{n+1}
\label{2.40} 
\end{eqnarray}
which reflects the properties (\ref{2.33}) of $\gamma_{n+1}$. By virtue
of (\ref{A.23a}) the Fourier transform of $G_{n+1}$ is the constant
function 
\begin{eqnarray}
\widetilde G_{n+1} (\rho)=-(-i)^{n(n-1)/2}\,\kappa^{n} \label{2.41}
\end{eqnarray}

In appendix A we defined the modified Hodge operator $\hodge$ for a
general Grassmann algebra and we showed that it is related to the
Fourier transformation via eq.(\ref{A.37}). Using the latter equation
together with the integral representation (\ref{2.14}) for the star
product it its not difficult to see that the application of $\hodge$ to
some $f\in {\cal W}^{\rm F}$ is essentially equivalent to a
star-multiplication with $G_{n+1}$ from the right. For a homogeneous
function of degree $p$,
\begin{eqnarray}
\hodge f^{(p)}(\theta)=-(-i)^{n(n-1)/2}\kappa^{2p-n}\,(f^{(p)}\circ
G_{n+1})(\theta) \label{2.42}
\end{eqnarray}
If we rescale $\theta$ we can write down a similar equation for
inhomogeneous functions even:
\begin{eqnarray}
\hodge f(\theta/\kappa)=-(-i)^{n(n-1)/2}\,(f\circ
G_{n+1})(\theta/\kappa) \label{2.43} 
\end{eqnarray}

Finally we remark that the chirality operator $\widehat G_{n+1}$ can be
expressed as an integral over the Weyl operators:
\begin{eqnarray}
\widehat G_{n+1}=-(-i)^{n(n-1)/2}\kappa^{n}\int d^{n}\rho\, \widehat \Omega
(\rho) \label{2.44}
\end{eqnarray}
This is a remarkable relation because contrary to the original
definition of $\gamma_{n+1}$ as the product of all Dirac matrices it
carries over to the symplectic case almost literally.

\subsection{${\cal W}^{\rm F}$ as an  Atiyah-K\"ahler algebra}

Let us come back to the abstract Atiyah-K\"ahler algebra ${\bf AK}(V,Q)$
discussed in the introduction. It is important to observe that the Weyl
algebra ${\cal W}^{\rm F}$ which we reviewed in the previous section
contains all the ingredients which make up an Atiyah-K\"ahler algebra:

\begin{itemize}
\item[{\bf (i)}] The vector space $V$ is spanned by the basis elements
  $\theta^{1},\cdots,\theta^{n}$ and the exterior algebra over this
  space, $\bigwedge(V)=\oplus_{p=0}^{n} \bigwedge^{p}(V)$, consists of
  monomials $\theta^{\mu_{1}}\cdots\theta^{\mu_{p}}\in \bigwedge^{p}(V)$.
  The exterior product ``$\wedge$'' on $\bigwedge(V)$ is the pointwise
  product of (inhomogeneous) functions $f(\theta)\in \bigwedge (V)$.
  
\item[{\bf (ii)}] By starting from the canonical anticommutation
  relations (\ref{2.1}) we have tacitly decided for an inner product
  $(\cdot | \cdot)$ on $\bigwedge(V)$. The quadratic form $Q$ is induced
  by $g_{\mu\nu}\equiv \delta_{\mu\nu}$, regarded as an inner product of
  $V$. On $\bigwedge^{1}(V)$ we have
\begin{eqnarray}
(\theta^{\mu}|\theta^{\nu})=\kappa^{-2}\,\delta^{\mu\nu} \label{2.45}
\end{eqnarray}
and similarly for $p>1$ (see below).
\item[{\bf (iii)}] The star product on ${\cal W}^{\rm F}$ provides a
  concrete realization of the abstract Clifford product. The Clifford
  product is associative and distributive over '+', and so is the star
  product.  Moreover, $\vee,\wedge$ and $(\cdot, \cdot)$ have to satisfy
  the consistency condition (\ref{1.2}). From eq.(\ref{2.12}) it follows
  that this relation is indeed satisfied by the star multiplication
  together with the pointwise multiplication and the inner product
  $(\cdot|\cdot)$:
\begin{eqnarray}
\theta^{\mu}\circ\theta^{\nu}=\theta^{\mu}\theta^{\nu}+
(\theta^{\mu}|\theta^{\nu})  
\label{2.46}
\end{eqnarray}
In particular, upon symmetrization,
$\theta^{\mu}\circ\theta^{\nu}+\theta^{\nu}\circ
\theta^{\mu}=2(\theta^{\mu}|\theta^{\nu})$. 
\end{itemize}

Thus we may conclude that {\it the fermionic Weyl algebra ${\cal W}^{\rm
    F}$ is a concrete realization of an Atiyah-K\"ahler algebra}.

Let us be more explicit about the inner product on $\bigwedge(V)$.
Within the symbol calculus, the standard inner product of DK-theory
\cite{graf} admits a very natural representation in terms of the star
product:
\begin{eqnarray}
(f_{1}| f_{2})=[\bar f_{1}\circ f_{2}](\theta=0) \label{2.47}
\end{eqnarray}
Here $f_{1}$ and $f_{2}$ are two arbitrary inhomogeneous functions. We
allow their expansion coefficients ${f_{\mu_{1}\cdots\mu_{p}}^{(p)}}$ to
become complex. Note, however, that the complex conjugation in
(\ref{2.47}) is necessary even if the coefficients are taken to be real,
see eq.(\ref{A.10}). Using the integral representation
\begin{eqnarray}
(f_{1}| f_{2})=(i\,\kappa^{2})^{-n}\epsilon_{n}\int
\overline{f_{1}(\theta_{1})}\, {\rm exp}
(\kappa^{2}\theta_{1}^{\mu}\theta_{2\mu})
\,f_{2}(\theta_{2})\,d^{n}\theta_{1}d^{n}\theta_{2}  
\label{2.48}
\end{eqnarray}
and expanding $f_{1}$ and $f_{2}$ according to 
\begin{eqnarray}
f(\theta)=\sum_{p=0}^{n}\frac{1}{p!}\,\kappa^{p}\,
f^{(p)}_{\mu_{1}\cdots\mu_{p}}\,\theta^{\mu_{1}}\cdots\theta^{\mu_{p}}  
\label{2.49}
\end{eqnarray}
with appropriate powers of $\kappa$ separated off from the expansion
coefficients, it is easy to derive that
\begin{eqnarray}
(f_{1}| f_{2})=\sum_{p=0}^{n}\frac{1}{p!}\,
\overline{f^{(p)}_{1\,\mu_{1}\cdots\mu_{p}}}\, 
f^{(p)\,\mu_{1}\cdots\mu_{p}}_{2} \label{2.50}
\end{eqnarray}

The inner products among the basis elements of $\bigwedge^{p}(V)$
(homogeneous functions of degree $p$) can be written down similarly. For
$p=1$ one recovers (\ref{2.45}), and for $p=2 $ one has, for instance
\begin{eqnarray}
(\theta^{\mu}\theta^{\nu}|\theta^{\rho}\theta^{{\sigma}})=
\kappa^{-4}(\delta^{\mu\rho}\delta^{\nu{\sigma}}-
\delta^{\mu{\sigma}}\delta^{\nu\rho})  
\label{2.51}
\end{eqnarray}

We note that $(\cdot|\cdot)$ has the important property of
making the star multiplication with $\theta^{\mu}$ a self-adjoint
operator. If we define
\begin{eqnarray}
C^{\mu}:{\cal W}^{\rm F}\rightarrow{\cal W}^{\rm F},\hspace{1cm}
(C^{\mu}f)(\theta)=\kappa\theta^{\mu}\circ f(\theta) \label{2.52}
\end{eqnarray}
then eq.(\ref{2.18}) tells us that $C^{\mu}$ is given by the first order
differential operator
\begin{eqnarray}
C^{\mu}=\kappa\theta^{\mu}+\frac{1}{\kappa} \frac{\partial}{\partial
  \theta_{\mu}} \label{2.53}
\end{eqnarray}
If one writes the inner product as in (\ref{2.47}), the self-adjointness
of $C^{\mu}$ is obvious:
\begin{eqnarray}
(C^{\mu}f_{1}| f_{2})& = & \kappa [\overline{(\theta^{\mu}\circ
  f_{1})}\circ f_{2}](0) \nonumber\\
& = & \kappa [(\overline{f_{1}}\circ \theta^{\mu})\circ f_{2}](0)
\nonumber\\ 
& = & \kappa [\overline{f_{1}}\circ (\theta^{\mu}\circ
f_{2})](0)\label{2.54}\\ 
& = & (f_{1}| C^{\mu} f_{2}) \nonumber
\end{eqnarray}
Here we exploited (\ref{2.13}) and the associativity of the star
product.

\subsection{Symbol-valued fields on space-time}

The most familiar application of the above symbol calculus is the
deformation theory approach \cite{flato,bm} to the quantization of
fermionic systems. In this context, the variables $\theta^{\mu}$ are
coordinates on the {\it phase-space} of the physical system under
consideration. If there are additional bosonic degrees of freedom
(such as the position of a spinning particle, say) this fermionic
phase-space is embedded in a larger graded phase-space, a supermanifold
with both commuting and anticommuting coordinates \cite{pct}.

In the present paper we are investigating a different setting. Rather
than phase-space, the physical arena here is {\it space-time}, an
ordinary Riemannian manifold $({\cal M}_{n},g)$, not a supermanifold.
The fermionic Weyl algebra ${\cal W}^{\rm F}$ enters the construction as
the {\it fiber} of certain bundles over space-time which we shall refer
to as ``Weyl algebra bundles'' \cite{fedbook}.

By definition, the base of a Weyl algebra bundle is $({\cal M}_{n},g)$
and the typical fiber is ${\cal W}^{\rm F}$, i.e., at each space-time
point $x$ we attach a local copy ${\cal W}^{\rm F}_{x}$ of ${\cal
  W}^{\rm F}$. The quadratic form $Q$ on ${\cal W}^{\rm F}_{x}$ is
provided by the metric $g$ evaluated at the point $x$. Local coordinates
on the total space are pairs $(x,f)$ where $x\equiv (x^{\mu})$ are
coordinates referring to some chart of ${\cal M}_{n}$, and $f$ is a
function of the Grassmann variables $\theta^{1},\cdots,\theta^{n}$. The
transition functions are defined in close analogy with the exterior
algebra bundle. A coordinate transformation $x\rightarrow \widetilde
x(x)$ on ${\cal M}_{n}$ is accompanied by $f\rightarrow \widetilde f$
with $\widetilde f$ such that $\widetilde f(\widetilde \theta)=f
(\theta)$ where $\widetilde \theta^{\mu}\equiv (\partial \widetilde
x^{\mu}/ \partial x^{\nu})\theta^{\nu}$, i.e. , $\theta^{\mu}$
transforms in the same manner as $dx^{\mu}$.

Sections through a Weyl algebra bundle are locally represented by
functions
\begin{eqnarray}
x\mapsto F(x,\cdot)\in {\cal W}^{\rm F}_{x} \label{2.55}
\end{eqnarray}
where
\begin{eqnarray}
F(x,\cdot):\theta\mapsto F(x,\theta) \label{2.56}
\end{eqnarray}
is a function of $n$ commuting and $n$ anticommuting variables. We
define a fiberwise star product of two such sections by
\begin{eqnarray}
(F_{1}\circ F_{2})(x,\theta)=\Big( F_{1}(x,\cdot)\circ
F_{2}(x,\cdot) \Big)(\theta) \label{2.56a}
\end{eqnarray}
for each point $x$.

We can apply the inverse symbol map to $F(x,\cdot)$ and thus obtain a
family of operators labelled by the space-time points $x$:
\begin{eqnarray}
\widehat F (x) \equiv {\rm symb}^{-1} F(x,\cdot) \label{2.57}
\end{eqnarray}
If we fix a concrete matrix representation of the fermionic operators on
some representation space ${\cal V}^{\rm F}$, then $\widehat F (x)$ acts
on a local copy ${\cal V}^{\rm F}_{x}$ of ${\cal V}^{\rm F}$, i.e.,
$\widehat F (x)\in {\cal L}({\cal V}^{\rm F}_{x}).$ We are particularly
interested in the case where ${\cal V}^{\rm F}$ carries the irreducible
$2^{n/2}$-dimensional representation of the Clifford algebra (for $n$
even). Then ${\cal V}^{\rm F}_{x}$ is a fiber of the usual spin bundle
over ${\cal M}_{n}$ whose sections are the familiar Dirac spinor fields.

In the present paper we shall not be concerned with the global
properties of Weyl algebra bundles. Our main interest is in the
metaplectic analog of the Dirac-K\"ahler construction, and for this
purpose it is sufficient to compare to the topologically trivial bundles
over the flat space-time ${\cal M}_{n}={\mathbf R}^{n}$. An analogous
discussion could be given for arbitrary curved space-times as well, but
we shall avoid the necessary technical complications here. Thus, in our
case, sections can be represented globally by functions $F(x,\theta)$.
We remark that there exists a natural inner product on the space of
these functions:
\begin{eqnarray}
\langle F_{1}| F_{2}\rangle=\int d^{n} x \,(F_{1}(x,\cdot)|
F_{2}(x,\cdot)) \label{2.58}
\end{eqnarray}

\subsection{Dirac-K\"ahler fields and symbol calculus}

Let us assume we are given an arbitrary DK field $\Phi$ on
flat euclidean space-time ${\mathbf R}^{n}$. It possesses an expansion
\begin{eqnarray}
\Phi(x)=\sum_{p=0}^{n}\frac{1}{p!}\,
F_{\mu_{1}\cdots\mu_{p}}^{(p)}(x)\,dx^{\mu_{1}}\wedge\cdots\wedge  
dx^{\mu_{p}} \label{2.59}
\end{eqnarray}
The (complex) coefficient functions are taken to be completely
antisymmetric in all $p$ indices so that there is a bijective
correspondence between forms $\Phi$ and sets
$\{F_{\mu_{1}\cdots\mu_{p}}^{(p)}\}$ of antisymmetric tensors.  Given these
tensors, we form the following matrix-valued field:
\begin{eqnarray}
\widehat F
(x)=\sum_{p=0}^{n}\frac{1}{p!}\,F_{\mu_{1}\cdots\mu_{p}}^{(p)}(x)\,
\gamma^{\mu_{1}}\gamma^{\mu_{2}}\cdots\gamma^{\mu_{p}}  
\label{2.60}
\end{eqnarray}
From now on we assume that $n$ is even and that the Dirac matrices are
in their irreducible representation. Hence $\widehat F (x)$ acts on a
local copy ${\cal V}^{\rm F}_{x}$ of the representation space ${\cal
  V}^{\rm F}={\mathbf C}^{k}, k=2^{n/2}$. By virtue of (\ref{2.35}) we may
regard $\widehat F (x)$ as a matrix realization of the abstract operator
\begin{eqnarray}
\widehat F
(x)=\sum_{p=0}^{n}\frac{\kappa^{p}}{p!}\,F_{\mu_{1}
  \cdots\mu_{p}}^{(p)}(x)\,\widehat\chi^{\mu_{1}}\widehat\chi^{\mu_{2}}
\cdots\widehat\chi^{\mu_{p}}  
\label{2.61}
\end{eqnarray}
For every point $x$, the symbol of this operator is $F(x,\theta)=[{\rm
  symb}\,\widehat F (x)](\theta)$, or
\begin{eqnarray}
F
(x,\theta)=\sum_{p=0}^{n}\frac{\kappa^{p}}{p!}\,
F_{\mu_{1}\cdots\mu_{p}}^{(p)}(x)\,\theta^{\mu_{1}}
\theta^{\mu_{2}}\cdots\theta^{\mu_{p}}  
\label{2.62}
\end{eqnarray}
Thus we have set up a linear one-to-one correspondence between
differential forms $\Phi(x)$ and symbol functions $F(x,\theta)$.
Schematically,
\begin{eqnarray}
\Phi(x)\in\bigwedge(T^{\ast}_{x}{\cal M}_{n}) 
\, \leftrightarrows \, 
\widehat F (x)\in {\cal L}({\cal V}^{\rm F}_{x})
\, \leftrightarrows \,
F(x,\cdot)\in {\cal W}^{\rm F}_{x} \label{2.63}
\end{eqnarray}
The first one of the two bijections in (\ref{2.63}) is the usual
``Dirac-K\"ahler correspondence'' $dx^{\mu}\leftrightarrows
\gamma^{\mu}$ which we mentioned already in the introduction, while the
second one is the Weyl symbol map. Taken in conjunction, these maps
relate DK-fields to symbols. In particular,
\begin{eqnarray}
dx^{\mu}\, \leftrightarrows \, \kappa \theta^{\mu} \label{2.64}
\end{eqnarray}
We shall use the notation $\Phi:F\mapsto \Phi[F]$ for the linear map
which yields the differential form belonging to a given symbol. For
instance,
\begin{eqnarray}
\Phi[\kappa\theta^{\mu}]=dx^{\mu} \label{2.65}
\end{eqnarray}

What makes the above construction particularly useful is that under the
map $\Phi$ many of the familiar linear and bilinear operations involving
differential forms naturally pass over to the symbol functions and vice
versa.  This is immediately obvious for the automorphism $\cal A$, the
antiautomorphism $\cal B$, the Hodge operator $\ast$ and the modified
Hodge operator $\hodge$. Comparing their definition for symbols in
appendix A to their standard definition in terms of differential
forms one sees that
\begin{eqnarray}
{\cal A} \Phi[F]&=&\Phi[{\cal A}F]\,,\qquad {\cal B} \Phi[F]=\Phi[{\cal
  B}F] \nonumber \\ 
\ast \Phi[F]&=&\Phi[\ast F]\,,\qquad {\bf \hodge} \Phi[F]=\Phi[{\bf
  \hodge}F]  
\label{2.66}
\end{eqnarray}
The exterior derivative $d=dx^{\mu}\partial_{\mu}$ translates into
$\kappa \theta^{\mu}\partial_{\mu}$,
\begin{eqnarray}
d\Phi[F]=\Phi[\kappa\theta^{\mu}\partial_{\mu}F] \label{2.67}
\end{eqnarray}
while the contraction ${\bf i} (v)$ with a vector field
$v=v^{\mu}\partial_{\mu}$ becomes a derivative with respect to $\theta$:
\begin{eqnarray}
{\bf i}(v)\Phi[F]=\Phi[\kappa^{-1}v^{\mu}\frac{\partial}{\partial
  \theta^{\mu}}F] \label{2.68}
\end{eqnarray}
In particular,
\begin{eqnarray}
e_{\mu}\neg \Phi[F]=\Phi[\kappa^{-1}\frac{\partial}{\partial
  \theta^{\mu}}F] \label{2.69}
\end{eqnarray}
The natural inner product on the space of DK-fields is\footnote{All
  terms which are not of degree $n$ are supposed to be discarded from
  the integrand in (\ref{2.70}).}
\begin{eqnarray}
\langle\Phi_{1},\Phi_{2}\rangle=\int \bar\Phi_{1}\wedge \ast \Phi_{2}
\label{2.70} 
\end{eqnarray}
Its counterpart at the symbol level is (\ref{2.58}) with (\ref{2.47}):
\begin{eqnarray}
\langle\Phi[F_{1}],\Phi[F_{2}]\rangle=\langle F_{1}| F_{2}\rangle
\label{2.71} 
\end{eqnarray}
The coderivative $d^{\dagger}$ is the formal adjoint of $d$ with respect
to $\langle \cdot , \cdot\rangle$. On flat space one has
\begin{eqnarray}
d^{\dagger}\Phi=-e^{\mu}\neg\partial_{\mu}\Phi \label{2.72}
\end{eqnarray}
whence
\begin{eqnarray}
d^{\dagger}
\Phi[F]=\Phi[-\kappa^{-1}\partial_{\mu}\frac{\partial}{\partial
  \theta_{\mu}}F] \label{2.73}
\end{eqnarray}

The wedge product of differential forms is mapped onto the pointwise
product of symbol functions:
\begin{eqnarray}
\Phi[F_{1}]\wedge\Phi[F_{2}]=\Phi[F_{1}F_{2}] \label{2.74}
\end{eqnarray}

The most important aspect of the form/symbol correspondence is that the
image of the Clifford product is precisely the fiberwise star product
(\ref{2.56a}):
\begin{eqnarray}
\Phi[F_{1}]\vee\Phi[F_{2}]=\Phi[F_{1}\circ F_{2}] \label{2.75}
\end{eqnarray}
This can be seen for instance by mapping K\"ahler's formula (\ref{1.4})
for the Clifford product on our representation (\ref{2.15}) of the
fermionic Weyl star product:
\begin{eqnarray}
\Phi[F_{1}]\!\!\!&\vee&\!\!\!\Phi[F_{2}]= \nonumber \\
&=&\!\!\!  \sum_{p=0}^{n}\frac{(-1)^{p(p-1)/2}}{p!}\Big( {\cal A}^{p}
  e_{\mu_{1}}\neg \cdots  
e_{\mu_{p}}\neg\Phi[F_{1}]\Big)\wedge \Big( e^{\mu_{1}}\neg \cdots
e^{\mu_{p}}\neg\Phi[F_{2}]\Big) \nonumber \\ 
&=&\!\!\! \sum_{p=0}^{n}\frac{(-1)^{p(p-1)/2}}{p!}\Phi[\kappa^{-p}{\cal
  A}^{p} \frac{\partial}{\partial 
  \theta^{\mu_{1}}} \cdots 
\frac{\partial}{\partial \theta^{\mu_{p}}} F_{1}]\wedge \Phi
[\kappa^{-p} \frac{\partial}{\partial 
    \theta_{\mu_{1}}} \cdots 
\frac{\partial}{\partial \theta_{\mu_{p}}} F_{2}]\nonumber \\
&=&\!\!\! \Phi \left[ \sum_{p=0}^{n}\kappa^{-2p}
  \frac{(-1)^{p(p-1)/2}}{p!}\left( {\cal A}^{p} \frac{\partial} 
{\partial \theta^{\mu_{1}}} \cdots \frac{\partial}{\partial
  \theta^{\mu_{p}}} F_{1}\right)\left(\frac{\partial} 
{\partial \theta_{\mu_{1}}} \cdots \frac{\partial}{\partial
  \theta_{\mu_{p}}} F_{2}\right) \right] \nonumber \\ 
&=&\!\!\! \Phi[ F_{1}\circ F_{2}] \label{2.76}
\end{eqnarray}
Here we used (\ref{2.66}), (\ref{2.69}) and (\ref{2.74}).

One also could prove eq.(\ref{2.75}) inductively. If we replace
$dx^{\mu}$ by $\kappa \theta^{\mu}$ and ''$\vee$'' by the star product
in the relations (\ref{1.3}) which define the Clifford product we obtain
exactly eqs.(\ref{2.12}) for the star product. Therefore
eq.(\ref{2.75}) is correct for zero-and one-forms. Its generalization
for arbitrary $p$-forms makes essential use of the associativity of both
the Clifford and the star product.

By virtue of our rules for the form/symbol correspondence also the
equations (\ref{1.9a}) and (\ref{2.18}) are now seen to be completely
equivalent.

In the DK-equation we need the Clifford product of $dx^{\mu}$ with an
arbitrary form :
\begin{eqnarray}
d x^{\mu}\vee\Phi[F]& = &\Phi[\kappa\theta^{\mu}]\vee\Phi[F] \nonumber\\
& = &\Phi[\kappa \theta^{\mu}\circ F]\nonumber \\
& = &\Phi[ C^{\mu}F] \label{2.77}
\end{eqnarray}
Here $C^{\mu}$ is the first order differential operator (\ref{2.53}). In
the introduction we discussed already that $dx^{\mu}\vee$, regarded as
an operator on the space of DK-fields, gives rise to the Clifford
algebra (\ref{1.10}).  For consistency the same should be true for the
star multiplication with $\kappa \theta^{\mu}$ and for $C^{\mu}$ on the
space of symbols. In fact, it is easy to see that
\begin{eqnarray}
(\kappa \theta^{\mu}\circ) (\kappa \theta^{\nu}\circ)+ (\kappa
\theta^{\nu}\circ) (\kappa \theta^{\mu}\circ)=2\delta^{\mu\nu} \label{2.78}
\end{eqnarray}
and
\begin{eqnarray}
C^{\mu} C^{\nu}+ C^{\nu} C^{\mu}=2\delta^{\mu\nu} \label{2.79}
\end{eqnarray}

The DK-operator acting on forms reads
\begin{eqnarray}
(d-d^{\dagger})\Phi& =
&dx^{\mu}\wedge\partial_{\mu}\Phi+e^{\mu}\neg\partial_{\mu}\Phi
\nonumber\\ 
& = & dx^{\mu}\vee \partial_{\mu}\Phi \label{2.80}
\end{eqnarray}
where the second equality follows from (\ref{1.9a}). Therefore
$d-d^{\dagger}$ becomes $\kappa \theta^{\mu}\circ\partial_{\mu}$ or
$C^{\mu}\partial_{\mu}$ at the symbol level:
\begin{eqnarray}
(d-d^{\dagger})\Phi[F]& =
&\Phi[\kappa\theta^{\mu}\circ\partial_{\mu}F]\nonumber\\ 
& = &\Phi[C^{\mu}\partial_{\mu}F] \label{2.81}
\end{eqnarray}
This converts the DK-equation to 
\begin{eqnarray}
\left[ \left( \frac{1}{\kappa} \frac{\partial}{\partial
      \theta_{\mu}}+\kappa
    \theta^{\mu}\right)\partial_{\mu}+m\right]F(x,\theta)=0 \label{2.82}
\end{eqnarray}

In closing we return to the chirality operator $\gamma_{n+1}$. Under the
map $\Phi$, the image of the delta-function is essentially the volume
form ${\rm Vol}\equiv dx^{1}\wedge dx^{2}\wedge\cdots\wedge dx^{n}$:
\begin{eqnarray}
\Phi[\kappa^{n}\delta(\cdot)]=(-1)^{n(n-1)/2}\,{\rm Vol}  \label{2.82a}
\end{eqnarray}
For the chirality operator this means that
\begin{eqnarray}
\Phi[G_{n+1}\circ F]=-i^{n(n-1)/2}\,{\rm Vol} \vee \Phi [F] \label{2.82c}
\end{eqnarray}
which, at the operator level, corresponds to 
\begin{eqnarray}
\widehat{G_{n+1}\circ F}=\gamma_{n+1}\widehat F \label{2.82d}
\end{eqnarray}
Thus we see that (up to unimportant constants) the fermionic
$\delta$-function, the volume form, and the chirality matrix
$\gamma_{n+1}$ are simply different variants of the same object.
Furthermore, by eq.(\ref{2.43}), star multiplication of $F$ by $G_{n+1}$
{\it from the right} amounts to applying the modified Hodge operator
$\cal \hodge$.

\subsection{Invariant subspaces of ${\cal W}^{\rm F}$}
The differential operators $C^{\mu}$ or the star left-multiplication by
$\kappa \theta^{\mu}$ define a representation of the Clifford algebra
(\ref{2.21}) in the space of symbols $f(\theta)$. As $f(\theta)$ has
$2^{n}$ independent (complex) components, this representation is
reducible. It can be decomposed into $2^{n/2}$ representations each of
which is isomorphic to the $2^{n/2}$-dimensional irreducible
representation provided by the matrices $\gamma^{\mu}$. (We assume $n$
even in this section.) As a consequence, a symbol-valued field
$F(x,\theta)$ describes $2^{n/2}$ ordinary Dirac spinor fields.

In the light of the form/symbol correspondence which we developed in the
previous section it is clear that the representation carried by
${\cal W}^{\rm F}$
could be decomposed simply by invoking the standard discussion at the
level of differential forms. However, as our main motivation for
studying the symbol formulation of DK-fields is to get some
understanding of their symplectic analogs we shall reformulate the
method of Becher and Joos \cite{bj} in symbol language and use this as a
guide in the symplectic case. As a by-product we shall find a very
elegant derivation of their matrix-valued form $Z$ which puts it in a
more general perspective.

We have to decompose the Weyl algebra in orthogonal subspaces, 
\begin{eqnarray}
{\cal W}^{\rm F}=\bigoplus_{\alpha=1}^{k} {\cal W}^{\rm
  F}_{(\alpha)}\qquad,\quad 
k\equiv 2^{n/2} \label{2.83}
\end{eqnarray}
such that ${\cal W}^{F}_{(\alpha)}$ is invariant under star
left-multiplication by $\theta^{\mu}$. Following a strategy similar to
the one described in the introduction we look for a $k\times k $-matrix
valued function $Z(\theta)$ with the property
\begin{eqnarray}
\kappa \theta^{\mu}\circ
Z_{\alpha\beta}(\theta)=\sum_{\gamma=1}^{k}\,(\gamma^{\mu
  T})_{\alpha\gamma}\,Z_{\gamma\beta}(\theta) \label{2.84}
\end{eqnarray}
The function $Z$ is readily found in our formalism. Since the star
product with $\theta^{\mu}$ involves first derivatives at most, 
eq.(\ref{2.84}) is reminiscent of the formulas for the derivative of the
Weyl operators $\widehat \Omega$ and $\check \Omega$ which we displayed
in section 2.1. In fact, using those formulas together with (\ref{2.18})
it is easy to show that there exists a rescaling of the arguments of
$\widehat \Omega$ and $\check \Omega$ in such a way that the star
multiplication by $\theta^{\mu}$ corresponds to an operator
multiplication by $\widehat \chi^{\mu}$ or $\Gamma^{\mu}$:
\begin{eqnarray}
\theta^{\mu}\circ\widehat\Omega (\pm i\kappa^{2}\theta)& = & \mp
\,\widehat \Omega (\pm i \kappa^{2} \theta) \,\widehat \chi^{\mu}
\label{2.85} \\ 
\theta^{\mu}\circ\widehat\Omega (\pm \kappa^{2}\theta)& = & \pm\, i\,
\widehat \chi^{\mu}\, \widehat \Omega (\pm  \kappa^{2} \theta)
\label{2.86}\\ 
\kappa \theta^{\mu}\circ\check\Omega (\pm i\kappa^{2}\theta)& = & \pm\, 
\Gamma^{\mu}  \,\check \Omega (\pm i \kappa^{2} \theta) \label{2.87} 
\end{eqnarray}
For the problem at hand, eq.(\ref{2.87}) is precisely what we need. If
$\Gamma_{\mu}$ constitutes a Clifford algebra, so does
$\Gamma_{\mu}^{T}$. Hence we may set $Z(\theta)=\check \Omega (i
\kappa^{2}\theta)$ with $\Gamma_{\mu}=\gamma^{T}_{\mu}$. Thus
\begin{eqnarray}
Z(\theta)={\rm exp}\left[\kappa \theta^{\mu}\gamma_{\mu}^{T}\right]
\label{2.88} 
\end{eqnarray}
or in expanded form
\begin{eqnarray}
Z(\theta)=\sum_{p=0}^{n}\frac{\kappa^{p}}{p!}\,
\gamma_{\mu_{1}}^{T}\cdots\gamma_{\mu_{p}}^{T}\,\theta^{\mu_{1}}
\cdots\theta^{\mu_{p}}  
\label{2.89}
\end{eqnarray}
Clearly (\ref{2.89}) is precisely the symbol corresponding to the form
(\ref{1.16}) which was found by Becher and Joos \cite{bj} using
different techniques. In the context of the present investigation it is
important to keep in mind that $Z$ is nothing but a rescaled fermionic Weyl
operator since the latter has a well-known bosonic analog.

Because of the completeness properties of the $\gamma$-matrices,
$\{Z_{\alpha\beta}; \\ \alpha,\beta=1,\cdots,k\}$ is a basis for ${\cal
  W}^{\rm F}$ and we may expand any symbol as
\begin{eqnarray}
F(x,\theta)& = &\sum_{\beta=1}^{k}\sum_{\alpha=1}^{k}\,
\psi^{(\beta)}_{\alpha}(x)\,Z_{\alpha\beta}(\theta) \nonumber \\
& = &\sum_{\beta=1}^{k}\,F^{(\beta)}(x,\theta) \label{2.90}
\end{eqnarray}
(Here we use already the notation appropriate for the role of ${\cal
  W}^{\rm F}$ as a fiber at the point $x$.) The rest of the argument
parallels our discussion in the Introduction. We obtain $k\equiv2^{n/2}$
invariant subspaces ${\cal W}^{\rm F}_{(\alpha)}$ (left ideals) which
are spanned by
\begin{eqnarray}
F^{(\beta)}(x,\cdot)& =
&\sum_{\alpha=1}^{k}\psi^{(\beta)}_{\alpha}(x)Z_{\alpha\beta}(\cdot) \quad \in
\,{\cal W}^{\rm F}_{x(\beta)} \label{2.91}
\end{eqnarray}
For every fixed value of $\beta$, the expansion coefficients
$\psi^{(\beta)}\equiv\{\psi^{(\beta)}_{\alpha};\alpha=1,\cdots,k\}$ can be
interpreted as an ordinary Dirac spinor.  Eq.(\ref{2.84}) shows that
acting with $\kappa\theta^{\mu}\circ$ on $F^{(\beta)}$ is equivalent to
applying $\gamma^{\mu}$ on $\psi^{(\beta)}$:
\begin{eqnarray}
\kappa\theta^{\mu}\circ F^{(\beta)}& = &\sum_{\alpha}\left(\sum_{\delta}
  \gamma_{\alpha\delta}^{\mu}\, \psi_{\delta}^{(\beta)}\right)\,
Z_{\alpha\beta}\nonumber \\  
& =
&\sum_{\alpha}\,[\gamma^{\mu}\psi^{(\beta)}]_{\alpha}\,Z_{\alpha\beta}
\label{2.92} 
\end{eqnarray}

Let us arrange the expansion coefficients $\psi_{\alpha}^{(\beta)}$ as a
$k \times k$-matrix: $(\widehat \psi)_{\alpha\beta}\equiv
\psi_{\alpha}^{(\beta)}$. Then,
\begin{eqnarray}
F(x,\theta)={\rm{Tr}}\left[\widehat \psi(x) Z(\theta)^{T}\right]
\label{2.92a} 
\end{eqnarray}
Denoting the $\widehat \psi$-matrix which belongs to a given section $F$
by $\widehat \psi[F]$ we obtain from (\ref{2.92})
\begin{eqnarray}
\widehat \psi [\kappa\theta^{\mu}\circ F]=\gamma^{\mu}\, \widehat \psi
[F] \label{2.93} 
\end{eqnarray}
which mirrors (\ref{1.22}) at the symbol level.

Given a symbol-valued field $F(x,\theta)$ we can immediately construct
the associated spinor matrix-valued field $\widehat F(x)$ of
(\ref{2.60}) by replacing $\theta^{\mu}\rightarrow
\kappa^{-1}\gamma^{\mu}$ in its series expansion (\ref{2.62}). In the
process of decomposing the reducible representation carried by $F$ we
discovered a second spinor-matrix, $\widehat \psi$, which is related to
$F$ in a canonical way, too. By essentially the same argument as in the
introduction it follows that the two matrices are equal up to a
constant:
\begin{eqnarray}
\widehat \psi [F](x)=2^{-n/2}\, \widehat F (x) \label{2.94}
\end{eqnarray}
If we insert the expansions (\ref{2.62}) and (\ref{2.89}) for
$F(x,\theta)$ and $Z(\theta)$, respectively, into eq.(\ref{2.92a}), we
obtain eq.(\ref{1.25}) for the set $\{F^{(p)}_{\mu_{1}\cdots\mu_{p}}\}$
expressed in
terms of $\widehat \psi$. Making an ansatz for $\widehat \psi$ in terms
of antisymmetrized products of $\gamma$-matrices and using the trace
identity (\ref{1.27}), one finds that the expansion coefficients of
$\widehat \psi$ and $\widehat F$ differ by an overall constant only.
  
  While this last step was straightforward for the ${\rm
    SO}(n)$-spinors, it will be much less trivial for metaplectic
  spinors where the representation space is infinite-dimensional and
  trace-identities such as (\ref{1.27}) are not likely to exist. It will
  be interesting to see how (\ref{2.94}) is modified then.

\renewcommand{\theequation}{3.\arabic{equation}}
\setcounter{equation}{0}
\section{Symplectic Dirac-K\"ahler fields}
In the previous section we reformulated the theory of standard
DK-fermions over space-time in terms of fields $F$ which assume values in
the fermionic Weyl algebra ${\cal W}^{\rm F}$. Now we are going to ask what
happens if we replace ${\cal W}^{\rm F}$ by its (actually much more
familiar) bosonic counterpart, the bosonic Weyl algebra ${\cal W}$.
Rather than space-time it is now a phase-space $({\cal M}_{2N},\omega)$
which plays the role of the base manifold. As we shall argue, replacing
the Riemannian structure by a symplectic one, the structure group
$SO(n)$ by ${\rm Sp}(2N)$, and, most importantly, fermionic Weyl symbols by
bosonic ones, we are led to the notion of a ``symplectic DK-field'' in a
very natural way.

In Subsection 3.1 we begin by working out some special properties of
bosonic Weyl symbols which will become important in our construction. In
this context, we are basically discussing the conventional quantum
mechanics of the auxiliary quantum system with canonical operators
$\widehat x ^{i}$ and $\widehat \pi ^{i}$ which results from quantizing
the flat ``auxiliary phase-space'' ${\mathbf R}^{2N}$. (Later on the
auxiliary phase-space will be identified with the tangent space to the
true (physical) phase-space ${\cal M}_{2N}$.) The operators $\widehat x
^{i}$ and $\widehat \pi ^{i}$ take over the role previously played by
$\widehat \chi ^{\mu}$.

Subsection 3.2 is devoted to the metaplectic $\gamma$-matrices. In
particular, we propose a symplectic analog of the chirality matrix
$\gamma_{5}$ there. The actual construction of the symplectic DK-fields
is performed in Section 3.3, and in Section 3.4 it is shown how they
relate to the metaplectic spinor fields.

\subsection{Bosonic Weyl symbols}

We consider a hamiltonian system with $N$ degrees of freedom whose
classical phase-space is the symplectic plane $({\mathbf
  R}^{2N},\omega)$. The associated quantum mechanical Hilbert space is
${\cal V}$ and ${\cal L(V)}$ denotes the space of linear
operators on ${\cal V}$. The Hilbert space ${\cal V}$ carries a
representation of the canonical commutation relations
\begin{eqnarray}
[\widehat \varphi^{a},\widehat \varphi^{b}]=i\,\hbar\,\omega^{ab};\qquad
a,b=1,\cdots,2N \label{3.1}
\end{eqnarray}
In a canonical operator basis we split $\widehat
\varphi^{a}\equiv(\widehat \pi ^{i},\widehat x ^{i}),\, i=1,\cdots,N$, so
that the only nonvanishing commutator is between the momenta $\widehat
\pi ^{i}$ and the positions $\widehat x ^{i}$ $:[\widehat \pi^ {i},\widehat
x ^{j}]=-i\hbar \delta^{ij}$. The matrix $(\omega^{ab})$ is the inverse
of the constant matrix $(\omega_{ab})$ formed from the
coefficients of the symplectic 2-form $\omega$:
$\omega_{ab}\omega^{bc}=\delta_{a}^{c}$. On $({\mathbf R}^{2N},\omega)$
we use canonical coordinates $y^{a}\equiv(y^{i}_{p},y_{q}^{i})$ such
that
\begin{displaymath}
(\omega_{ab})=
\left( \begin{array}{cc}
0 & +I \\
-I & 0 \end{array} \right), \qquad
(\omega^{ab})=
\left( \begin{array}{cc}
0 & -I \\
+I & 0 \end{array} \right) \label{3.2}
\end{displaymath}
For the natural skew-symmetric inner product on the symplectic plane we
write
\begin{eqnarray}
\omega(y_{1},y_{2})\equiv y^{a}_{1}\,\omega_{ab}\,y_{2}^{b} \label{3.3}
\end{eqnarray}
The Weyl (or Heisenberg) operators \cite{lj}
\begin{eqnarray}
\widehat T (y)={\rm exp} (\frac{i}{\hbar} y^{a}\omega_{ab}\widehat
\varphi^{b}) \label{3.4} 
\end{eqnarray}
implement the translations on phase-space in the Hilbert space $\cal V$:
\begin{eqnarray}
\widehat T (y)^{\dagger}\,\widehat \varphi ^{a} \,\widehat T (y)=\widehat
\varphi ^{a} +y^{a} \label{3.5}
\end{eqnarray}
This is a projective representation of the translation group since 
\begin{eqnarray}
\widehat T (y_{1}) \,\widehat T (y_{2})={\rm exp} [\frac{i}{2\hbar}
\omega(y_{1},y_{2})]\,\widehat T (y_{1}+y_{2}) \label{3.6}
\end{eqnarray}
The Weyl operators are orthogonal and complete in the sense that
\begin{eqnarray}
{\rm \rm{Tr}}[\widehat T (y_{1})^{\dagger}\, \widehat T (y_{2})]\!\!\!& =
&\!\!\!(2\pi\hbar)^{N}\,\delta^{(2N)}(y_{1}-y_{2}) \label{3.7} \\
\int d^{2N} y \,\langle \alpha| \widehat T (y)^{\dagger}| \alpha '
\rangle \langle\beta| \widehat T (y)| \beta '\rangle\!\!\! & = &\!\!\!
(2\pi\hbar)^{N}\,\delta^{(N)}(\alpha-\beta ')\,\delta^{(N)}(\beta-\alpha ') 
\label{3.8}
\end{eqnarray}
Here $\{ | \alpha \rangle \}$ is the basis which diagonalizes the
position operators: 
\begin{eqnarray}
\widehat x ^{i} | \alpha \rangle =\alpha^{i}|\alpha\rangle, \qquad
\alpha\equiv(\alpha^{1},\cdots,\alpha^{N}) \label{3.9}
\end{eqnarray}

Sometimes it will be more suggestive to use a tensor notation instead of
the bra-ket formalism; for instance, one writes $\widehat b
^{\alpha}_{\,\,\beta}\equiv \langle \alpha | \widehat b |\beta\rangle$
for the matrix elements of some arbitrary $\widehat b \in {\cal L(V)}$
or $\delta^{\alpha}_{\,\,\beta}\equiv\delta^{(N)}(\alpha-\beta)$ for the
identity operator. The eigenvalues $\alpha \in \mathbb{R}^N$ should be
thought of as a continuous analog of a spinor index. In the $\widehat
x$-eigenbasis, the Weyl operators are given by
\begin{eqnarray}
\widehat T (y)^{\alpha}_{\,\,\beta}={\rm
  exp}\left[\frac{i}{\hbar}(y_{p}\alpha-\frac{1}{2}y_{p}y_{q})\right]
\,\delta^{(N)}(\alpha-\beta-y_{q}) \label{3.10} 
\end{eqnarray}
with $y_{p}\alpha\equiv y^{i}_{p}\alpha^{i}$, etc., where the summation
over $i=1,\cdots,N$ is understood.

>From the completeness relation (\ref{3.8}) it follows that every
operator $\widehat b$ can be represented as
\begin{eqnarray}
\widehat b =(2\pi\hbar)^{-N}\int d^{2N} y \,\widetilde b (y)\, \widehat
T 
(y) \label{3.11}
\end{eqnarray}
with the complex-valued function $\widetilde b$ given by
\begin{eqnarray}
\widetilde b  (y)={\rm \rm{Tr}} \left[\widehat T (y)^{\dagger}
  \,\widehat b \right] \label{3.12} 
\end{eqnarray}
The function $\widetilde b$ (referred to as the {\it alternative}
Weyl symbol \cite{lj}) is closely related to the Weyl
symbol of $\widetilde b$. In fact, $b(y)\equiv [{\rm symb} (\widehat b
)](y)$ is the Fourier transform of $\widetilde b$:
\begin{eqnarray}
b(y)=(2\pi\hbar)^{-N}\int d^{2N} y_{0}\, \widetilde b (y_{0})\, {\rm exp}
[\frac{i}{\hbar}\omega(y_{0},y)] \label{3.13}
\end{eqnarray}
The inverse transformation reads
\begin{eqnarray}
\widetilde b (y)=(2\pi\hbar)^{-N}\int d^{2N} y_{0} \,b (y_{0})\, {\rm exp}
[\frac{i}{\hbar}\omega(y_{0},y)] \label{3.14}
\end{eqnarray}
i.e., the symplectic Fourier transformation is an exact involution,
${\widetilde {\widetilde b}}(y) =b(y)$ (and not only an involution up to a
reflection of the argument).

Eqs. (\ref{3.11})-(\ref{3.13}) define the (bosonic) Weyl symbol map
``symb'' from ${\cal L(V)}$ to the space of (generalized) functions over
the symplectic plane, as well as its inverse. The classical
phase-function $b(y)$ uniquely represents an operator $\widehat b$ which
is Weyl ordered. In particular, the monomial $y^{a_{1}} y^{a_{2}}\cdots
y^{a_{p}}$ stands for the completely symmetrized operator product
$\widehat \varphi^{(a_{1}}\widehat \varphi^{a_{2}}\cdots\widehat
\varphi^{a_{p})}$. Conversely,
\begin{eqnarray}
[{\rm symb} \{\widehat \varphi^{(a_{1}}\cdots\widehat
  \varphi^{a_{p})}\}](y)= y^{a_{1}} y^{a_{2}}\cdots y^{a_{p}} \label{3.15}
\end{eqnarray}
The symmetrization in (\ref{3.15}) is crucial, otherwise commutator
terms would occur. For instance,
\begin{eqnarray}
[{\rm symb} \{\widehat \varphi^{a} \widehat \varphi^{b}\}](y)= y^{a}
y^{b}+i\frac{\hbar}{2}\omega^{ab} \label{3.16}
\end{eqnarray}

An important special class of symbols are those which admit a power
series expansion
\begin{eqnarray}
b(y)=\sum_{p=0}^{\infty}\frac{\kappa^{p}}{p!}\, b^{(p)}_{a_{1}\cdots a_{p}}
\, y^{a_{1}} y^{a_{2}}\cdots y^{a_{p}} \label{3.17}
\end{eqnarray}
By the symbol map, they are bijectively related to the operators
\begin{eqnarray}
\widehat b=\sum_{p=0}^{\infty}\frac{\kappa^{p}}{p!}\,b^{(p)}_{a_{1}\cdots
  a_{p}} \,\widehat \varphi ^{a_{1}}\cdots \widehat \varphi ^{a_{p}}
\label{3.18} 
\end{eqnarray}
provided the tensors $ b^{(p)}_{a_{1}\cdots a_{p}}$ are completely
symmetric. If $b$ is a power series, the ``alternative Weyl symbol''
$\widetilde b$ is a sum of derivatives of $\delta$-functions:
\begin{eqnarray}
\widetilde b (y)=(2\pi\hbar)^{N}
\,b\left(i\hbar\omega^{ac}\frac{\partial}{\partial
    y^{c}}\right)\,\delta^{(2N)}(y) \label{3.19} 
\end{eqnarray}

As in every symbol calculus, the pertinent star product is required to
satisfy ${\rm symb}(\widehat b_{1} \widehat b_{2} )=b_{1}\circ b_{2}$
where $b_{1}$ and $b_{2}$ are the symbols of $\widehat b_{1}$ and
$\widehat b_{2}$, respectively. At least for power series, the bosonic
Weyl star product is uniquely determined by its associativity, the
distributivity over '+', and the basic relations
\begin{eqnarray}
1\circ 1=1,\quad&\,& \quad 1\circ y^{a}=y^{a}\circ 1=y^{a} \nonumber \\
y^{a}\circ y^{b}&=& y^{a} y^{b}+i\frac{\hbar}{2} \omega^{ab} \label{3.20}
\end{eqnarray}
which follow from (\ref{3.15}), (\ref{3.16}) and ${\rm symb}(I)=1$.
Explicit formulas for the star product \cite{ber,winf} of arbitrary
symbols include
\begin{eqnarray}
(b_{1}\circ b_{2})(y)=b_{1}(y) \,{\rm exp} \left[ i\frac{\hbar}{2}
\stackrel{\longleftarrow}{\frac{\partial}{\partial y^{a}}} \omega^{ac}
\stackrel{\longrightarrow}{\frac{\partial}{\partial y^{c}}}  
\right]b_{2}(y) \label{3.21}
\end{eqnarray}
and
\begin{eqnarray}
(b_{1}\circ b_{2})(y)\!\!\!&=& 
 (\pi\hbar)^{-2N} \int\!\! d^{2N}y_{1}\, d^{2N} y_{2} \,{\rm
  exp}\left[\right. -2 i
  \{\omega(y,y_{1})\nonumber \\
&\,&+\omega(y_{1},y_{2})+\omega(y_{2},y)\}
/\hbar \left. \right]\, b_{1}(y_{1})
b_{2}(y_{2}) \label{3.22}
\end{eqnarray}

The differential operators which effect the star left-multiplication
with $\kappa y^{a}$, 
\begin{eqnarray}
(C^{a}b)(y)=\kappa y^{a}\circ b(y), \label{3.23}
\end{eqnarray}
are easily read off from eq.(\ref{3.21}):
\begin{eqnarray}
C^{a}=\kappa y^{a}+\frac{i}{\kappa}\,\omega^{ab}\frac{\partial}{\partial
  y^{b}} \label{3.24}
\end{eqnarray}

On the space of symbols with an appropriate fall-off behavior we would
like to introduce a sesquilinear inner product $(\cdot| \cdot)$ with
respect to which $C^{a}$ is selfadjoint,
\begin{eqnarray}
\left( C^{a} b_{1} | b_{2}\right)=\left(b_{1}| C^{a} b_{2}\right)
\label{3.25} 
\end{eqnarray}
It is clear from our earlier discussion that the choice 
\begin{eqnarray}
\left( b_{1} | b_{2}\right)=[\bar b_{1} \circ b_{2} ] (y=0)  \label{3.26}
\end{eqnarray}
meets this requirement. Since $\overline{b_{1} \circ b_{2}}=\bar b_{2}
\circ \bar b_{1}$ also here, the proof is the same as in (\ref{2.54}).
If $b_{1}$ and $b_{2}$ are power series of the type (\ref{3.17}),
eq.(\ref{3.26}) boils down to
\begin{eqnarray}
\left( b_{1} | b_{2}\right)=\sum_{p=0}^\infty \frac{i^{p}}{p!}\, \bar
b_{1,a_{1}\cdots a_{p}}^{(p)}\,
\omega^{a_{1}c_{1}}\cdots\omega^{a_{p}c_{p}} \,b_{2,c_{1}\cdots 
  c_{p}}^{(p)}  \label{3.27}
\end{eqnarray}

It is instructive to look at various alternative ways of representing
this inner product. There exists the integral representation
\begin{eqnarray}
\left( b_{1} | b_{2}\right)=(2\pi)^{-2N}\int d^{2N}y_{1} \,
d^{2N}y_{2}\, \bar
b_{1} (y_{1}/\kappa)\,e^{-i\omega(y_{1},y_{2})}\,b_{2}(y_{2}/\kappa)
\label{3.28} 
\end{eqnarray}
which can be reexpressed in terms of a symplectic Fourier transform:
\begin{eqnarray}
\left( b_{1} | b_{2}\right)=(2\pi\hbar)^{-N}\int d^{2N}y \,\bar b_{1}
(\frac{1}{2}y)\,\widetilde b_{2}(y) \label{3.29}
\end{eqnarray}
Furthermore, if (\ref{3.19}) can be applied,
\begin{eqnarray}
\left( b_{1} | b_{2}\right)=b_{2}\left(-i\kappa^{-2}
\omega^{ac}\frac{\partial}{\partial y^{c}}\right)\,\bar b_{1} (y)
|_{y=0} \label{3.30} 
\end{eqnarray}

\vspace{4mm}
The above formulae should be compared to their counterparts in the
fermionic symbol calculus. Bosonic symbols admitting a power series
expansion are characterized by sets $\{b_{a_{1}\cdots
    a_{p}}^{(p)},p=0,1,2,\cdots\}$ consisting of infinitely many {\it
  symmetric} tensors. Fermionic symbol functions are equivalent to a
finite set $\{f^{(p)}_{\mu_{1}\cdots\mu_{p}},p=0,1,\cdots,n\}$ of {\it
  antisymmetric} tensors instead. 

We saw that the (modified) Hodge
operator is essentially the same operation as the Grassmannian Fourier
transformation. Omitting all sign factors (which anyhow have no bosonic
analog) we have, schematically,
\begin{eqnarray}
\ast f(\theta)\propto {\bf \hodge} f(\theta) \propto \widetilde f
(\theta) \propto f(\frac{\partial}{\partial \theta})\delta(\theta)
\label{3.31} 
\end{eqnarray}
Thus one is tempted to define a bosonic version of the Hodge operator
simply by setting $(\ast b)(y)=\widetilde b (y)$ so that $\ast \ast =1$
on any $b$. If $b$ is a power series, eq.(\ref{3.19}) is indeed formally
analogous to (\ref{A.21a}) for the fermionic Fourier transformation.
However, the difference is that the derivative of the fermionic
delta-function, $f(\partial /\partial \theta)\delta (\theta)$, again is
a powers series in the $\theta$'s, while this is of course not true for
the derivatives of the bosonic delta-function, $\delta^{(2N)}(y)$. In
the former case, the monomials
$\theta^{\mu_{1}}\cdots\theta^{\mu_{p}}$are mapped onto monomials of the
same type. Therefore one set of antisymmetric tensors $\{f^{(p)}_{a_{1}\cdots
    a_{p}}\}$ is mapped onto another set of such tensors. In the latter
case, the space of symmetric tensors $b^{(p)}_{a_{1}\cdots a_{p}}$ is
not mapped onto itself. The image of $y^{a_{1}} y^{a_{2}}\cdots
y^{a_{p}}$ is a singular symbol $\propto
\partial^{a_{1}}_{y}\cdots\partial^{a_{p}}_{y}\delta^{(2N)}(y),\quad
\partial^{a}_{y}\equiv\omega^{ab}\,\partial/\partial y^{b}$.
    
    Nevertheless it will be helpful to think of the symplectic Fourier
    transformation as the bosonic (symmetric tensor) analog of the
    Hodge operator. For instance, by (\ref{2.48}) with (\ref{A.22}) and
    (\ref{A.36}) the fermionic inner product has the same general
    structure as (\ref{3.29}):
\begin{eqnarray}
\left( f_{1} | f_{2}\right) & \propto &\int \bar f_{1} (\theta) \,
\widetilde f_{2} ( \theta)\, d^{n}\theta \nonumber \\ 
& \propto &\int \bar f_{1} (\theta)\, (\ast f_{2} )( \theta)
\,d^{n}\theta  \label{3.32} 
\end{eqnarray}
In the language of differential forms this is nothing but the familiar
inner product $\ast (\bar \Phi_{1} \wedge \ast\Phi_{2})$ in disguise.
The product $\left( b_{1}| b_{2}\right)$ introduced above is analogous
to it, but refers to symmetric rather than antisymmetric tensors.

\vspace{3mm}
The space of symbols $b(y)$ equipped with the pointwise product of
functions, the star product, and the inner product constitutes the
bosonic Weyl algebra ${\cal W}$. It is the counterpart of the algebra
${\cal W}^{\rm F}$ which, endowed with analogous structures, had turned out
to be an Atiyah-K\"ahler algebra. Because $\left( y^{a}| y^{b}\right)=i
\hbar\omega^{ab}/2$ we see that the three product structures on ${\cal
  W}$ satisfy the consistency condition
\begin{eqnarray}
y^{a}\circ y^{b}= y^{a} y^{b}+ \left( y^{a}| y^{b}\right) \label{3.32a}
\end{eqnarray}
This relation is completely analogous to eq.(\ref{2.46}) which
had been identified with the defining property of an Atiyah-K\"ahler
algebra, eq.(\ref{1.2}). This supports our point of view that {\it the
bosonic Weyl algebra is the natural analog of an Atiyah-K\"ahler algebra
if one works in a symplectic rather than a Riemannian setting}.

\subsection{$\gamma$-Matrices for ${\rm Mp}(2N)$ and the analog of
  $\gamma_{5}$} 
 
The generators of ${\rm Mp}(2N)$ in the spinor representation are
obtained as symmetrized bilinears $\Sigma_{\rm
  meta}^{ab}=(\gamma^{a}\gamma^{b}+\gamma^{b}\gamma^{a})/4$ built from
$2N$ ``$\gamma$-matrices'' satisfying
\begin{eqnarray}
\gamma^{a}\gamma^{b}-\gamma^{b}\gamma^{a}=2i\, \omega^{ab} \label{3.33}
\end{eqnarray}
Upon identifying
\begin{eqnarray}
\gamma^{a}=\kappa\, \widehat \varphi ^{a} \label{3.34}
\end{eqnarray}
it is clear that the relations (\ref{3.33}) coincide precisely with
(\ref{3.1}). Hence, what in the language of group theory is called a
``symplectic Clifford algebra'' is nothing but the canonical
commutation relations of a bosonic quantum system with the canonical
operators $\widehat \varphi ^{a}=(\widehat \pi ^{j},\widehat x ^{j})$.
For $N$ finite, all irreducible representations of the canonical
commutation relations are unitarily equivalent, so the same is true for
the symplectic Clifford algebra. All these representations are infinite
dimensional.

We consider representations where $\gamma^{a}$ is a hermitian operator
on the Hilbert space $\cal V$. Frequently $\cal V$ is taken to be the
Fock space of $N$ independent harmonic oscillators \cite{pct,frad}. Then
the $\gamma^{a}$'s are linear combinations of the corresponding creation
and annihilation operators. Here we shall employ another representation
which is particularly natural in the gauge theory approach to quantum
mechanics \cite{metaqm}. We pick the $\widehat x$-eigenbasis (\ref{3.9})
with respect to which $\langle \alpha | \widehat x ^{j} | \beta
\rangle=\alpha^{j} \delta^{(N)}(\alpha-\beta)$ and $\langle \alpha |
\widehat \pi ^{j} | \beta \rangle=-i\hbar\partial_{j}
\delta^{(N)}(\alpha-\beta)$. Therefore, in a symbolic matrix notation
with $\langle \alpha | \gamma^{a} | \beta \rangle\equiv
(\gamma^{a})^{\alpha}_{\,\,\beta}$,
\begin{eqnarray}
(\gamma^{j})^{\alpha}_{\,\,\beta}&=&-(2i/\kappa)\, \partial^{j}
\delta^{(N)}(\alpha-\beta)\nonumber \\ 
(\gamma^{N+j})^{\alpha}_{\,\, \beta}&=&\kappa \alpha^{j} 
\delta^{(N)}(\alpha-\beta)\quad ;\quad j=1,\cdots,N \label{3.35}
\end{eqnarray}
The Hilbert space $\cal V$ is the space of square integrable functions
$\psi(\alpha)\equiv \langle \alpha|\psi\rangle \equiv \psi^{\alpha}$
with its usual inner product. The generators $\Sigma^{ab}_{\rm meta}$
act on $\cal V$ as second order differential operators (Schr\"odinger
Hamiltonians with a quadratic potential; see refs. \cite{meta1,meta2}
for further details).

Any attempt at putting metaplectic spinors on a similar footing as the
$SO(n)$-spinors faces the problem that $\cal V$ is infinite dimensional
and that a metaplectic spinor formally is an object $\psi^{\alpha}\equiv
\psi(\alpha)$ with infinitely many components. As an immediate
consequence, trace identities such as (\ref{1.27}) have no direct
counterpart for the metaplectic $\gamma$-``matrices''. In the $\widehat
x$-basis, for instance, the trace of an operator $\widehat b \in {\cal
  L(V)}$ reads ${\rm Tr} (\widehat b )=\int d^{N}\alpha \langle \alpha |
\widehat b | \alpha \rangle$, and it is clear that monomials such as
$\gamma^{a_{1}}\cdots\gamma^{a_{p}}$ do not possess a trace. Remarkably
enough, it turns out that there exist identities similar to (\ref{1.27})
even in the infinite dimensional case which, however, involve the ${\rm
  Sp}(2N)$-analog of $\gamma_{n+1}$.

We are familiar with the fact that when we are dealing with spinors on
an even-dimensional space-time there exists a chirality matrix
$\gamma_{n+1}$, a generalization of $\gamma_{5}$ in 4 dimension, which
anticommutes with any $\gamma^{\mu}$. Its eigenvalues are $-1$ and $+1$,
and the corresponding eigenspaces are the left- and right-handed Weyl
spinors, respectively. It is quite interesting that we can introduce an
analogous concept for metaplectic spinors and that the pertinent
``chirality operator'' has a very natural interpretation even. Let us
try to find an operator $\gamma_{P}\in {\cal L(V)}$ which anticommutes
with all $\gamma^{a}$'s,
\begin{eqnarray}
\gamma_{P}\gamma^{a}+\gamma^{a}\gamma_{P}=0 \label{3.36}
\end{eqnarray}
and satisfies
\begin{eqnarray}
\gamma_{P}^{\dagger}=\gamma_{P}^{-1}=\gamma_{P} \label{3.37}
\end{eqnarray}
Thus $\gamma_{P}$ has the same algebraic properties as $\gamma_{5}$, its
eigenvalues are $\pm 1$ and, provided it actually exists, we can use it
to form the ``chiral'' projections
\begin{eqnarray}
\psi_{\pm}=\Pi_{\pm}\psi,\qquad\qquad\Pi_{\pm} \equiv \frac{1}{2} (1\pm
\gamma_{P}) \label{3.38}
\end{eqnarray}
of any $\psi\in{\cal V}.$ Since $\Sigma^{ab}_{\rm meta}$ commutes with
$\gamma_{P}$, the ${\rm Mp}(2N)$-transformations leave the subspaces
with $\gamma_{P}=+1$ and $\gamma_{P}=-1$ invariant, so that the
representation of ${\rm Mp}(2N)$ implied by the $\gamma$-matrices
(\ref{3.35}) decomposes accordingly.

Looking at the ``metaplectic
$\gamma_{5}$-matrix'' from the point of view of the auxiliary quantum
mechanics with the $\widehat \varphi$-degrees of freedom it becomes
clear that we may identify $\gamma_{P}$ with the standard parity
operator $P$ in this context. By definition, $P$ changes the sign of
both the positions $\widehat x ^{i}$ and the momenta $\widehat \pi
^{i} :
P \widehat x ^{i} P=-\widehat x ^{i}, \quad P \widehat \pi ^{i} P=-
\widehat \pi ^{i} \label{3.39}$.
Hence $P\gamma^{a}P=-\gamma^{a}$ for $\gamma^{a}=\kappa (\widehat \pi
^{i},\widehat x ^{i})$, which is exactly (\ref{3.36}) with
$\gamma_{P}\equiv P$. The operator $P$ acts on the wave functions
$\psi\in {\cal V}$ as
$(P\psi)(\alpha)\equiv(\gamma_{P}\psi)(\alpha)=\psi(-\alpha)$. This
means that in the $\widehat x$-representation
\begin{eqnarray}
\gamma_{P}| \alpha\rangle =| -\alpha\rangle \label{3.40}
\end{eqnarray}
so that the matrix elements of $\gamma_{P}$ are given by 
\begin{eqnarray}
(\gamma_{P})^{\alpha}_{\,\,\beta} \equiv \langle
\alpha|\gamma_{P}|\beta\rangle=\delta ^{(N)}(\alpha+\beta) \label{3.41}
\end{eqnarray}
Thus, ``metaplectic chirality'' is nothing but ``fiberwise parity'', and
the projections $\Pi_{\pm}{\cal V}$ are simply the subspaces of even and 
odd wave functions, respectively.

The operator $\gamma_{P}$ can be written in a manifestly basis
independent way\footnote{Eq.(\ref{3.42}) shows that $\gamma_{P}$ belongs to
  the family of parity-type operators discussed by Grossmann \cite{gro}
  and Royer \cite {roy}.}:
\begin{eqnarray}
\gamma_{P}=(4\pi\hbar)^{-N}\int d^{2N} y \, \widehat T (y) \label{3.42}
\end{eqnarray}
The general properties of the Weyl operators imply that (\ref{3.42}) has
the desired properties (\ref{3.36}), (\ref{3.37}) and using the matrix
elements (\ref{3.10}) one finds that (\ref{3.42}) coincides with
(\ref{3.41}). Eq.(\ref{3.42}) is strikingly similar to eq.(\ref{2.44})
for $\widehat G_{n+1}$ which confirms our interpretation that {\it  the
fiberwise parity transformation is the analog of} $\gamma_{5}$.

The operator $\gamma_{P}$ has a well defined finite trace:
\begin{eqnarray}
{\rm Tr}[\gamma_{P}]=2^{-N} \label{3.43}
\end{eqnarray}
This follows from (\ref{3.42}) with (\ref{3.7}) or simply by noting that
\begin{eqnarray}
{\rm Tr}[\gamma_{P}]=\int d^{N} \alpha\, (\gamma_{P})^{\alpha}_{
 \,\, \alpha}=\int d^{N} \alpha \,\delta^{(N)} (2 \alpha)=2^{-N}
\label{3.44} 
\end{eqnarray}
While the very existence of this trace is remarkable, we see the first
major difference between the bosonic and the fermionic case here. Both
$\gamma_{5}$ and $\gamma_{P}$ have eigenvalues $\pm 1$, but the pairing
of positive and negative eigenvalues which leads to
$\rm{Tr}(\gamma_{5})=0$ does not happen for $\gamma_{P}$.

Finite products of $\gamma^{a}$-matrices and in particular the unit
operator do not possess a well defined trace.  On the other hand, traces
with a $\gamma_{P}$-insertion,
\begin{eqnarray}
{\rm Tr}\left[\widehat b\, \gamma_{P}\right]=\int d^{N} \alpha \,\langle
\alpha | 
\widehat b| -\alpha\rangle \label{3.45}
\end{eqnarray}
are much better behaved because the reflection $\alpha \mapsto -\alpha$
removes possible ``short distance singularities'' (reminiscent of
ultraviolet divergences in field theory) which would plague $\langle
\alpha | \widehat b | \alpha \rangle$.

This situation is quite similar to what one encounters in quantum field
theory in the computation of chiral anomalies or, from a mathematical
point of view, of the analytical index of the Dirac operator
\cite{heat,bert}. There one considers ${\rm Tr} (I)$ and
${\rm Tr}(\gamma_{5})$ where the trace is over the infinite dimensional
Hilbert space of Dirac spinor fields. While ${\rm Tr}(I)$ does not exist,
${\rm Tr}(\gamma_{5})$ can be interpreted as the index of the Dirac
operator.

An important trace of the type (\ref{3.45}) is
\begin{eqnarray}
{\rm Tr}[\gamma^{(a_{1}}\cdots\gamma^{a_{p})}\,\gamma_{P}\,
\gamma_{(b_{1}}\cdots\gamma_{b_{q})}]= i^{p} \,2^{-N}p!\, \delta^{pq}
\,\delta^{a_{1}}_{(b_{1}}\delta^{a_{2}}_{b_{2}}
\cdots\delta^{a_{p}}_{b_{p})}  
\label{3.46}
\end{eqnarray}
with the convenient abbreviation $\gamma_{a}\equiv
\omega_{ab}\gamma^{b}, \, \omega^{ab}\gamma_{b}=\gamma^{a}$.
Eq.(\ref{3.46}) is similar to (\ref{1.27}) for the $SO(n)$
$\gamma$-matrices, but contains an additional factor of $\gamma_{P}$
without which the trace would not exist. Eq.(\ref{3.46}) follows from
the properties for the $\widehat T$-operators. First one uses
(\ref{3.42}) with (\ref{3.7}) to show that
\begin{eqnarray}
{\rm Tr}[\widehat T (y) \gamma_{P}]=2^{-N} \label{3.47}
\end{eqnarray}
is independent of $y$. Next one writes
\begin{eqnarray}
{\rm Tr}[\widehat T (y_{1}) \gamma_{P} \widehat T (y_{2})]& = & {\rm
  Tr}[\widehat T 
(y_{1}) \widehat T (-y_{2}) \gamma_P]\nonumber \\
& = &
{\rm exp}[\frac{i}{2\hbar}\omega(y_{1},-y_{2})]\, {\rm Tr}[\widehat T
(y_{1}-y_{2}) \gamma_P] \nonumber \\ & = & 
2^{-N} 
{\rm exp}[-\frac{i}{2\hbar} y^{a}_{1} \omega_{ab} y^{b}_{2}] \label{3.48}
\end{eqnarray}
If one now expands the first and the last expression of (\ref{3.48}) in
powers of $y_{1}$ and $y_{2}$ and equates equal powers, the result is
precisely eq.(\ref{3.46}).

Some important special cases of (\ref{3.46}) include
\begin{eqnarray}
{\rm Tr}[\gamma^{a}\gamma_{P}]&=&0 \nonumber \\
{\rm Tr}[\gamma^{(a_{1}}\cdots\gamma^{a_{p})}\gamma_{P}]&=&0 \\
{\rm Tr}[\gamma^{a}\gamma^{b}\gamma_{P}]&=&2^{-N}i\,\omega^{ab}
\nonumber  \label{3.49} 
\end{eqnarray}
The reader is invited to check some of these relations by using the
matrix elements of $\gamma^{a}$ in the $| \alpha \rangle$-basis. It is
instructive to see that these calculations involve only well defined
manipulations of distributions and that no additional ad hoc
regularization is needed. This is different from the derivation of the
closely related dimension-counting formulas for the spinors of $OSp(n|
2N)$ which appear in certain approaches to the covariant
quantization of superstrings \cite{green}, for instance.

\subsection{The Dirac-K\"ahler construction on phase-space}

Let $({\cal M}_{2N},\omega)$ denote an arbitrary $2N$-dimensional
symplectic manifold which serves as the phase-space of some hamiltonian
system. Let us consider the Weyl algebra bundle \cite{fedbook,frad} over
${\cal M}_{2N}$. Its typical fiber is the bosonic Weyl algebra ${\cal
  W}$, i.e. the space of symbols $b(\cdot)$ equipped with the pointwise
product of functions, the star product, and the inner product $(\cdot |
\cdot )$.  At each point $\phi$ of ${\cal M}_{2N}$ we attach a local
copy ${\cal W}_{\phi}$ of ${\cal W}$. The matrix $(\omega_{ab})$ which
enters the definition of the Weyl algebra ${\cal W}_{\phi}$ are the
coefficients of the symplectic 2-form $\omega$ evaluated at the point
$\phi$. By virtue of Darboux's theorem, there exist local coordinates
$(\phi^{a})$ such that those coefficients assume their canonical form
on the entire $(\phi^{a})$-chart. Local coordinates on the
total space are pairs $(\phi,b)$ with $b$ a function $b: {\mathbf
  R}^{2N}\rightarrow {\mathbf C}, y\mapsto b(y)$. The transition
functions of the bundle are defined in such a way the variables
$(y^{1},\cdots,y^{2N})$ on which $b$ depends are the components of a
vector $y\in T_{\phi}{\cal M}_{2N}$, i.e. $y^{a}=d\phi^{a}(y)$. A
symplectic change of coordinates $\phi^{a}\rightarrow \widetilde
\phi^{a}(\phi)$ (canonical transformation) is to be combined with a
transformation in the fiber, $b\rightarrow \widetilde b$, such that
$\widetilde b (\widetilde y ) =b(y)$ with $\widetilde y ^{a}=(\partial
\widetilde \phi ^{a}/\partial \phi^{b}) y^{b}$.

Along with the Weyl algebra bundle we also consider the metaplectic
spinor bundle over $({\cal M}_{2N},\omega)$ which we described in the
Introduction. Its fiber at $\phi$, ${\cal V}_{\phi}$, is a copy of the
Hilbert space ${\cal V}$ on which we already constructed a
representation of the metaplectic Clifford algebra and, as a
consequence, of the structure group ${\rm Mp}(2N)$.

Let us look at sections through the Weyl algebra bundle. Locally they
are specified by functions $\phi\mapsto B (\phi,\cdot) \in {\cal
  W}_{\phi}$ where $B(\phi,\cdot):{\mathbf R}^{2N}\rightarrow {\mathbf
  C}, y\mapsto B(\phi,y)$ is a Weyl symbol ``living'' in the fiber at
$\phi$. In this context, the flat ``auxiliary phase-space'' ${\mathbf
  R}^{2N}$ is identified with the tangent space $T_{\phi}{\cal
  M}_{2N}$. Hence the function $B(\cdot, \cdot)$ is a map from (a part
of ) the total space of the tangent bundle into ${\mathbf C}$.

Many of the concepts which we developed in Section 3.1 for symbols $b\in
{\cal W}$ naturally pass over to the sections $B$. At every point $\phi$
of ${\cal M}_{2N}$ we can apply the inverse symbol map to
$B(\phi,\cdot)\in {\cal W}_{\phi}$ and obtain a unique operator
$\widehat B (\phi)={\rm symb}^{-1}B(\phi,\cdot)$ which acts on the local
copy ${\cal V}_{\phi}$ of the Hilbert space ${\cal V}$. Thus a section
$B$ gives rise to a family of operators $\widehat B (\phi) \in {\cal L}
({\cal V}_{\phi})$ labeled by the points of phase-space. Its matrix
elements with respect to a given basis in ${\cal V}$ will be denoted
$\widehat B (\phi)^{\alpha}_{\,\,\beta}\equiv \langle \alpha | \widehat
B (\phi) | \beta \rangle$. Globally speaking, $\widehat B$ is a section
through the bundle of (1,1)-multispinors \cite{meta1,metaqm}.

The fiberwise star product of two sections is defined by
\begin{eqnarray}
\left(B_{1} \circ B_{2}\right) (\phi,y)=B_{1}(\phi,y)\, {\rm exp} 
\left[
  \frac{i\hbar}{2} 
\stackrel{\longleftarrow}{\frac{\partial}
{\partial y^{a}}} 
\omega^{ab} 
\stackrel{\longrightarrow}{\frac{\partial}{\partial y^{b}}} 
\right] B_{2} (\phi,y) \label{3.50}
\end{eqnarray}
This star product has to be carefully distinguished from the
$\ast_{{\cal M}}$-product whose associated Moyal bracket $\lbrace
f,g\rbrace_{M}=(f\ast_{{\cal M}} g - g \ast_{{\cal M}} f) / i\hbar$
replaces the classical Poisson bracket in the deformation quantization
approach \cite{flato,kvich}, and which involves derivatives with respect
to $\phi^{a}$ rather than $y^{a}$. In general, the $\ast_{\cal
  M}$-product is much more complicated than the $\circ$-product. It can
be constructed iteratively by Fedosov's method
\cite{fed,emm,fedrec,rom,gel}, but we shall not need it in the present
context.

The fiberwise inner product of two sections is given by $(B_{1} |
B_{2})(\phi)=(\bar B _{1} \circ B_{2} )(\phi,0)$. The natural
sesquilinear form on the space of sections is $\langle B_{1}|
B_{2}\rangle =\int d\mu_{L} (B_{1}| B_{2}) $ where $d\mu_{L}$ is the
Liouville measure.

\vspace{4mm}
After these preparations we are now able to construct an analog of the
Dirac-K\"ahler fields on phase-spaces.

Let $\bigotimes^{p}_{\rm sym} (T^{\ast}{\cal M}_{2N})$ denote the
$p$-fold symmetrized
tensor power of the cotangent bundle. A section ${\bf {\bf \Sigma}}^{(p)}$
through this bundle is a symmetric tensor field of rank $p$. We shall
also consider the direct sum
\begin{eqnarray} {\scriptsize \bigotimes} _{\rm sym} (T^{\ast}{\cal
    M}_{2N})=\bigoplus_{p=0}^{\infty} 
{\small \bigotimes}\,_{\rm sym}^p (T^{\ast}{\cal M}_{2N}) \label{3.51}
\end{eqnarray}
Its sections ${\bf {\bf \Sigma}}=\sum_{p=0}^{\infty} {\bf {\bf \Sigma}}
^{(p)}$ are 
analogous to the inhomogeneous differential forms, but with symmetric
rather than antisymmetric tensor fields. In local (Darboux) coordinates
$\phi^{a}$, ${\bf {\bf \Sigma}}$ can be expanded as
\begin{eqnarray}
{\bf \Sigma}(\phi)=\sum^{\infty}_{p=0}
\frac{1}{p!}\,B^{(p)}_{a_{1}\cdots
  a_{p}}(\phi)\,d\phi^{a_{1}}\otimes_{\rm sym}  
d\phi^{a_{2}}\otimes_{\rm sym}\cdots \otimes_{\rm sym}d\phi^{a_{p}}
\label{3.52} 
\end{eqnarray}
with $\otimes_{\rm sym}$ denoting the symmetric counterpart of the wedge
product; for instance, $d\phi^{a}\otimes_{\rm sym}
d\phi^{b}=d\phi^{a}\otimes d\phi^{b} +d\phi^{b}\otimes d\phi^{a}$. The
complex-valued coefficients $B^{(p)}_{a_{1}\cdots a_{p}}$ are taken to
be completely symmetric in all $p$ indices. We shall refer to ${\bf
  {\bf \Sigma}}$ as an ``inhomogeneous symmetric tensor'' (IST).

Guided by the corresponding construction in the fermionic case we shall
now associate an operator $\widehat B (\phi)\in {\cal L} ({\cal
  V}_{\phi})$ to ${\bf \Sigma}(\phi)$ by replacing in eq.(\ref{3.52}) the
differentials $d\phi^{a}$ with the Gamma-matrices $\gamma^{a}$:
\begin{eqnarray}
\widehat B (\phi)& = &\sum_{p=0}^{\infty} \frac{1}{p!}\,
B^{(p)}_{a_{1}\cdots a_{p}}(\phi)
\,\gamma^{a_{1}}\gamma^{a_{2}}\cdots\gamma^{a_{p}} \nonumber \\ 
& = &\sum_{p=0}^{\infty} \frac{\kappa^{p}}{p!}\, B^{(p)}_{a_{1}\cdots
  a_{p}}(\phi) \,\widehat \varphi ^{a_{1}} \widehat \varphi
^{a_{2}}\cdots\widehat \varphi ^{a_{p}} \label{3.53}
\end{eqnarray}
(As discussed earlier, we interpret the $\gamma$-matrices as $\kappa$
times the canonical operators $\widehat \varphi ^{a}$ of the auxiliary
quantum system in the fiber). Conversely, every Weyl ordered operator on
${\cal V}_{\phi}$ which admits a power series expansion gives rise to a
unique IST. The operator can be expanded in the symmetrized monomials
$\widehat \varphi ^{(a_{1}}\cdots\widehat \varphi ^{a_{p})}$ with
coefficients which are symmetric tensors and define an IST therefore.

Now we form the symbol of $\widehat B$ : $B(\phi,y)=[ {\rm symb}
\widehat B (\phi) ] (y)$, i.e.
\begin{eqnarray}
B (\phi,y) = \sum_{p=0}^{\infty} \frac{\kappa^{p}}{p!}\,
B^{(p)}_{a_{1}\cdots a_{p}}(\phi) \,y^{a_{1}} y^{a_{2}}\cdots y^{a_{p}}
\label{3.54} 
\end{eqnarray}

Taking both steps together we arrive at a one-to-one correspondence
between IST's and symbols with a power series expansion in $y$:
\begin{eqnarray}
{\bf \Sigma} (\phi) \in {\scriptsize \bigotimes} _{\rm sym}
(T^{\ast}_{\phi}{\cal M}_{2N})  
\,\leftrightarrows\, 
\widehat B (\phi) \in  \widetilde {\cal L} ({\cal V}_{\phi})
\,\leftrightarrows \,
B(\phi,\cdot)\in \widetilde {\cal W} _{\phi} \label{3.55} 
\end{eqnarray}
This chain of bijections is similar to (\ref{2.63}). However, the
difference is that in the fermionic setting every symbol or every Weyl
ordered operator gives rise to an inhomogeneous tensor field. This is
not true in the bosonic case. We have to explicitly restrict the symbols
and operators to those which allow for a power series expansion in
$y^{a}$ or $\widehat \varphi^{a}$, respectively. (This is indicated by
the notation $\widetilde {\cal L} ({\cal V}_{\phi})$ and $\widetilde
{\cal W} _{\phi}$.)  Nevertheless we shall continue to consider also
symbols $B$ which are not analytic in $y$ because they will play a
central role in the reduction of symplectic DK-fields.

Now let us look at the rules of the symbol/tensor-correspondence in the
bosonic case. Clearly the differentials $d\phi^{a}$ correspond to
$\kappa y^{a} : d\phi^{a} \leftrightarrows \kappa y^{a}$. Hence, if we
write the (linear, invertible ) map from the symbols to the IST's as
$B\mapsto {\bf \Sigma} [B]$, we have ${\bf \Sigma} [\kappa
y^{a}]=d\phi^{a}$ or more generally
\begin{eqnarray}
{\bf \Sigma} [\kappa^{p} y^{a_{1}}\cdots
y^{a_{p}}]=d\phi^{a_{1}}\otimes_{\rm 
  sym}\cdots\otimes_{\rm sym} d\phi^{a_{p}} \label{3.56}
\end{eqnarray}
Up to this point the situation is the same as in Section 2.4 with the
commuting $y$'s replacing the anticommuting $\theta$'s. This converts
the wedge product to the symmetric tensor product. Differences become
manifest when we look at the list of natural operations for IST's and
their realization at the symbol level.

The automorphism ${\cal A}$ and the antiautomorphism ${\cal B}$, while
important for dealing with the ubiquitous sign factors in exterior
algebra computations, are unnecessary for symmetric tensors. As we
argued already, the Hodge operator has a natural bosonic translation,
the symplectic Fourier transformation. However it does not leave the
space $\widetilde {\cal W}$ invariant and, as a consequence, does not
induce a map of one IST onto another. Furthermore, the exterior
derivative is a derivation on the exterior algebra which does not
require a connection for its definition. Also this concept has no analog
on the bosonic side.

However, every vector field $v=v^{a}(\phi)\partial_{a}$ on ${\cal
  M}_{2N}$ gives rise to a contraction operator ${\bf i}(v)$. By
definition, it is a linear operator on the space of IST's, depending
linearly on $v$, and satisfying ${\bf i}
(\partial_{a})1=0, {\bf i} (\partial_{a}) d\phi^{b}=\delta_{a}^{b}$ as
well as
\begin{eqnarray}
{\bf i} (v)[{\bf \Sigma}_{1} \otimes_{\rm sym} {\bf \Sigma}_{2}]=[{\bf
  i} (v) {\bf \Sigma}_{1}] \otimes_{\rm sym} {\bf \Sigma}_{2}  
+{\bf \Sigma}_{1} \otimes_{\rm sym} [{\bf i} (v) {\bf \Sigma}_{2}]
\label{3.57} 
\end{eqnarray}
Its realization on $\widetilde {\cal W} $ reads
\begin{eqnarray}
{\bf i} (v)\, {\bf \Sigma} [B]={\bf \Sigma} [\kappa^{-1} v^{a}
\frac{\partial}{\partial y^{a}} B] \label{3.58}
\end{eqnarray}
We also define the operators 
\begin{eqnarray}
e_{a} \neg \,\equiv \,{\bf i} (\partial_{a}), \qquad e^{a}\neg \,\equiv\,
\omega^{ab}\, {\bf i} (\partial_{b}) \label{3.59}
\end{eqnarray}
with the basis vectors $\partial_{a}\equiv \partial /\partial \phi^{a}$
referring to a system of Darboux local coordinates.

The most important properties of the fermionic Weyl algebra ${\cal
  W}^{\rm F}$ were the three different product structures with which it
is endowed and which make it an Atiyah-K\"ahler algebra. The bosonic
Weyl algebra ${\cal W}$ is equipped with three analogous products
(pointwise multiplication, star product, inner product) which satisfy
the basic consistency condition (\ref{3.32a}). At the end of Section 3.1
this led us to the conclusion that ${\cal W}$ is the symplectic
counterpart of an Atiyah-K\"ahler algebra. In the same sense the IST's
${\bf \Sigma}$ are analogous to the Dirac-K\"ahler fields $\Phi$.

The product structures on ${\cal W}_{\phi}$ give rise to related
products on the space of symmetric tensor fields. One easily verifies
that the pointwise product of bosonic symbols is tantamount to the
symmetric tensor product:
\begin{eqnarray}
{\bf \Sigma} [B_{1}]\otimes_{\rm sym} {\bf \Sigma}[B_{2}]={\bf
  \Sigma}[B_{1}B_{2}] \label{3.60} 
\end{eqnarray}

Furthermore, guided by our experience with the fermionic case, we now
{\it define} the Clifford product for symmetric tensor field as the
image of the bosonic star product under the symbol/tensor correspondence
(\ref{3.55}):
\begin{eqnarray}
{\bf \Sigma} [B_{1}] \vee {\bf \Sigma}[B_{2}]={\bf \Sigma}[B_{1}\circ
B_{2}] \label{3.61} 
\end{eqnarray}
By construction, the ``symplectic Clifford product'', also denoted
`$\vee$', is associative and distributive (but not commutative). From
eqs. (\ref{3.50}), (\ref{3.58}) and (\ref{3.59}) one obtains the
following explicit representation for the product of two IST's:
\begin{eqnarray}
{\bf \Sigma}_{1} \vee {\bf \Sigma}_{2}=\sum_{p=0}^{\infty}
\frac{i^{p}}{p!} \left[ e_{a_{1}}\neg e_{a_{2}}\neg\cdots e_{a_{p}} 
\neg {\bf \Sigma}_{1}\right] \otimes_{\rm sym} \left[e^{a_{1}}\neg
e^{a_{2}}\neg \cdots e^{a_{p}}\neg {\bf \Sigma}_{2} \right] \label{3.62}
\end{eqnarray}
This equation is strikingly similar to K\"ahler's formula (\ref{1.4})
for the ordinary Clifford product. We emphasize that while eq.(\ref{3.62})
might look complicated it is uniquely determined by the fundamental
relations
\begin{eqnarray}
1\vee 1=1\quad&,& \quad 1\vee d\phi^{a}=d\phi^{a}\vee 1=d\phi^{a}
\nonumber \\ 
d\phi^{a}\vee d\phi^{b}&=&d\phi^{a}\otimes_{\rm sym}d\phi^{b} +i 
\omega^{ab} \label{3.63}
\end{eqnarray}
if associativity and distributivity are imposed.

Turning to the last product structure on ${\cal W}$, there is an obvious
choice for a fiberwise inner product $(\cdot,\cdot)$ of symmetric tensor
fields: $({\bf \Sigma}_{1},{\bf \Sigma}_{2})= (B_{1}| B_{2})$ where
${\bf \Sigma}_{1,2}$ is related to $B_{1,2}$ via (\ref{3.55}). Thus it
is clear that the IST's may be regarded as sections through a
``symplectic Atiyah-K\"ahler bundle''.

The left-multiplication by the basis element $d\phi^{a}$ reads
explicitly
\begin{eqnarray}
d\phi^{a} \vee {\bf \Sigma}=d\phi^{a}\otimes_{\rm sym} {\bf \Sigma} +i\,
e^{a}\neg \,
{\bf \Sigma} \label{3.64}
\end{eqnarray}
It defines a representation of the symplectic Clifford algebra in the
space of inhomogeneous symmetric tensor fields:
\begin{eqnarray}
d\phi^{a}\vee d\phi^{b} - d\phi^{b} \vee d\phi^{a}=2 i\,\omega^{ab}
\label{3.65} 
\end{eqnarray}
Comparing (\ref{3.65}) to (\ref{3.33}), $d\phi^{a} \vee$ takes the place
of the metaplectic Dirac-matrix $\gamma^{a}$. Since $d\phi^{a}={\bf \Sigma}
[\kappa y^{a} ],\, d\phi^{a} \vee$ applied to tensors is the same as
$\kappa y^{a}\circ$ applied to symbols:
\begin{eqnarray}
d\phi^{a}\vee {\bf \Sigma} [B]& = &{\bf \Sigma} [\kappa y^{a}\circ B]
\nonumber \\ 
& = & {\bf \Sigma} [C^{a} B] \label{3.66}
\end{eqnarray}
The differential operators $C^{a}$ were introduced in eq.(\ref{3.24}).
They are formally self-adjoint with respect to the inner product
$(\cdot|\cdot)$. They constitute a representation of the symplectic
Clifford algebra in space of bosonic Weyl symbols:
\begin{eqnarray}
C^{a}\, C^{b} -C^{b} \,C^{a}=2 i\, \omega^{ab} \label{3.67}
\end{eqnarray}
Since $\kappa y^{a}$ is the symbol of $\kappa \widehat \varphi
^{a}=\gamma^{a}$, the operator associated to $C^{a}B$ is
$\gamma^{a}\widehat B $ with $\widehat B ={\rm symb}^{-1} (B)$. In
summary, we have the chain of correspondences
\begin{eqnarray}
d\phi^{a}\vee {\bf \Sigma} \quad
\leftrightarrows\quad \gamma^{a} \widehat B \quad
\leftrightarrows\quad C^{a} B \label{3.68}
\end{eqnarray}

Thus we managed to implement the essence of the Dirac-K\"ahler idea in a
symplectic rather than a Riemannian setting. We constructed a
representation of the corresponding Clifford algebra on the space of
symmetric tensor fields over a phase-space manifold rather than on the
exterior algebra over space-time.

Up to this point our considerations focused on the kinematic aspects of
the theory. We have not yet found an analog of the DK-equation. Since
$d$ and $d^{\dagger}$ do not exist for symmetric tensors, the
DK-operator $d-d^{\dagger}$ has no direct counterpart. Still it is
possible to write down a ``symplectic DK-equation'' with the necessary
covariance properties:
\begin{eqnarray}
[d\phi^{a} \vee \nabla_{a} +m] {\bf \Sigma} =0 \label{3.69}
\end{eqnarray}
(Here $\nabla$ is a symplectic connection.) This equation could be
rewritten as a set of ``metaplectic Dirac equations'' in the same way as
the ordinary DK-equation can be decomposed into a set of ordinary Dirac
equations. Metaplectic Dirac operators have been investigated in the
mathematical literature recently \cite{hab} but no physical application
has emerged so far. In Section 4 we shall see that from a kinematical and
representation theory point of view the symplectic DK-fields indeed do
play an important role in the gauge theory approach to quantization. The
interpretation of field equations such as eq.(\ref{3.69}),
if any, will remain an open problem though.

We close this section with a few comments on the ``metaplectic
$\gamma_{5}$-matrix'' in relation to the DK-fields. In the $SO(n)$-case
we saw that $\gamma_{n+1}$, the volume form, and the $\delta$-function
are different guises of the same object. Some properties of
$\gamma_{n+1}$ are similar in the symplectic case, others are quite
different. The symbol of $\gamma_{P}$, too, is proportional to a
$\delta$-function,
\begin{eqnarray}
G_{P}\equiv {\rm symb} (\gamma_{P}),\qquad
G_{P}(y)=(\pi\hbar)^{N}\,\delta^{(2N)}(y) \label{3.70}
\end{eqnarray}
This symbol is completely unrelated to the volume form, however. In the
$SO(n)$-case we know that the Clifford right multiplication by $G_{n+1}$
is equivalent to the modified Hodge operator (${\bf \hodge} f \propto
f\circ G_{n+1} \propto f \circ \delta^{(n)} $). This property has a
partial analog since by virtue of (\ref{3.22}) the symplectic Fourier
transformation which corresponds to ${\bf \hodge}$ is essentially the
same operation as the star multiplication by $G_{P}$ from the right:
\begin{eqnarray}
\widetilde b (y)=2^{-N} \,(b\circ G_{P})(y/2) \label{3.71}
\end{eqnarray}
However, this statement on the space of symbol functions (including
distributions) does not imply a corresponding relation for symmetric
tensors. The symbol $G_{P}$ has no IST associated to it.

The matrix $\gamma_{P}$ makes its appearance also in the natural inner
product on ${\cal L}({\cal V}_{\phi})$. By virtue of the identity
\begin{eqnarray}
(B_{1}| B_{2})=2^{N}{\rm Tr}\,\left[\widehat B _{1} ^{\dagger}\,
  \widehat B _{2} 
\,\gamma_{P} \right] \label{3.72}
\end{eqnarray}
the inner product on ${\cal W}_{\phi}$ induces a corresponding product
for the operators. The latter differs form the familiar Hilbert-Schmidt
inner product by the additional $\gamma_{P}$-matrix which tends to
improve the regularity properties of the trace. Eq.(\ref{3.72}) is most
easily proven as follows:
\begin{eqnarray}
(B_{1}| B_{2})& = & (\bar B_{1} \circ B_{2})(y=0) \nonumber \\
& = & (\pi \hbar)^{-N} \int d^{2N} y \,(\bar B_{1} \circ B_{2})(y)
\,G_{P}(y) \nonumber \\ 
& = & (\pi \hbar)^{-N} \int d^{2N} y \,(\bar B_{1} \circ B_{2} \circ
G_{P})(y) \nonumber \\ 
& = & (\pi \hbar)^{-N} \int d^{2N} y \,[{\rm symb} \{ \widehat B _{1}
^{\dagger} \widehat B _{2} \gamma_{P}\}](y) \nonumber \\ 
& = &2^{N}{\rm Tr}[\widehat B _{1} ^{\dagger} \widehat B _{2} \gamma_{P}
] \label{3.73}
\end{eqnarray}
Here (\ref{3.70}) was used along with the standard results \cite{flato}
$\int d^{2N} y b_{1}(y)b_{2}(y)= \int d^{2N} y (b_{1}\circ b_{2})(y)$
and ${\rm Tr}(\widehat b )=(2\pi\hbar)^{-N} \int d^{2N} y \,b(y)$.

\subsection{Decomposition of the symplectic \\ DK-representation}
We have seen that $d\phi^{a}\vee$ and $\kappa y^{a} \circ$ induce a
representation of the symplectic Clifford algebra on the space of
symmetric tensors and their symbols, respectively. We also saw that the
corresponding representations in the $SO(n)$-case are reducible, so it
is natural to ask if the same is true in the symplectic setting. We shall
demonstrate that {\it at the level of the symbols} the representation is
indeed reducible. However, in contradistinction to the $SO(n)$-case, the
decomposition of ${\cal W}$ does not induce a concomitant decomposition
of the (symmetric) tensor algebra.  

We shall see that the representation of the symplectic Clifford algebra
carried by the symbol-valued fields 
$B(\phi, y)$ can be decomposed into infinitely many irreducible
representations each of which is equivalent to the one defined by the
metaplectic $\gamma$-matrices (\ref{3.35}).  (We recall that this is the
representation of the Heisenberg algebra used in conventional canonical
quantization.) As a consequence, every field $B(\phi,y)$ amounts to a
collection of infinitely many metaplectic spinor fields $\psi^{\alpha}
(\phi)$.  Now we discuss the question of the (ir)reducibility for the
symbols $B$, the operators $\widehat B$ and the tensors ${\bf \Sigma}$
separately.

\vspace{5mm}
{\noindent \normalsize  \bf (a)  Symbols\qquad}
\vspace{5mm}

\noindent We are going to show that the bosonic Weyl algebra $\cal W$ admits an
orthogonal decomposition 
\begin{eqnarray}
{\cal W}=\bigoplus_{\alpha\in {\mathbf R}^{N}} {\cal W}_{(\alpha)}
\label{3.74} 
\end{eqnarray}
such that the subspaces ${\cal W}_{(\alpha)}$ are invariant under
star-left multiplication by $y^{a}$, i.e.  $y^{a}\circ b \in {\cal
  W}_{(\alpha)}$ if $b\in {\cal W}_{(\alpha)}$. To this end we use an
infinite dimensional generalization of the Becher-Joos method \cite{bj}.
We look for a $2N$-parameter family of operators $\widehat Z (y), y\in
{\mathbf R}^{2N}$, with the property
\begin{eqnarray}
y^{a}\circ \widehat Z (y) =\widehat Z (y)\, \widehat \varphi ^{a}
\label{3.75} 
\end{eqnarray}
One should think of $\widehat Z (\cdot)$ as an operator-valued symbol,
i.e. the `$y^{a} \circ$' in (\ref{3.75}) is given by $\kappa^{-1} C^{a}$
as if $\widehat Z$ was an ordinary symbol. With our experience from the
fermionic case we suspect that $\widehat Z$ should be closely related to
the Weyl operators. It turns out that this is indeed the case. The
derivative of the Weyl operators reads
\begin{eqnarray}
\frac{\partial}{\partial y^{a}}\, \widehat T (y) & = &\frac{i}{\hbar}
\omega_{ab} [\widehat \varphi ^{b} -\frac{1}{2} y^{b}] \, \widehat T (y)
\nonumber \\ 
& = &\frac{i}{\hbar} \omega_{ab} \,\widehat T (y)\, [\widehat \varphi ^{b}
+\frac{1}{2} y^{b}] \label{3.76}
\end{eqnarray}
This equation entails that the argument of $\widehat T$ can be rescaled
in such a way that left multiplication with $y^{a}$ is equivalent to the
operator multiplication by $\widehat \varphi ^{a}$, either from the left
or from the right:
\begin{eqnarray}
y^{a}\circ \widehat T (\pm 2 i y)&=&\mp\, i\, \widehat \varphi ^{a}\,
\widehat T (\pm 2 i y) \nonumber \\ 
y^{a}\circ \widehat T (\pm 2 y)&=&\mp  \, \widehat T (\pm 2  y)\,
\widehat \varphi ^{a} \label{3.77}
\end{eqnarray}
Hence 
\begin{eqnarray}
\widehat Z (y) =\widehat T (-2 y)={\rm exp}(-i\kappa\, y^{a} \omega_{ab}
\gamma^{b}) \label{3.78}
\end{eqnarray}
is a solution to our problem. In the $\widehat x$-eigenbasis the matrix
elements $\widehat Z (y)^{\alpha}_{\,\,\beta}= \langle \alpha | \widehat
Z | \beta \rangle$ are given by
\begin{eqnarray}
\widehat Z (y)^{\alpha}_{\,\,\beta}={\rm exp} [-\frac{i}{\hbar}
y_{p}(\alpha +\beta)] \,\delta^{(N)} (\alpha -\beta + 2 y_{q}) \label{3.79}
\end{eqnarray}
They can be used in order to verify that 
\begin{eqnarray}
\langle \alpha| y_{q}\circ \widehat Z (y)| \beta \rangle&=&\beta\,
\langle \alpha| \widehat Z (y)| \beta \rangle \nonumber \\ 
\langle \alpha| y_{p}\circ \widehat Z (y)| \beta \rangle&=&i \hbar
\frac{\partial}{\partial \beta} \,\langle \alpha| \widehat Z (y)| \beta
\rangle \label{3.80}
\end{eqnarray}
which is (\ref{3.75}) in the ``position representation''. 

We shall need
the star product of two different $\widehat Z$ matrix elements. After
some algebra one finds the remarkably simple result
\begin{eqnarray}
\widehat Z (y)^{\alpha}_{\,\,\beta} \circ \widehat Z (y)^{\bar
  \alpha}_{\,\,\bar \beta}=2^{-N}\, (\gamma_{P})^{\alpha}_{\,\,\bar \beta}\,
\,\widehat Z (y)^{\bar \alpha}_{\,\,\beta} \label{3.81}
\end{eqnarray}
When combined with the identity $\widehat Z ^{\dagger}=\gamma_{P} \widehat Z
\gamma_{P}$ the above equation at $y=0$ gives rise to the inner product
\begin{eqnarray}
(\widehat Z ^{\alpha}_{\,\,\beta} | \widehat Z ^{\bar \alpha}_{\,\,\bar
  \beta})=2^{-N}\, (\gamma_{P})^{\alpha\bar \alpha} \,\delta_{
  \beta\bar\beta} \label{3.81a}
\end{eqnarray}

The orthogonality and completeness relations (\ref{3.7}), (\ref{3.8}) for
$\widehat T (y)$ imply similar relations for $\widehat Z (y)$. As a
consequence, $\{ \widehat Z (\cdot) ^{\alpha}_{\,\, \beta}| \alpha, \beta
  \in{\mathbf R}^{N}\}$ is a basis in the space of symbol functions
$b(\cdot)$. Every $b\in {\cal W}$ has an expansion of the form
$b(y)=\int d^{N} \alpha d^{N} \beta\, \psi^{\alpha}_{(\beta)} \widehat Z
(y)^{\beta}_{\,\, \alpha}$ where the ``coefficients''
$\psi^{\alpha}_{(\beta)}$ are actually functions ${\mathbf R}^N\times
{\mathbf R}^N\rightarrow {\mathbf C}$.  

We continue the discussion
directly for the case when ${\cal W}$ is the fiber ${\cal W}_{\phi}$ and
the symbols $b(\cdot)$ are the ${\cal W}$-valued fields $B(\cdot, \phi)$
evaluated at a given point $\phi$. Eqs.  (\ref{3.7}) and (\ref{3.8})
imply that $B$ can be expanded as
\begin{eqnarray}
B(\phi,y)=\int d^{N} \alpha \int d^{N} \beta \,\psi^{\alpha}_{(\beta)} (\phi)
\,\widehat Z (y)^{\beta}_{\,\,\alpha} \label{3.82}
\end{eqnarray}
and that the expansion coefficients are given by
\begin{eqnarray}
\psi^{\alpha}_{(\beta)}(\phi)=(\pi \hbar /2)^{-N} \int d^{2N} y\, B (\phi,
y) \,\widehat Z ^{\dagger} (y)^{\alpha}_{\,\,\beta} \label{3.83}
\end{eqnarray}
In a sense which we shall make precise later on, $\psi_{(\beta)}\equiv
\{\psi^{\alpha}_{(\beta)}; \alpha\in {\mathbf R}^{N}\}$ are the components
of infinitely many metaplectic spinors labeled by the ``index'' $\beta$.
If we define
\begin{eqnarray}
B_{(\beta)}(\phi,y)\equiv \int d^{N} \alpha \,\psi^{\alpha}_{(\beta)}
(\phi) \,\widehat Z (y)^{\beta}_{\,\,\alpha} \label{3.84}
\end{eqnarray}
so that $B(\phi,y)=\int d^{N} \beta\, B_{(\beta)} (\phi,y)$ then
eq.(\ref{3.75}) implies that the invariant subspace ${\cal W}_{(\beta)}$
is 
spanned by precisely the symbols of the type (\ref{3.84}):
\begin{eqnarray}
y^{a}\circ B_{(\beta)}& = &\int d^{N} \alpha \int d^{N} \bar \alpha\,
\psi^{\alpha}_{(\beta)} \,\widehat Z ^{\beta}_{\,\,\bar
  \alpha}\,(\widehat \varphi ^{a})^{\bar \alpha}_{\,\,\alpha} \nonumber
\\ 
& = &\int d^{N} \bar \alpha \,(\widehat \varphi ^{a}
\psi_{(\beta)})^{\bar\alpha} \,\widehat Z ^{\beta}_{\,\, \bar \alpha}
\label{3.85} 
\end{eqnarray}
Here $(\widehat \varphi ^{a} \psi_{(\beta)})^{\bar \alpha}\equiv \int
d^{N} \alpha \,(\widehat \varphi ^{a} )^{\bar \alpha}_{
\,\,\alpha}\psi^{\alpha}_{(\beta)}$. We see that if the symbol
$B_{(\beta)}$ is related to the spinor $\psi_{(\beta)}$ by (\ref{3.84})
then $y^{a}\circ B_{(\beta)}$ and $\widehat \varphi ^{a}\psi_{(\beta)}$
are related in the same way. Likewise $\kappa y^{a} 
\circ $ corresponds to a
multiplication by $\gamma^{a}$.  

Given an arbitrary symbol in ${\cal W}$
we can project it on any of the subspaces ${\cal W}_{(\beta)}$. We
introduce projection operators ${\cal P}_{(\beta)}$ by
$B_{(\beta)}={\cal P}_{(\beta)} B$. If we combine eqs. (\ref{3.83}) and
(\ref{3.84}) it follows that
\begin{eqnarray}
B_{(\beta)}(\phi,y) = \int d^{2N} y' \,{\cal
  P}_{(\beta)}(y,y')\,B(\phi,y') \label{3.86} 
\end{eqnarray}
where the integral kernel of the projector is given by
\begin{eqnarray}
{\cal P}_{(\beta)} (y,y')=(\pi \hbar/2)^{-N} \,\langle \beta | \widehat Z
(y) \widehat Z ^{\dagger} (y')| \beta\rangle \label{3.87}
\end{eqnarray}
Upon using (\ref{3.78}), (\ref{3.6}) and (\ref{3.10}) we obtain explicitly 
\begin{eqnarray}
{\cal P}_{(\beta)} (y,y')=(\pi \hbar)^{-N} \,{\exp} [ -\frac{2i}{\hbar}
(\beta + y_{q})(y_{p}-y'_{p})] \,\delta^{(N)} (y_{q}-y'_{q}) \label{3.88}
\end{eqnarray}
The projectors $\{{\cal P}_{(\beta)} ; \beta \in {\mathbf R}^{N}\}$ are
orthogonal and complete in the sense that
\begin{eqnarray}
\int d^{2N} y' \,{\cal P}_{(\beta)} (y,y') \,{\cal P}_{(\bar \beta)}
(y',y'')
& = &\delta^{(N)}(\beta-\bar\beta)\, {\cal P}_{(\beta)} (y,y'') \nonumber \\
\int d^{N} \beta \,{\cal P}_{(\beta)} (y,y')
& = &\delta^{(2N)} (y-y') \label{3.89}
\end{eqnarray}
Furthermore, as a consequence of eq.(\ref{3.81}), the inner product of
two different projections reads
\begin{eqnarray}
(B_{(-\beta_{1})}|
B_{(\beta_{2})})=2^{-N}\,\delta^{(N)}(\beta_{1}-\beta_{2}) \int
d^{N}\alpha\, \bar \psi ^{\alpha}_{(\beta_{1})} \psi
^{\alpha}_{(\beta_{2})} \label{3.89a}
\end{eqnarray}
Note the sign flip on the LHS of this equation. Obviously $B_{(-\beta)}$
is the natural dual of $B_{(\beta)}$ (similar to a spinor adjoint).  

To summarize: Every symbol-valued field $B(\phi,y)$ gives rise to
infinitely many projections $B_{(\beta)} (\phi,y)$ each of which is
equivalent to a metaplectic spinor field $\psi_{(\beta)}(\phi)$ with
components $\psi^{\alpha}_{(\beta)}(\phi)$ given by (\ref{3.83}). This
is to mean that the fields $\psi_{(\beta)}$ carry an irreducible
representation of the Clifford algebra: $\kappa y^{a} \circ B_{(\beta)}$
corresponds  to the spinor multiplied by a $\gamma$-matrix, $\gamma^{a}
\psi_{(\beta)}$.  

Up to this point the situation is similar to the $SO(n)$-case, but
differences will show up shortly.

\vspace{5mm}
{\noindent \normalsize   \bf (b)  Operators \qquad } 
\vspace{5mm}

As in the fermionic case, it proves advantageous to combine the
expansion coefficients $\psi^{\alpha}_{(\beta)}$ as a matrix $\widehat
\psi$ : 
\begin{eqnarray}
\widehat \psi ^{\alpha}_{\,\, \beta}\equiv \psi
^{\alpha}_{(\beta)}\equiv \langle \alpha | \widehat \psi|
\beta\rangle \label{3.89b}
\end{eqnarray} 
(We suppress the argument $\phi$ for the time being.) We
shall need some properties of the linear, invertible map $B\mapsto
\widehat \psi [B]$ which relates the symbols to the new operator
$\widehat \psi$. 

By definition, $B(y)$ is the {\it ordinary} Weyl symbol
of the operator $\widehat B$ introduced earlier. Remarkably enough, this
symbol plays a dual role: the same function but with its argument
rescaled, $B(\frac{1}{2} y)$, turns out to be the {\it alternative} Weyl
symbol of the new operator $\widehat \psi$. This is most easily seen if
one uses $\widehat Z ^{\dagger}(y)=\widehat T (2y)$ in
\begin{eqnarray}
\widehat \psi& = &(\pi \hbar/2)^{-N} \int d^{2N} y\, B(y) \,\widehat Z
^{\dagger} (y) \label{3.90} \\
B(y)& = &{\rm Tr} [\widehat Z (y) \widehat \psi ] \label{3.91}
\end{eqnarray}
which follows from the equations in subsection (a), and then compares
(\ref{3.90}), (\ref{3.91}) to eqs.(\ref{3.11}), 
(\ref{3.12}). Thus,
\begin{eqnarray}
[{\rm symb} \{ \widehat B \} ](y)=B(y) \quad \Leftrightarrow \quad 
[{\rm alt}\mbox{-}{\rm symb} \{ \widehat \psi \} ](y)=B(\frac{1}{2}y)
\label{extra} 
\end{eqnarray} 
 This dual role played by $B$ is another hint at the very
natural relationship between the Dirac-K\"ahler idea and the Weyl symbol
calculus.  

Regarding $\widehat \psi$ as a functional of $B$ it is not
difficult to establish that
\begin{eqnarray}
\widehat \psi [1]& = &2^{N} \,\gamma_{P} \label{3.92} \\
\widehat \psi [\kappa y^{a}] & = &2^{N} \,\gamma^{a} \gamma_{P}
\label{3.93}\\ 
\widehat \psi [\kappa y^{a}\circ B]& = &\gamma^{a} \widehat \psi [B]
\label{3.94}\\ 
\widehat \psi [B_{1}\circ B_{2}]& = &2^{-N}\, \widehat \psi [B_{1}]
\, \gamma_{P}\, \widehat \psi [B_{2}] \label{3.95}\\
\widehat \psi [\kappa^{p} y^{a_{1}}\circ y^{a_{2}}\circ \cdots\circ
y^{a_{p}}]& = &2^{N}\, \gamma^{a_{1}} \gamma^{a_{2}} \cdots
\gamma^{a_{p}}\gamma_{P} \label{3.96}\\
\widehat \psi [\kappa^{2} y^{a} y^{b}]& = &2^{N}\, \gamma^{a} \gamma^{b} 
\gamma_{P}- 2^{N} i \omega^{ab} \gamma_{P} \label{3.97}
\end{eqnarray}
Eq.(\ref{3.92}) follows directly from the definition of $\gamma_{P}$ and
eq.(\ref{3.94}) is our earlier result (\ref{3.85}), eq.(\ref{3.93})
being a special case. The most important relation is (\ref{3.95}). It
can be proven by using (\ref{3.81}) and (\ref{3.91}) in order to show
that $B_{1}\circ B_{2}=2^{-N} {\rm Tr}\{\widehat Z \widehat \psi [B_{1}]
\gamma_{P} \widehat \psi [B_{2}]\}$. When compared to eq.(\ref{3.91}),
this equation implies (\ref{3.95}).

Above we had
introduced the projectors ${\cal P}_{(\beta)}$ which project any symbol
on the invariant subspaces ${\cal W}_{(\beta)}$. The map $B\mapsto
\widehat \psi [B]$ given by (\ref{3.90}) induces a corresponding
projection on the space of operators $\widehat \psi$.  In the language
of our auxiliary quantum mechanical system this projection has a very
natural interpretation: it is simply the projection on the position
eigenstate $| \beta \rangle$. From eq.(\ref{3.84}) we can read off that
$B_{(\beta)}$ has the structure of an expectation value in the state $|
\beta\rangle$
\begin{eqnarray}
B_{(\beta)}(y)& = &\langle \beta| \widehat Z (y) \widehat \psi [B] |
\beta \rangle \nonumber \\ 
& =  & {\rm Tr} \left[ \widehat Z (y) \,\widehat \psi [B] \,\widehat P
  _{(\beta)} 
\right] \label{3.98}
\end{eqnarray}
Here $\widehat P _{(\beta)}\equiv | \beta \rangle\langle \beta |$ is the
corresponding projector on the Hilbert space. It follows from (\ref{3.98})
that symbols $B\in {\cal W}_{(\beta)}$ are associated to operators of
the form $\widehat \psi P_{(\beta)}$:
\begin{eqnarray}
\widehat \psi [{\cal P}_{(\beta)} B]=\widehat \psi [B]\, \widehat P
_{(\beta)} \label{3.99}
\end{eqnarray}

Finally we have to address the important question of how the operator
$\widehat \psi$ is related to the operator $\widehat B$ which was the
central building block in the Dirac-K\"ahler construction. Imitating the
$SO(n)$-case, we had obtained $\widehat B$ in eq.(\ref{3.53}) by
replacing $d\phi^{a} \rightarrow \gamma^{a}$ in the tensor field
${\bf \Sigma}$. In Section 2 we have seen that for ordinary DK-fields
$\widehat \psi$ and $\widehat F$ coincide up to a constant factor. It is
quite remarkable that, with a minor modification, the same
identification is possible in the symplectic situation where ${\cal V}$
is infinite dimensional. It turns out that
\begin{eqnarray}
\widehat \psi [B]=2^{N} \,\widehat B\, \gamma_{P} \qquad {\rm or} \qquad
\widehat B=2^{-N} \,\widehat \psi [B]\, \gamma_{P} \label{3.98a}
\end{eqnarray}

This relationship can be proven in a variety of ways. For instance, we
can take advantage of the following very compact representation of
operators $\widehat b$ in terms of their symbols $b$ \cite{gro}:
\begin{eqnarray}
\widehat b =2^{-N} (2\pi \hbar)^{-N} \int d^{2N} y \,\,b(\frac{1}{2} y)
\,\widehat T (y)\, \gamma_{P} \label{3.99a}
\end{eqnarray}
The advantage of (\ref{3.99a}) as compared to the old representation
(\ref{3.11}) is that no Fourier transformation is involved any longer.
Eq.(\ref{3.99a}) is easily established by inserting the integral
representation for $\gamma_{P}$ on its RHS and then combining the two
Weyl operators with the help of (\ref{3.6}). From eq.(\ref{3.99a}) we
infer that if $B(y)$ is the ordinary symbol of $\widehat B$ then
$B(\frac{1}{2} y)$ is the alternative Weyl symbol of $2^{N}\widehat B
\gamma_{P}$. Moreover, we saw already that $B(\frac{1}{2} y)$ is the
alternative Weyl symbol of $\widehat \psi$. As a consequence, $\widehat
\psi$ must coincide with $2^{N} \widehat B \gamma_{P}$.  

It is instructive to give a different proof when $B(y)$ is a power
series. This is the case for instance when the symbol originates from an
IST via the DK-construction. For $\widehat B$ or $B(y)$ given, the task
is to solve $B(y)={\rm Tr} [\widehat Z (y) \widehat \psi ]$ for the
unknown operator $\widehat \psi$. Using the expansion (\ref{3.54}) for
$B$ and the (expanded) exponential (\ref{3.78}) for $\widehat Z$, this
equation turns into
\begin{eqnarray}
B^{(p)}_{a_{1}\cdots a_{p}}=i^{-p}\,
{\rm Tr}[\gamma_{(a_{1}}\cdots\gamma_{a_{p})} \widehat \psi ] \label{3.100}
\end{eqnarray}
In the corresponding calculation for the $SO(n)$-case we made an ansatz
for $\widehat \psi$ as a power series in $\gamma^{\mu}$ and used the
$\gamma^{\mu}$-trace identities in order to project on its coefficients.
Because of the additional matrix $\gamma_{P}$ in the analogous
identities (\ref{3.46}) for the metaplectic $\gamma$-matrices the
appropriate ansatz for the symplectic $\widehat \psi$ is a power series
in $\gamma^a$ 
(with coefficients $\psi^{(p)}_{a_{1}\cdots a_{p}}$) times an explicit
factor of $\gamma_{P}$. With this ansatz in (\ref{3.100}), the trace
identities imply $\psi^{(p)}_{a_{1}\cdots a_{p}}=2^{N}
B^{(p)}_{a_{1}\cdots a_{p}}$ which proves (\ref{3.98a}). In this manner
we see that the factor of $\gamma_{P}$ connecting $\widehat \psi$ to
$\widehat B$ is simply a reflection of the corresponding factor in the
trace identities. We discussed already that in the infinite dimensional
situation the $\gamma_{P}$ under the traces is crucial in order to make
them well defined.

The $\gamma_{P}$-matrix in (\ref{3.98a}) has the consequence that
$\widehat \psi$ does not admit a power series expansion even if
$\widehat B$ does so . This has important implications for the
Dirac-K\"ahler program. As we are going to discuss next it means that
the decomposition of ${\cal W}$ into subspaces ${\cal W}_{(\alpha )}$
which are invariant under star left multiplication does {\it not}
translate into a corresponding decomposition of the symmetric tensor
algebra into subspaces invariant under (symplectic) Clifford left
multiplication. In this respect the $SO(n)$ and the ${\rm Sp}(2N)$-cases are
quite different.  

Let us first look at how the space of operators
$\widehat B$ decomposes under ${\cal W}=\bigoplus {\cal W}_{(\beta)}$.
Eqs. (\ref{3.99}) and (\ref{3.98a}) imply that
\begin{eqnarray}
\widehat{ {\cal P}_{(\beta)} B} & = &\widehat B \,\gamma_{P}\, \widehat
P _{(\beta)}\, \gamma_{P} \nonumber \\ 
& = & \widehat B \,\widehat P _{(-\beta)} \label{3.101}
\end{eqnarray}
Hence, at the level of the $\widehat B$-operators, the projection ${\cal
  P}_{(\beta)}$ amounts to a right multiplication by $\widehat P
_{(-\beta)}$.  

>From eq.(\ref{3.101}) we can obtain a very useful by-product. If we take
the symbol on both sides of this equation and abbreviate
$P_{(\alpha)}\equiv {\rm symb} \left[\widehat P _{(\alpha)}\right]$ then
the result is the following compact formula for the projection
$B_{(\beta)}$:
\begin{eqnarray}
{\cal P}_{(\beta)} \,B \equiv B_{(\beta)} =B\circ P_{(-\beta)}
\label{3.102} 
\end{eqnarray}
More explicitly, because $P_{(\alpha)}(y)=\delta^{(N)}(y_{q}-\alpha)$,
this means that 
\begin{eqnarray}
B_{(\beta)}(y)=B(y)\circ \delta^{(N)}(y_{q}+\beta) \label{3.102b}
\end{eqnarray} 
By
virtue of (\ref{3.22}) the latter equation can be brought to the
following form which is the most convenient one for practical
calculations:
\begin{eqnarray}
B_{(\beta)}(y_{p},y_{q}) =B \left(y_{p}-\frac{i\hbar}{2}
  \frac{\partial}{\partial \bar y_{q}},\, y_{q}\right)\, \delta^{(N)} 
(\bar y_{q} + \beta) \left. \right|_{\bar y_{q} =y_{q}} \label{3.103}
\end{eqnarray}
As usual, $y\equiv (y_{p},y_{q})$ consists of $N$-component momentum-
and position-type variables $y_{p}$ and $y_{q}$. 

The structure of $B_{(\beta)}$ is particularly transparent if
$B(y)\equiv B(y_{q})$ does not depend on the momenta. Then its
projection on ${\cal W}_{(\beta)}$ reads
\begin{eqnarray}
B_{(\beta)} (y) = B(y_{q})\,\delta^{(N)} (y_{q}+\beta), \label{3.104}
\end{eqnarray}
i.e., it is sharply localized at $y_{q}=-\beta$. If $B$ depends also on
$y_{p}$ there are additional terms involving derivatives of
$\delta^{(N)}(y_{q}+\beta)$. Nevertheless, {\it as long as $B$ depends
  on $y$ polynomially, the projected symbol $B_{(\beta)}$ has support
  only on the hyperplane $y_{q}=-\beta$}. This localization of the
symbols makes it very easy to visualize the $\beta$-subspace of ${\cal
  W}$. In fact, this intuitive interpretation of ${\cal W}_{(\beta)}$ is
the reason why we are using the $\widehat x$-eigenbasis on ${\cal V}$
rather than the harmonic oscillator (Fock space) basis which yields the
traditional representation of the $\gamma^{a}$-matrices.

\vspace{5mm}
{ \noindent \normalsize  \bf (c)  Inhomogeneous symmetric tensors} 
\vspace{3mm}

We know
that every symbol-valued field $B(\phi,y)$ gives rise to infinitely many
projections $B_{(\beta)}\in {\cal W}_{(\beta)}$ each of which is
equivalent to a spinor $\psi_{(\beta)}$. On ${\cal W}_{(\beta)},\quad
\kappa y^{a}\circ B_{(\beta)}$ corresponds to $\gamma^{a}\psi_{(\beta)}
$ and it represents the Clifford algebra irreducibly. On the other hand,
in Section 3.3 we defined the symplectic Clifford product as the image
of the star product under the symbol/tensor-correspondence
(\ref{3.55}). It is a natural question therefore whether the
representation of the Clifford algebra provided by ``$d\phi^{a}\vee$''
on the space of symmetric tensors is reducible as well.  

At this point we have to recall that the symbol/tensor-correspondence
(\ref{3.55}) is a bijection between tensors ${\bf \Sigma} (\phi)$ and
symbols $B(\phi,y)$ which are {\it analytic in} $y$.  Only if $B$ allows
for a power series expansion in $y$ the substitution $\kappa y^{a}
\rightarrow d\phi^{a}$ yields a tensor field. As for the question of the
reducibility, the crucial observation is that {\it even if $B(\phi,y)$
  is analytic in $y$, the projections $B_{(\beta)}(\phi,y)$ are not in
  general}. This is obvious from eq.(\ref{3.103}) which shows that
$B_{(\beta)}$ is typically a distribution with a sharp localization (in
the auxiliary phase-space) on the plane $y_{q}=-\beta$.

Therefore we must conclude that the
decomposition of the bosonic Weyl algebra ${\cal W}=\bigoplus {\cal
  W}_{(\beta)}$ does not imply a corresponding decomposition of the
space of IST's. This was different in the fermionic case where the
analyticity of $F(x,\theta)$ comes for free and where ``symbol-valued
fields'' and ``inhomogeneous differential forms'' are two completely
equivalent concepts.  

From these observations we can learn what the correct notion of a
``symplectic Dirac-K\"ahler field'' actually is.  Traditionally, in the
$SO(N)$-case, a DK-field meant a set of (antisymmetric) tensor fields.
This is a historic accident, however, and one could have talked equally
well about ${\cal W}^{\rm F}$-valued fields over space-time.  When we go
from space-time to phase-space and from $SO(n)$ to ${\rm Sp}(2N)$ we see
that the notion which generalizes is not that of a collection (of now
symmetric) tensor fields but rather the idea of Weyl symbol-valued
fields.  On phase-space the fields $B(\phi,y)$, with a not necessarily
analytic dependence on $y$, play a role which is completely analogous to
that of $F(x,\theta)$ on space-time. The former is equivalent to a set
of ${\rm Mp}(2N)$-spinors in very much the same way as the latter gives
rise to a multiplet of ${\rm Spin}(n)$-spinors.

\renewcommand{\theequation}{4.\arabic{equation}}
\setcounter{equation}{0}
\section{Summary and Discussion}
In the first part of this paper we have shown that the theory of
space-time DK-fermions allows for a remarkably simple and natural
reinterpretation in the framework of the symbol calculus. More
precisely, it is the fermionic Weyl symbol which is to be used here.
This symbol was employed in the context of first quantized
particle and string theory occasionally, but so far it has not reached the
popularity of the Wick symbol which is commonly chosen for fermionic
systems. 

We have set up a one-to-one correspondence between DK-fields
$\Phi(x)$ and symbol-valued fields $F(x,\theta)$ by associating a family
of auxiliary quantum systems, with canonical operators $\widehat \chi
^{\mu}$ and anticommuting phase-space coordinates $\theta^{\mu}$, to
each point $x$ of space-time. The fermionic operators $\widehat \chi
^{\mu}$ and Grassmann variables $\theta^{\mu}$ replace the Dirac
matrices $\gamma^{\mu}$ and the differentials $d x^{\mu}$, respectively.
The nontrivial aspect of this correspondence is that it maps all the
natural operations which we know for differential forms onto equally
natural and well known operations for symbols. For instance, the star
product which is at the heart of every symbol calculus turned out to be
related to the Clifford product, a pivotal concept in standard
DK-theory, in precisely this manner. More generally, we were able to
identify all the defining structures of an Atiyah-K\"ahler algebra on
the space of fermionic Weyl symbols.  

Our approach provides some new
computational tools for calculations involving DK-fields, an integral
representation of the Clifford product, for example.  More importantly,
it sheds new light on the geometrical meaning of various constructions
in the standard approach. For instance, the matrix-valued form $Z$ has
turned out to be nothing but a fermionic Weyl operator.  

\vspace{3mm}

In the second
part of this paper we developed a symplectic analog of DK-theory. We
replaced space-time by phase-space, the ``Lorentz group'' $SO(n)$ by
${\rm Sp}(2N)$, Dirac fields by metaplectic spinors, and we then asked if
there exists a corresponding notion of a DK-field. The answer turned out
to be in the affirmative, but with some qualifications. The crucial step
in our construction was switching from the fermionic auxiliary quantum
system to a bosonic one whose basic operators $\widehat \varphi ^{a}$
satisfy canonical commutation relations and thus realize the symplectic
Clifford algebra. Using the Riemannian situation as a guide line we
formulated the auxiliary quantum theory in terms of (now bosonic) Weyl
symbols. We argued that it is the symbol-valued fields $B(\phi,y)$ which
deserve the name of a ``symplectic Dirac-K\"ahler field''. The fields
with an analytic dependence on $y$ are equivalent to a set of symmetric
tensor fields, the symplectic counterpart of an inhomogeneous
differential form. We described in detail which properties of the
standard DK-fields pass over to the symplectic case and which don't. We
discovered for example that all the defining structures of an
Atiyah-K\"ahler algebra have analogs in the symplectic setting. In
particular, the bosonic Weyl star product gives rise to a ``Clifford
product''. 

It is an interesting feature of this method that both the
ordinary and the symplectic Clifford product arise as a quantum
deformation (in the sense of \cite{flato}) of the corresponding tensor
product (wedge product and $\otimes_{\rm sym}$), the deformation
parameter being $\hbar$ or $\kappa^{-2}$.\footnote{In order to make
  this explicit at the tensor level one should refrain from the
  convenient rescaling of the tensor components by factors of $\kappa$.}

The most important differences between the Riemannian and the symplectic
case occur when it comes to decomposing the representation of the
Clifford algebra carried by the symbol-valued fields. While the
decomposition of the bosonic Weyl algebra into left invariant subspaces
can be carried out along the same lines as for the fermionic algebra, it
does not induce a corresponding decomposition of the space of
inhomogeneous symmetric tensor fields. The reason is that the projection
on the invariant subspaces does not respect the analyticity of
$B(\phi,y)$ which is necessary for a tensor interpretation. We take this
as a hint that it is actually the concept of a Weyl-algebra-valued field
which is at the heart of DK-theory, both on space-time and on
phase-space, rather than the idea of inhomogeneous (anti)symmetric
tensors. The fields $B(\phi,y)$ are equivalent to a multiplet of
metaplectic spinors in the same way an ordinary DK-field is equivalent
to a multiplet of Dirac spinors.  
\vspace{3mm}

Let us close with a few additional comments.  

We begin with a remark on what it precisely means that a DK-field is
``equivalent'' to a set of spinor fields. This remark applies to $SO(n)$
and ${\rm Sp}(2N)$ DK-fields alike.  Strictly speaking, a metaplectic
spinor is defined by its transformation properties under local ${\rm
  Sp}(2N)$-transformations, the phase-space analog of the local
Lorentz-transformations\footnote{See ref.\cite{metaqm} for a detailed
  discussion of those transformation properties and of the vielbein
  formalism for phase-spaces.}. Let us fix some point $\phi$ of ${\cal
  M}_{2N}$ and let us change the basis in its tangent space
$T_{\phi}{\cal M}_{2N}$ by means of a symplectic matrix $S(\phi)\equiv
[S(\phi)^{a} _{\,\,b}].$ This induces a corresponding unitary
transformation $M(S)\in {\rm Mp}(2N)$ in the local Hilbert space ${\cal
  V}_{\phi}$. The components of a vector and a spinor transform as
$y^{a}\rightarrow (S^{-1})^{a} _{\,\,b} y^{b}$ and $\psi^\alpha
\rightarrow M(S)^{\alpha} _{\,\,\beta}\psi^{\beta}$, respectively. It is
important to observe that the spinors contained in a DK-field $B(\phi,
y)$ do not individually transform in this manner. In fact, as a direct
consequence of (\ref{V.1a}) the $\widehat Z$-operator transforms
according to
\begin{eqnarray}
M(S)^{\dagger}\, \widehat Z (y)\, M(S) = \widehat Z (S^{-1} y) \label{4.1}
\end{eqnarray}
Therefore eq.(\ref{3.91}) reads in the rotated basis
\begin{eqnarray}
B(\phi,S^{-1} y)= {\rm Tr}\left[\widehat Z (y)\, M(S)\, \widehat \psi
  (\phi) 
\,M(S)^{\dagger}\right] \label{4.2}
\end{eqnarray}
This means that $\widehat \psi \equiv (\psi^{\alpha}_{(\beta)})$ does
not transform as a set of independent spinors labeled by the index
$\beta$. The index $\beta$, too, is acted upon by a spin matrix:
\begin{eqnarray}
\psi^{\alpha}_{(\beta)}\rightarrow M(S)^{\alpha} _{\,\,\gamma}
\,\psi^{\gamma}_{(\delta)}\,M^{\dagger}(S)^{\delta} _{\,\,\beta}
\label{4.3} 
\end{eqnarray}

For space-time DK-fields this is a well known phenomenon which is
referred to as ``flavor mixing'' \cite{bj}.  Among other things it
implies that DK-fermions have a nonstandard coupling to gravity
\cite{banks,dipl,tuck}.  The curved-space Dirac equation for a massless
DK-field reads
\begin{eqnarray}
\gamma^{\mu}\left(\partial_{\mu}\widehat \psi - i \omega^{IJ}_{\mu}
[\sigma_{IJ},\widehat \psi]\right)=0 \label{4.4}
\end{eqnarray}
If $\widehat \psi$ was a set of independent spinors the spin connection
$\omega_{\mu}^{IJ}\sigma_{IJ}$ multiplied $\widehat \psi$ from the left
only. The flavor mixing caused by the commutator is weak in the
Newtonian limit of gravity and most probably cannot be excluded on
experimental grounds \cite{dipl}.  

\vspace{4mm}
Finally let us comment on the
relation of the symplectic DK-fields to the gauge theory formulation of
quantum mechanics \cite{metaqm} which was proposed recently. Its basic
ingredient is a family of local Hilbert spaces\footnote{ For different
  formulations of quantum (field) theory using local Hilbert spaces see
  refs. \cite{local,frad,mack}} ${\cal V}_{\phi}$ attached to the
points of phase-space. This theory resulted from an attempt at
understanding the principles of canonical quantization at a perhaps
deeper or at least physically and geometrically more natural level. 

The
theory is a Yang-Mills-type gauge theory on phase-space. Its ``matter
fields'' are metaplectic spinors $\psi^{\alpha}(\phi)$. Canonical
quantization is replaced by two new rules.  The first one is that in
order to go from classical mechanics to semiclassical quantum mechanics
we must switch from the vector representation of ${\rm Sp}(2N)$ to its spinor
representation. The second rule is a consistency condition which tells
us how to sew together local semiclassical expansions so as to recover
exact quantum mechanics. It is formulated as symmetry principle: the
Yang-Mills theory must be invariant under a new type of
background-quantum split symmetry. As it turns out, this implies that
the gauge field is a universal, nondynamical abelian connection
$\widetilde \Gamma$.\footnote{The gauge group is the group of all
  unitary transformations on ${\cal V}$ and the connection components
  $\widetilde \Gamma_{a}$ are hermitian operators.  Being abelian means
  that the curvature of $\widetilde \Gamma$ is proportional to the unit
  operator on ${\cal V}$.} 

The upshot of this construction is the
following two-step procedure for the quantization of physical systems on
arbitrary curved phase-spaces ${\cal{M}}_{2N}$: 
\begin{enumerate}
\item[(1)] Find an abelian spin
connection $\widetilde \Gamma$ on ${\cal M}_{2N}$. It is guaranteed to
exist on any symplectic manifold and can be constructed iteratively by
Fedosov's method \cite{fed,fedbook,fedrec}. 
\item[(2)] Construct (multi) spinor fields which are covariantly
  constant (possibly up to a phase) with respect to the connection
  $\widetilde \Gamma$. They are local generalizations of states and
  observables.
\end{enumerate}

In particular, states are represented by a covariantly
constant spinor field $\psi^{\alpha}(\phi)$. If the value of this field
is known at a fixed reference point $\phi_{0}$ it is known everywhere in
phase-space (up to a physically irrelevant phase). The wave function
${\bf \Psi}$ of conventional quantum mechanics is identified with
$\psi^{\alpha}(\phi_{0})\equiv {\bf \Psi} (\alpha)$. For further details
we refer to \cite{metaqm}.  

This approach reveals that, in a sense,
classical mechanics is related to quantum mechanics in the same way
tensor fields (integer spin) relate to spinor fields (half-integer spin)
or space-time bosons relate to fermions. What is at the heart of the
quantization process is changing the representation of ${\rm Sp}(2N)$, the
``Lorentz group'' of phase-space.  

According to the proposal of ref.
\cite{metaqm} this change of representation, while very natural from a
particle physics point of view, still has to be done ``by hand'' in the
same sense as in the standard approach the canonical commutation
relations are imposed ``by hand''. One might wonder if there is more
natural way of describing this change of representation, and it is here
that Dirac-K\"ahler theory comes into play. DK-theory certainly cannot
{\it explain} why nature has decided to pick the spinor representation
of ${\rm Sp}(2N)$ but it can put this question into a novel and perhaps
somewhat unexpected perspective.  

The symplectic DK-fields give a precise meaning to the idea that
classical mechanics ``contains'' the basic building blocks of quantum
mechanics, namely the metaplectic spinor fields. On the one hand, the
DK-fields $B(\phi,y)$ belong to the realm of classical mechanics in the
sense that they are c-number functions on the classical tangent bundle.
On the other hand, $B(\phi,y)$ is equivalent to a family of spinor
fields $\psi_{(\beta)}=(\psi^{\alpha}_{(\beta)})$ whose members are
labelled by the ``flavor index'' $\beta$. Quantum mechanics is a theory
whose basic ingredient is a {\it single} metaplectic spinor field. This
leads us to conclude that {\it the process of quantization can be
  understood as the elimination of all but one flavor of metaplectic
  spinors, i.e. as a projection on a fixed
  $\beta$-subspace}\footnote{Note, however, that covariantly constant
  DK-fields do not amount to covariantly constant projected spinor
  fields. The reason is the flavor mixing: the condition $\nabla
  (\widetilde \Gamma )B=0$ involves a commutator of $\widetilde \Gamma$
  with $B$, while $\nabla (\widetilde \Gamma ) \psi =0$ contains only a
  left-multiplication by $\widetilde \Gamma$.}.  

In the same sense as
above, this projection has to be done ``by hand''\footnote{We mention
  that also the approach of ref. \cite{gozzi} constructs quantum
  mechanics from functions on the classical tangent bundle by imposing
  certain constraints. This approach does not involve metaplectic
  spinors, however, and the DK-construction seems not to answer the
  questions raised there.}. However, with this interpretation, there is
an almost perfect analogy between the following two problems which are
usually thought of as belonging to rather different branches of physics:
the construction of a lattice theory which describes a single species of
fermions, and the quantization of physical systems in general. On the
Riemannian (or space-time) side, the question is how to avoid the
fermion replication which results from the Kogut-Susskind action, and
the corresponding symplectic (or phase-space) problem is how to obtain a
quantum theory from classical structures. At a heuristic level, the solution
to both problems is exactly the same: one must project out a single
spinor from a Dirac-K\"ahler field. Whether this is merely a formal
similarity or whether space-time fermions can teach us something about
the general structure of quantum mechanics remains to be seen.
\vspace{4mm}

\newpage
\noindent {\bf Acknowledgements:} It is a pleasure to thank Professor
H. Joos for introducing me to the 
Dirac-K\"ahler equation long ago and for many interesting discussions
ever since. I am also grateful to E. Gozzi for numerous helpful
conversations and for the hospitality at the Department of Theoretical
Physics, University of Trieste, while this work was in progress. This
project was supported in part by a NATO traveling grant.

\newpage
\begin{appendix}
\renewcommand{\theequation}{A.\arabic{equation}}
\setcounter{equation}{0}
\section*{Appendix A}

In this appendix we collect a number of definitions and identities
related to Grassmann algebras which are needed in the main body of the
paper. In particular, the main automorphism ${\cal A}$, the main
antiautomorphism ${\cal B}$, the Hodge operator $\ast$ and the modified
Hodge operator ${\bf \hodge}$ are discussed and our conventions are
specified. 

We consider a Grassmann algebra with the real generators
${\theta^{1},\cdots,\theta^{n}}$, i.e.
$\theta^{\mu}\theta^{\nu}+\theta^{\nu}\theta^{\mu}=0$ for all
$\mu,\nu=1,\cdots,n$, and introduce functions
\begin{eqnarray}
f(\theta)\equiv f(\theta^{1},\cdots,\theta^{n})=\sum_{p=0}^{n}
f^{(p)}(\theta) \label{A.1}
\end{eqnarray}
where $f^{(p)}$ is homogeneous of degree $p$:
\begin{eqnarray}
f^{(p)}(\theta)=\frac{1}{p!}\, f^{(p)}_{\mu_{1}\cdots\mu_{p}}\,
\theta^{\mu_{1}}\cdots\theta^{\mu_{p}}  
\label{A.2}
\end{eqnarray}
The complex-valued constants $f_{\mu_{1}\cdots\mu_{p}}$ are completely
antisymmetric in all indices. By definition, the main automorphism
${\cal A}$ and the main antiautomorphism ${\cal B}$ act on these
functions according to
\begin{eqnarray}
({\cal A}f)(\theta)& = &\sum_{p=0}^{n}(-1)^{p}\,f^{(p)}(\theta)
\label{A.3} \\ 
({\cal B}f)(\theta)& = &\sum_{p=0}^{n} (-1)^{p(p-1)/2}\,f^{(p)}(\theta)
\label{A.4} 
\end{eqnarray}
Their main properties are
\begin{eqnarray}
{\cal A}^{2}={\cal B}^{2}=1&,&\qquad {\cal A B}={\cal B A} \nonumber \\
{\cal A}(f g)&=&({\cal A} f)({\cal A} g) \nonumber \\
{\cal B}(f g)&=&({\cal B} g)({\cal B} f) \label{A.5}
\end{eqnarray}
where $(fg)(\theta)\equiv f(\theta)g(\theta)$ is the pointwise product.
Some useful identities involving ${\cal A}$ and ${\cal B}$ include
\begin{eqnarray}
\theta^{\mu}f(\theta)& = &({\cal A} f) (\theta) \,\theta^{\mu}
\label{A.6}\\ 
\theta^{\mu_{1}}\cdots\theta^{\mu_{p}} f(\theta)& = &({\cal
  A}^{p}f)(\theta)\,\theta^{\mu_{1}}\cdots\theta^{\mu_{p}} \label{A.7} \\
\theta^{\mu_{p}}\theta^{\mu_{p-1}}\cdots\theta^{\mu_{1}}& =
&(-1)^{p(p-1)/2}\,\theta^{\mu_{1}}
\theta^{\mu_{2}}\cdots\theta^{\mu_{p}}\nonumber  
\\
& = & {\cal B} \,\theta^{\mu_{1}}\theta^{\mu_{2}} \cdots\theta^{\mu_{p}}
\label{A.8} 
\end{eqnarray}
Denoting complex conjugation by an overbar we assume $\bar \theta
^{\mu}=\theta ^{\mu}$ and set
\begin{eqnarray}
\overline{f g}=\bar g \bar f \label{A.9}
\end{eqnarray}
for any two functions $f$ and $g$. If one makes the additional assumption
that the coefficients ${f^{(p)}_{\mu_{1}\cdots\mu_{p}}}$ are real, then
eq.(\ref{A.8}) shows that
\begin{eqnarray}
\bar f (\theta)=({\cal B} f)(\theta) \label{A.10}
\end{eqnarray}
Usually we shall allow the coefficients to be complex though. The
automorphism ${\cal A}$ can be used in order to convert
right-derivatives $\stackrel{\longleftarrow}{\frac{\partial}{\partial
    \theta^{\mu}}}$ to left-derivatives
$\stackrel{\longrightarrow}{\frac{\partial}{\partial
    \theta^{\mu}}}\equiv \frac{\partial}{\partial \theta^{\mu}}$:
\begin{eqnarray}
f(\theta) 
\stackrel{\longleftarrow}{\frac{\partial}{\partial \theta^{\mu}}}
={\cal A} \frac{\partial}{\partial \theta^{\mu}} f(\theta) \label{A.11}
\end{eqnarray}
More generally, one has
\begin{eqnarray}
f(\theta) 
\stackrel{\longleftarrow}{\frac{\partial}{\partial \theta^{\mu_{p}}}}\cdots
\stackrel{\longleftarrow}{\frac{\partial}{\partial \theta^{\mu_{1}}}}
={\cal A}^{p} \frac{\partial}{\partial \theta^{\mu_{p}}}\cdots 
\frac{\partial}{\partial \theta^{\mu_{1}}}f(\theta) \label{A.12}
\end{eqnarray}
which is easily proven by induction. Since ${\cal A}$ anticommutes with
$\frac{\partial}{\partial \theta^{\mu}}$, it follows that
$(p,q=0,1,2,\cdots)$
\begin{eqnarray}
{\cal A}^{q} \frac{\partial}{\partial
  \theta^{\mu_{1}}}\cdots\frac{\partial}{\partial
  \theta^{\mu_{p}}}f(\theta)=(-1)^{pq}\frac{\partial}{\partial
  \theta^{\mu_{1}}}\cdots\frac{\partial}{\partial \theta^{\mu_{p}}}{\cal
  A}^{q}f(\theta) \label{A.13}
\end{eqnarray}
In particular,
\begin{eqnarray}
{\cal A}^{p}\frac{\partial}{\partial
  \theta^{\mu_{1}}}\cdots\frac{\partial}{\partial
  \theta^{\mu_{p}}}f(\theta)=(-1)^{p} \frac{\partial}{\partial
  \theta^{\mu_{1}}}\cdots\frac{\partial}{\partial \theta^{\mu_{p}}}{\cal
  A}^{p}f(\theta) \label{A.14}
\end{eqnarray}
These identities will be needed in order to establish the equivalence of
the Clifford product and the fermionic star product.

Our conventions for the integration are $\int d\theta^{\mu}=0$ and $\int
\theta^{\mu}d\theta^{\mu}=1 $ ($\mu$ not summed). We define
\begin{eqnarray}
d^{n}\theta\equiv d\theta^{1}d\theta^{2}\cdots d\theta^{n} \label{A.15}
\end{eqnarray}
so that
\begin{eqnarray}
\int
\theta^{\mu_{n}}\theta^{\mu_{n-1}}\cdots
\theta^{\mu_{1}}\,d^{n}\theta=\epsilon^{\mu_{1}\mu_{2}\cdots\mu_{n}}  
\label{A.16}
\end{eqnarray}
with $\epsilon^{12\cdots n}=+1$. Using (\ref{A.11}) one can show that
\begin{eqnarray}
\int f(\theta)\,\frac{\partial}{\partial \theta^{\mu}}\, g(\theta)
\,d^{n}\theta=\int f(\theta)
\stackrel{\longleftarrow}{\frac{\partial}{\partial \theta^{\mu}}}
g(\theta) \,d^{n} \theta \label{A.17}
\end{eqnarray}
for arbitrary inhomogeneous functions $f$ and $g$. 

In our conventions,
the delta-function is defined to satisfy
\begin{eqnarray}
\int f(\theta) \,\delta (\theta-\xi)\,d^{n}\theta=f(\xi) \label{A.18}
\end{eqnarray}
(Note the order of the factors.) It is given by
\begin{eqnarray}
\delta(\theta-\xi)=(\theta^{n}-\xi^{n})(\theta^{n-1}-
\xi^{n-1})\cdots(\theta^{1}-\xi^{1})  
\label{A.19}
\end{eqnarray}
or by the Fourier representation
\begin{eqnarray}
\delta(\theta-\xi)=(-1)^{n(n-1)/2}\int
e^{(\theta^{\mu}-\xi^{\mu})\rho_{\mu}}\,d^{n}\rho \label{A.20}
\end{eqnarray}
Here $\{\xi^{1},\cdots,\xi^{n}\}$ and $\{\rho_{1},\cdots,\rho_{n}\}$ are
two additional sets of real Grassmann variables which anticommute among
themselves and with the $\theta$'s. (Indices are raised and lowered with
the flat metric $g_{\mu\nu}=\delta_{\mu\nu}$.) Depending on the value of
$n$, $\delta$ is either Grassmann-real or purely imaginary:
\begin{eqnarray}
\overline{\delta(\theta)}=(-1)^{n(n-1)/2}\,\delta(\theta) \label{A.21}
\end{eqnarray}

The Fourier transform $\widetilde f$ is defined
according to
\begin{eqnarray}
\widetilde f (\rho)=\epsilon^{-1}_{n} \int e^{i\theta^{\mu}
  \rho_{\mu}}\,f(\theta)\, d^{n}\theta \label{A.22}
\end{eqnarray}
with\footnote{All formulas given in this appendix are valid for both $n$
  even and $n$ odd.}
\begin{eqnarray}
\epsilon_n\,\equiv 
\left\{ \begin{array}{r@{\quad:\quad}l}
1& {\rm for\quad \mbox{$n$}\quad even} \\
-i& {\rm for\quad \mbox{$n$}\quad odd}\end{array} \right. \label{A.23}
\end{eqnarray}
The advantage of our conventions is that they give rise to a simple
formula for $\widetilde f$ in terms of multiple derivatives of the
$\delta$-function which is free from explicit sign factors and powers of
$i$. One obtains
\begin{eqnarray}
\widetilde f (\rho)=f(i \frac{\partial}{\partial \rho})\delta (\rho)
\label{A.21a} 
\end{eqnarray}
because with (\ref{A.20})
\begin{eqnarray}
\widetilde f (\rho)& = & \epsilon^{-1}_{n} \int f (i
\frac{\partial}{\partial \rho})\,e^{i
  \theta^{\mu}\rho_{\mu}}\,d^{n}\theta \nonumber \\
& = & f(i \frac{\partial}{\partial \rho})\, \epsilon_{n}^{-1}
(-1)^{n(n-1)/2} \,\delta(-i\rho) \\
& = & f(i \frac{\partial}{\partial \rho})\, \delta(\rho)\nonumber
\label{A.22a} 
\end{eqnarray}
In particular,
\begin{eqnarray}
f(\theta)=1 \,\Longrightarrow\, \widetilde f (\rho)=\delta (\rho)
\label{A.23a} 
\end{eqnarray}
The inverse transformation reads
\begin{eqnarray}
f(\theta)=\int e^{-i \theta^{\mu}\rho_{\mu}}\, \widetilde f (\rho)\,
d^{n} 
\rho \label{A.24}
\end{eqnarray}
The Grassmann Fourier transformation has the involutive property
\begin{eqnarray}
\widetilde{\widetilde{f}} (\theta)=\epsilon^{-1}_{n} f(\theta),
\label{A.25} 
\end{eqnarray}
i.e. for $n$ even it is an exact involution. Derivative and multiplication
operators are conjugate in the sense that
\begin{eqnarray}
\widetilde{[\theta^{\mu}f(\theta)]} (\rho)&=&i \frac{\partial}{\partial
  \rho_{\mu} }  \,\widetilde f (\rho) \\
\widetilde{[i \frac{\partial}{\partial \theta ^{\mu}} f(\theta)]}
(\rho)&=& \rho_{\mu}\, \widetilde f (\rho) \label{A.26}
\end{eqnarray}
Using either (\ref{A.22}) or (\ref{A.21a}) one can work out the Fourier
transform of a product of $\theta$'s. The result is
\begin{eqnarray}
\widetilde{[\theta^{\mu_{1}}\theta^{\mu_{2}}\cdots
\theta^{\mu_{p}}]}(\rho)=\frac{C_{np}}{(n-p)!}\,
\epsilon^{\mu_{1}\cdots\mu_{p}\nu_{1}\cdots\nu_{n-p}}
\,\rho_{\nu_{1}}\rho_{\nu_{2}}\cdots\rho_{\nu_{n-p}} 
\label{A.27}
\end{eqnarray}
with the constants
\begin{eqnarray}
C_{np}\equiv i^{p}\, (-1)^{p(p-1)/2}\,(-1)^{n(n-1)/2} \label{A.28}
\end{eqnarray}

Identifying $dx^{\mu}\equiv \theta^{\mu}$, the exterior algebra $\bigwedge
(T_{x}^{\ast}{\mathbf R}^{n})$ endowed with the inner product coming
from $g_{\mu\nu}=\delta_{\mu\nu}$ provides a special example of a
Grassmann algebra. In this context we are familiar with the notion of a
Hodge star operator which maps $p$-forms onto $(n-p)$-forms. In the case
at hand we introduce a corresponding linear map $\ast:f(\theta)\mapsto
(\ast f)(\theta)$ which generalizes this concept. On the basis elements,
the Hodge operator acts according to
\begin{eqnarray}
\ast
(\theta^{\mu_{1}}\cdots\theta^{\mu_{p}})=
\frac{1}{(n-p)!}\,\epsilon^{\mu_{1}\cdots\mu_{p}}_{\qquad  
  \nu_{1}\cdots\nu_{n-p}}\,\theta^{\nu_{1}}\cdots \theta^{\nu_{n-p}}
\label{A.29} 
\end{eqnarray}
and it is extended to arbitrary functions $f(\theta)$ by linearity. Writing
$(\ast f)(\theta)=\sum_{p=0}^{n}\frac{1}{p!}[\ast
f]^{(p)}_{\mu_{1}\cdots\mu_{p}}\theta^{\mu_{1}}\cdots\theta^{\mu_{p}}$
one finds for the components\footnote{Note that in parts of the
  literature a different definition of $\ast$ is used which amounts to
  interchanging the transformation laws of the basis vectors and the
  components, respectively.}
\begin{eqnarray}
[\ast
f]^{(n-p)}_{\mu_{1}\cdots\mu_{n-p}}=\frac{1}{p!}\,
f^{(p)}_{\nu_{1}\cdots\nu_{p}}  
\, \epsilon^{\nu_{1}\cdots\nu_{p}}_{\qquad\mu_{1}\cdots\mu_{n-p}} 
\label{A.30}
\end{eqnarray}
Acting twice with $\ast$ on a homogeneous function of degree $p$ the
result is
\begin{eqnarray}
\ast \ast f^{(p)}=(-1)^{p(n-p)}\,f^{(p)} \label{A.31}
\end{eqnarray}
Because of the $p$-dependent sign factor on the RHS of (\ref{A.31}) the
star operator does not give rise to an involution on the space of all
(i.e., inhomogeneous) functions. This motivates us to introduce the
operator
\begin{eqnarray}
{\bf \hodge}\equiv \ast {\cal B} \label{A.32}
\end{eqnarray}
which we shall refer to as the ``modified Hodge operator''. For
homogeneous functions,
\begin{eqnarray}
\hodge f^{(p)}=(-1)^{p(p-1)/2}\,\ast f^{(p)} \label{A.33}
\end{eqnarray}
which implies
\begin{eqnarray}
{\bf \hodge}{\bf \hodge} f^{(p)}=(-1)^{n(n-1)/2}\,f^{(p)} \label{A.34}
\end{eqnarray}
with a sign factor independent of $p$. Hence, for any inhomogeneous
function $f$,
\begin{eqnarray}
{\bf \hodge}{\bf \hodge} f=(-1)^{n(n-1)/2}\,f \label{A.35}
\end{eqnarray}
For $n=4$, say, ${\bf \hodge}{\bf \hodge}=1$ so that ${\bf \hodge}$ is an
exact involution.

The (modified) Hodge operator is closely related to the Grassmann
Fourier transformation. Comparing (\ref{A.27}) to (\ref{A.29}) shows
that for homogeneous functions
\begin{eqnarray}
\ast f^{(p)}&=&(-i)^{p}(-1)^{p(p-1)/2}(-1)^{n(n-1)/2}\,
\widetilde{f^{(p)}} \label{A.36} \\ 
{\bf \hodge} f^{(p)}&=&(-i)^{p} (-1)^{n(n-1)/2}\, \widetilde{f^{(p)}}
\label{A.37} 
\end{eqnarray}
Using (\ref{A.21a}) we may express the Fourier transform by the
derivative of a $\delta$-function:
\begin{eqnarray}
\ast f^{(p)}(\theta)&=&(-1)^{p(p-1)/2}(-1)^{n(n-1)/2}\,
f^{(p)}(\frac{\partial}{\partial \theta})\,\delta(\theta) \label{A.38}\\ 
{\bf \hodge}
f^{(p)}(\theta)&=&(-1)^{n(n-1)/2}\,f^{(p)}(\frac{\partial}{\partial
  \theta})\,\delta(\theta) \label{A.39}
\end{eqnarray}
Note that the sign factor on the RHS of (\ref{A.39}) is independent of
$p$. Hence it follows that for arbitrary inhomogeneous functions
\begin{eqnarray}
{\bf \hodge} f(\theta)=(-1)^{n(n-1)/2}\,f(\frac{\partial}{\partial
  \theta})\,\delta(\theta) \label{A.40}
\end{eqnarray}
This is an interesting representation of the Hodge operator, because in
contrast to (\ref{A.29}), eq.(\ref{A.40}) continues to be meaningful if
we regard $\theta^{\mu}$ as a {\it commuting} variable. This fact will
become important in the construction of the metaplectic DK-fields.

\renewcommand{\theequation}{B.\arabic{equation}}
\setcounter{equation}{0}
\section*{Appendix B}

In this appendix we derive several important representations of the
fermionic star product, eqs. (\ref{2.15}), (\ref{2.16}) and
(\ref{2.17}), from the integral representation (\ref{2.14}).

We start by shifting $\theta_{1}$ and $\theta_{2}$ in eq.(\ref{2.14}):

\begin{eqnarray}
(f_{1}\circ f_{2})(\theta)=\epsilon_{n}\left( \frac{\hbar}{2i}
\right)^{n} \int 
{\rm exp} \left( \frac{2}{\hbar} \theta_{1} \theta_{2} \right)
f_{1}(\theta_{1}+\theta)f_{2}(\theta_{2}+\theta)\,d^{n}\theta_{1}d^{n} 
\theta_{2} \label{B.1}
\end{eqnarray}
Next we Taylor-expand $f_{1}$ and $f_{2}$ with respect to $\theta_{1}$
and $\theta_{2}$. Because the exponential produces only terms with equal
numbers of $\theta_{1}$'s and $\theta_{2}$'s, only those terms in the
product of the two Taylor series survive the integration which contain
equal numbers as well:
\begin{eqnarray}
(f_{1}\circ f_{2})(\theta) =\epsilon_{n}\left( \frac{\hbar}{2i}
\right)^{n}&&
\!\!\!\!\!\!\!\!\! \sum_{p=0}^{n} \left( \frac{1}{p!} \right)^{2}\int {\rm
  exp} \left(\frac{2}{\hbar} \theta_{1}\theta_{2}\right)
 [\theta_{1}^{\mu_{1}}\cdots\theta_{1}^{\mu_{p}} \,\widetilde \partial
_{\mu_{1}}\cdots \widetilde \partial _{\mu_{p}}
\,f_{1}(\theta)]\nonumber \\ &&
\times[\theta_{2}^{\nu_{1}}\cdots\theta_{2}^{\nu_{p}} \widetilde 
\partial _{\nu_{1}}\cdots \widetilde \partial _{\nu_{p}} f_{2}(\theta)]
d^{n}\theta_{1}d^{n}\theta_{2} \label{B.2}
\end{eqnarray}
Here $\widetilde \partial _{\mu} \equiv 
\frac{\partial}{\partial \theta ^{\mu}}$.
Because
\begin{eqnarray}
(\theta^{1}\theta^{2}\cdots\theta^{p})(\xi^{1}\xi^{2}
\cdots\xi^{p})=(-1)^{p}(\xi^{1}\xi^{2}\cdots\xi^{p})
(\theta^{1}\theta^{2}\cdots\theta^{p})  
\label{B.3}
\end{eqnarray}
for two arbitrary sets of mutually anticommuting Grassmann-odd objects,
we may use eqs. (\ref{A.7}) and (\ref{A.14}) to write
\begin{eqnarray}
\theta_{1}^{\mu_{1}}\cdots\theta_{1}^{\mu_{p}}\, \widetilde \partial
_{\mu_{1}}\cdots\widetilde \partial _{\mu_{p}}\,f_{1}(\theta) 
& = & 
(-1)^{p}\,\widetilde \partial _{\mu_{1}}\cdots\widetilde \partial
_{\mu_{p}}  \,\theta_{1}^{\mu_{1}}\cdots\theta_{1}^{\mu_{p}}\,f_{1}
(\theta) \nonumber \\& = &  (-1)^{p}\,\widetilde \partial
_{\mu_{1}}\cdots\widetilde \partial _{\mu_{p}} \,({\cal
  A}^{p}f_{1})\,(\theta) 
\theta_{1}^{\mu_{1}}\cdots\theta_{1}^{\mu_{p}}\\ & = & [ {\cal A}^{p}\,
\widetilde \partial _{\mu_{1}}\cdots\widetilde \partial  
_{\mu_{p}}\, f_{1} (\theta)]\,
\theta_{1}^{\mu_{1}}\cdots\theta_{1}^{\mu_{p}}\nonumber \label{B.4}
\end{eqnarray}
Thus we arrive at 
\begin{eqnarray}
&&(f_{1}\circ f_{2})(\theta)= \nonumber \\ 
&&\epsilon_{n}\left( \frac{\hbar}{2i} \right)^{n} \sum_{p=0}^{n}
\left(\frac{1}{p!}\right)^{2} [ {\cal A}^{p} \widetilde \partial
_{\mu_{1}}\cdots\widetilde \partial _{\mu_{p}} f_{1} (\theta)]
\,I_{\nu_{1}\cdots\nu_{p}}^{\mu_{1}\cdots\mu_{p}}\, [\widetilde \partial
^{\nu_{1}}\cdots\widetilde \partial ^{\nu_{p}} f_{2} (\theta)] \label{B.5}
\end{eqnarray}
with
\begin{eqnarray}
I_{\nu_{1}\cdots\nu_{p}}^{\mu_{1}\cdots\mu_{p}}=\int {\rm exp}
(\frac{2}{\hbar} \theta_{1}\theta_{2}) \,\theta_{1}^{\mu_{1}}\cdots
\theta_{1}^{\mu_{p}}\,\theta_{2\nu_{1}}\cdots\theta_{2\nu_{p}}\, 
d^{n}\theta_{1}d^{n}\theta_{2} 
\label{B.6}
\end{eqnarray}
Upon expanding the exponential, only the term of order $n-p$ can
contribute to the integral:
\begin{eqnarray}
I_{\nu_{1}\cdots\nu_{p}}^{\mu_{1}\cdots\mu_{p}}&=&\frac{1}{(n-p)!} \left(
  \frac{2}{\hbar} \right)^{n-p}
J_{\nu_{1}\cdots\nu_{p}}^{\mu_{1}\cdots\mu_{p}}  \label{B.7}\\
J_{\nu_{1}\cdots\nu_{p}}^{\mu_{1}\cdots\mu_{p}}&\equiv& \int
(\theta_{1}^{\alpha}\theta_{2 \alpha})^{n-p}\, \theta_{1}^{\mu_{1}}
\cdots\theta_{1}^{\mu_{p}}\, \theta_{2 \nu_{1}}\cdots\theta_{2
  \nu_{p}}\,d^{n}\theta_{1} d^{n}\theta_{2} \label{B.8}
\end{eqnarray}
For symmetry reasons the tensor $J$ must have the structure
\begin{eqnarray}
J_{\nu_{1}\cdots\nu_{p}}^{\mu_{1}\cdots\mu_{p}}& = & \lambda (n,p)\,
\delta^{[\mu_{1}}_{\nu_{1}}\cdots \delta^{\mu_{p}]}_{\nu_{p}} \nonumber \\
& = &\frac{\lambda(n,p)}{p!} \sum_{\pi \in {\cal S}_{p}} {\rm sign}(\pi)
\, \delta^{\pi(\mu_{1})}_{\nu_{1}}\delta^{\pi(\mu_{2})}_{\nu_{2}}
\cdots\delta^{\pi(\mu_{p})}_{\nu_{p}}  
\label{B.9}
\end{eqnarray}
where ${\cal S}_{p}$ is the symmetric group of $p$ objects. The
constants $\lambda(n,p)$ are most easily determined by choosing the
special index combination $J_{12\cdots p}^{12\cdots p}$ for which only
the identical permutation contributes in (\ref{B.9}). Furthermore, the
summation over $\alpha$ in (\ref{B.8}) is restricted to $\alpha > p$
then:
\begin{eqnarray}
&&\!\!\!\!\!\!\!\!\!\!\!\!\!\!\!\!\!\!\lambda (n,p)= \\ 
=&& \!\!\!\!\!\!\!\!\! p! \int [ \theta_{1}^{p+1} 
\theta_{2}^{p+1}+\cdots+ \theta_{1}^{n} \theta_{2}^{n}]^{n-p} 
\,( \theta_{1}^{1}\theta_{1}^{2}\cdots\theta_{1}^{p})\,
(\theta_{2}^{1}\theta_{2}^{2}\cdots\theta_{2}^{p})\,d^{n} 
\theta_{1} d^{n}\theta_{2} \nonumber \\
 =&&\!\!\!\!\!\!\!\!\!  p! (n-p)! \int 
(\theta_{1}^{1}\theta_{1}^{2}\cdots\theta_{1}^{p})\,
[\theta_{1}^{p+1}\theta_{2}^{p+1}\theta_{1}^{p+2}
\theta_{2}^{p+2}\cdots\theta_{1}^{n}\theta_{2}^{n}]\,
(\theta_{2}^{1}\theta_{2}^{2}\cdots\theta_{2}^{p})\,d^{n}\theta_{1}
d^{n}\theta_{2} \label{B.10}\nonumber
\end{eqnarray}
Commuting the $\theta$'s next to the corresponding $d\theta$'s produces
various sign factors so that finally
\begin{eqnarray}
\lambda (n,p)=(-1)^{n}(-1)^{n(n-1)/2}(-1)^{p(p-1)/2}\,p!(n-p)! \label{B.11}
\end{eqnarray}
If we note that $\epsilon_{n}i^{n}(-1)^{n(n-1)/2}=1$ both for $n$ even
and $n$ odd, we see that
\begin{eqnarray}
\epsilon_{n} \left( \frac{\hbar}{2i} \right)^{n} \frac{1}{p!}\,
I_{\nu_{1}\cdots\nu_{p}}^{\mu_{1}\cdots\mu_{p}}=(-1)^{p(p-1)/2} \left(
  \frac{\hbar}{2}\right)^{p} \delta^{[\mu_{1}}_{\nu_{1}}\cdots
\delta^{\mu_{p}]}_{\nu_{p}} \label{B.12}
\end{eqnarray}
Inserting (\ref{B.12}) into (\ref{B.5}) we obtain precisely the final
result given in eq.(\ref{2.15}) of the main text.

The representation (\ref{2.16}) follows from (\ref{2.15}) by using
(\ref{A.12}) in order to convert the left derivatives which act on
$f_{1}$ to right derivatives. One also needs (\ref{A.8}) to switch from
the index sequence $(\mu_{1},\mu_{2},\cdots,\mu_{p})$ to
$(\mu_{p},\mu_{p-1},\cdots,\mu_{1})$.

The last representation, eq.(\ref{2.17}), follows from (\ref{2.16}) by
commuting left and right derivatives with the same index next to each
other. No sign factor is picked up during this reshuffling.
\end{appendix}


\begin{thebibliography}{99}
\bibitem{ka} E. K\"ahler, Rend. Math. Ser. V, 21 (1962) 425;
\bibitem{banks} T. Banks, Y. Dothan, D. Horn, Phys. Lett. 117B (1982)
  413; \bibitem{bj} P. Becher, H. Joos, Z. Phys. C 15 (1982) 343;
\bibitem{joos} H. Joos, Helv. Phys. Acta 63 (1990) 670;\\
  Nucl. Phys. B. (Proc. Suppl.) 17 (1990) 704;\\ H. Joos, S.I. Azakov,
  Helv. Phys. Acta 67 (1994) 723; \bibitem{ara} H. Aratyn, A. H.
  Zimerman, Phys. Lett. 137B (1984) 392; \bibitem{tuck} I.M. Benn, R.W.
  Tucker, Commun. Math. Phys. 89 (1983) 341;
\bibitem{linha} C.A. Linhares, J.A. Mignaco, M.A. Rego Monteiro,\\
  Lett. Math. Phys. 10 (1985) 79; \bibitem{shim} M. Shimono, Prog.
  Theor. Phys. 84 (1990) 331; \bibitem{dil} H. Dilger, H. Joos, Nucl.
  Phys. B (Proc. Suppl.) 34 (1994) 195; \bibitem{at} M. Atiyah, {\it
    Vector Fields on Manifolds}, Arbeitsgemeinschaft f\"ur Forschung\\ 
  des Landes Nordrhein-Westfalen, Heft 200, 1970; \bibitem{graf} W.
  Graf, Ann. Inst. H. Poincar\'e, A29 (1978) 85; \bibitem{mm} I.
  Montvay, G. M\"unster, {\it Quantum Fields on a Lattice}, \\ Cambridge
  University Press, 1994;
\bibitem{kosu} J. Kogut, L. Susskind, Phys. Rev. D11 (1975) 395;\\
  L. Susskind, Phys. Rev. D16 (1977) 3031; \bibitem{kost} B. Kostant,
  Symposia Mathematica 14 (1974) 139; \bibitem{wood} N. Woodhouse, {\it
    Geometric Quantization}, Clarendon Press, Oxford, 1992; \bibitem{lj}
  R.G. Littlejohn, Phys. Rep. 138 (1986) 193; \bibitem{pct} E. Gozzi, M.
  Reuter, Nucl. Phys. B325 (1989) 356; \bibitem{green} M.B. Green, C.M.
  Hull, Phys. Lett. B 225 (1989) 57 and B 229 (1989)\\ 215; I. Bars, R.
  Kallosh, Phys. Lett. B233 (1989) 117; \bibitem{compean} H.
  Garc\'ia-Compe\'an, Preprint IASSNS-HEP-98/36, hep-th/9804188;
\bibitem{castano} D.J. Castano, Phys. Lett. B 269 (1991) 345;
\bibitem{metaqm} M. Reuter, Int. J. Mod. Phys. A 13 (1998) 3835 and
  hep-th/9804036 \bibitem{meta1} E. Gozzi, M. Reuter, J. Mod. Phys. A:
  Math. Gen. 26 (1993) 6319; \bibitem{meta2} M. Reuter, Int. J. Mod.
  Phys. A 10 (1995) 65; \bibitem{flato} F. Bayen, M. Flato, C. Fronsdal,
  A. Lichnerowicz, D. Sternheimer,\\ Ann. Phys. (NY) 111 (1978) 64 and
  111; \bibitem{ber} F.A. Berezin, Sov. Phys. Usp. 23 (1980) 763;
\bibitem{bm} F.A. Berezin, M.S. Marinov, Ann. Phys. (NY) 104 (1977) 336;
\bibitem{ht} M. Henneaux, C. Teitelboim, {\it Quantization of Gauge
    Systems}, \\ Princeton University Press, 1992; \bibitem{winf} E.
  Gozzi, M. Reuter, Int. J. Mod. Phys. A 9 (1994) 5801; \bibitem{fed} B.
  Fedosov, J. Diff. Geom. 40 (1994) 213; \bibitem{fedbook} B. Fedosov,
  {\it Deformation Quantization and Index Theory}, \\ Akademie Verlag,
  Berlin, 1996; \bibitem{frad} E.S. Fradkin, V.Y. Linetsky, Nucl. Phys.
  B 431 (1994) 569, \\ Nucl. Phys. B444 (1995) 577; \bibitem{gro} A.
  Grossmann, Commun. Math. Phys. 48 (1976) 191; \bibitem{roy} A. Royer,
  Phys. Rev. A 15 (1977) 449; \bibitem{heat} N. Berline, E. Getzler, M.
  Vergne, {\it Heat Kernels and Dirac Operators}, Springer, New York,
  1992; \bibitem{bert} R. A. Bertlmann, {\it Anomalies in Quantum Field
    Theory}, \\ Oxford University Press, 1996; \bibitem{kvich} M.
  Kontsevich, Preprint q-alg/9709040; \bibitem{emm} C. Emmrich, A.
  Weinstein, Preprint hep-th/9311094; \bibitem{fedrec} C. Emmrich, H.
  R\"omer, Acta Phys. Polon. B 27 (1996) 2393; \\ Preprint
  hep-th/9701111
\bibitem{rom} M. Bordemann, S. Waldmann, q-alg/9605012;\\
  M. Bordemann, q-alg/9605038; \\
  M. Bordemann, N. Neumaier, S. Waldmann, Commun. Math. Phys. \\
  198 (1998) 363; q-alg/9711016 \bibitem{gel} I. Gelfand, V. Retakh, M.
  Shubin, Preprint dg-ga/9707024;
\bibitem{hab} K. Habermann, Commun. Math. Phys. 184 (1997) 629 \\
  and references therein; \bibitem{dipl} J. Kr\"uger, Diploma Thesis,
  University of Hamburg, 1985;
\bibitem{local} T. Wyrozumski, Phys. Rev. D42 (1990) 1152;\\
  E. Prugovecki, Class. Quantum Grav. 13 (1996) 1007 and references
  therein; \\
  W. Drechsler, P. Tuckey, Class. Quantum Grav. 13 (1996) 611;\\ D.
  Graudenz, gr-qc/9412013 and hep-th/9604180;\\ B. Iliev,
  quant-ph/9804062 and references therein; \bibitem{mack} G. Mack, V.
  Schomerus, in: {\it Infinite dimensional geometry, \\ noncommutative
    geometry, operator algbras, fundamental interactions,} \\ R.
  Coquereaux, M. Dubois-Violette, P. Flad
  (Eds.), \\
  World Scientific, Singapore, 1995;\\ G. Mack, in: {\it
    Salamfestschrift}, A. Ali, J. Ellis, S. Randjbar-Daemi (Eds.), World
  Scientific, Singapore, 1994;
\bibitem{gozzi} E. Gozzi, Phys. Lett. A 202 (1995) 330;\\
  Nucl. Phys. (Proc. Suppl.) 57 (1997) 223;
\end{thebibliography}
\end{document}